\documentclass [aps,prb,showpacs,amsfonts,eqsecnum,twocolumn]{revtex4}

\usepackage{epsfig}
\usepackage{bm}

\newcommand{\smfrac}[2]{\mbox{$\frac{#1}{#2}$}}

\begin{document}
\title{The thermoballistic approach to charge carrier transport in
semi\-con\-duc\-tors\\[0.0cm]}

\author{R.\ Lipperheide}
\author{U.\ Wille}
\email{wille@helmholtz-berlin.de}

\affiliation{Helmholtz-Zentrum Berlin f\"{u}r Materialien und Energie
(formerly Hahn-Meitner-Institut Berlin), Lise-Meitner-Campus Wannsee,\\
Hahn-Meitner-Platz 1, D-14109 Berlin, Germany}

\date{\today}

\begin{abstract}

A comprehensive survey is given of the thermoballistic approach to charge 
carrier transport in semiconductors.  This semiclassical approach bridges the 
gap between the
drift-diffusion and ballistic (``thermionic'') models of carrier transport,
whose validity is limited to the range of very small and very large values,
respectively, of the carrier mean free path.  The physical concept underlying
the thermoballistic approach, while incorporating basic features of the
drift-diffusion and ballistic descriptions, constitutes a novel, unifying
scheme.  It is based on the introduction of ``ballistic configurations''
defined by a random partitioning of the length of a semiconducting sample into
ballistic transport intervals.  The points linking adjacent ballistic intervals
are assumed to be points of local thermodynamic equilibrium characterized by a
local chemical potential.  Carriers thermally emitted at any such point are
ballistically transmitted across either interval, while at the same time
carriers transmitted from the equilibrium points next to it are ``absorbed'' at
that point, i.e., they are assumed to be instantaneously equilibrated there.
During their transmission, the carriers face, in general, potential energy
barriers arising from internal and external electrostatic potentials in the
sample.  The lengths of the ballistic intervals are stochastic variables, with
associated probabilities given by the probabilities for carriers to traverse an
interval without collisions with the scattering centers randomly distributed
over the sample.  These probabilities are controlled by the carrier mean free
path, whose magnitude is arbitrary.  By averaging the ballistic carrier
currents over all ballistic configurations, a position-dependent
thermoballistic current is derived, which is the key element of the
thermoballistic concept and forms the point of departure for the calculation of
various transport properties.

The present article starts out with a preparatory account of the standard
drift-diffusion and ballistic transport models which form the cornerstones of
the thermoballistic concept, and of a prototype model which paves the way for
the fully developed form of that concept.  In the main body of the article, a
coherent exposition of the thermoballistic approach is given within a
general formulation that takes into account arbitrarily shaped, spin-split
potential energy profiles and spin relaxation during the carrier motion across
ballistic intervals.  The calculational procedures devised for implementing the
thermoballistic concept are described.  Specific examples relevant to
present-day semiconductor and spintronics research are considered.

\end{abstract}


\maketitle

\tableofcontents

\section{Introduction}

\label{sec:intro}

The phenomenon of electric conduction in metals and semiconductors has been a
prominent research topic ever since the early days of solid-state physics.  The
idea that electric currents flowing inside solid materials are effected by the
transport of ``small'' charged particles (``charge carriers'') was first
conceived by Weber.\cite{web75,web94} Following the discovery of the electron
by Thomson,\cite{tho97} Weber's idea quickly found its concrete expression in
attempts to understand electric (as well as thermal) conduction in solids as a
manifestation of electron transport.

The basic concept for a theoretical treatment of conduction in terms of the
motion of individual carriers was outlined by Riecke.\cite{rie98} Relying to
some extent on this concept, Drude\cite{dru00} formulated his celebrated
transport model, which subsequently was refined by Lorentz.\cite{lor09} In
Drude's model, the atomistic picture of matter and the kinetic theory of gases
are combined to describe conduction in terms of a homogeneous gas of
non-interacting, mobile charge carriers in thermodynamic equilibrium, which are
assumed to move against a background of spatially fixed, heavy atoms.  When an
external electric field is applied, the interplay of field-induced acceleration
and subsequent thermalizing collisions with the heavy atoms gives rise to a
``drift current'' of the carriers. The magnitude of this current is determined 
by the ``mean free path'', i.e., the average distance the carriers  
travel between two collisions. For the drift current to be a valid concept, the
mean free path must be very small as compared with typical dimensions of the 
sample. While originally conceived to describe carrier transport in metals,
Drude's model later on has been frequently used in qualitative, or
semi-quantitative, analyses of transport properties of semiconductors as well.

To overcome the shortcomings of Drude's model in the quantitative description
of carrier transport in semiconductors, particularly in inhomogeneously doped
systems, the drift current was supplemented with a diffusion
current,\cite{wag33,fre35} whereby Drude's model has been extended to the
``drift-diffusion model'' of transport.  The latter model, while representing a
substantial improvement over Drude's model and serving as a benchmark of
semiclassical transport theories even in modern times, is again valid in the
range of very small carrier mean free paths only.  In the opposite case of very
large mean free paths, carrier transport in semiconductors can be described in
terms of the ballistic (``thermionic'') 
model,\cite{ric03,ric29,som28,som33,bet42} in which carriers thermally emitted
at the ends of a sample are assumed to traverse it without collisions with the 
background atoms.

Until recently, no systematic attempts had been made to bridge the gap between
the limiting cases of the drift-diffusion and ballistic descriptions within a
unified approach.  In view of this situation, we set out to develop the
``thermoballistic approach''\cite{foot1,has71} to carrier transport in 
semiconductors.  Apart from perceiving the challenge to fill a long-standing
gap in the theory of carrier transport in semiconductors, we found that recent
progress in device physics, spintronics, and photovoltaics called for an
extension of the theoretical framework hitherto available for analyzing
experimental data in these fields. 

The physical concept underlying the thermoballistic approach rests on a random
partitioning of the length of a semiconducting sample into ``ballistic
transport intervals''.  Here, ``random'' implies (i) an arbitrary
number of intervals and (ii) arbitrary positions of the end-points, and thus
arbitrary lengths, of the intervals.  [The drift-diffusion model may be viewed
as implying a partitioning of the sample length into infinitesimally short
ballistic intervals;\ in the ballistic transport model, on the other hand, the
sample length constitutes a single ballistic interval.] An individual partition
defines what we call a ``ballistic configuration''.  The points linking
adjacent intervals in a ballistic configuration are assumed to be points of
local thermodynamic equilibrium characterized by a local chemical potential.
Any such point acts both as a source of, and a sink for, carriers.  That is, on
the one hand, carriers are thermally emitted there, with a velocity
distribution determined by the local temperature;\ subsequent to their
emission, the carriers are ballistically transmitted across either interval
to the left and right, facing, in general, potential energy barriers arising
from the internal and external potentials inside the sample.  On the other 
hand, carriers emitted at the two equilibrium points neighboring 
the point under consideration on either side and transmitted to it are 
``absorbed'' there, i.e., they return to thermodynamic equilibrium 
instantaneously.  This equilibration is assumed to result from  
collisions of the carriers with spatially fixed scattering centers 
(``impurities'') randomly distributed over the sample, a view adopted from 
Drude's transport model.

In the manner by which the ballistic intervals have been introduced, the
lengths of these intervals are stochastic variables, with associated
probabilities given by the probabilities for carriers to traverse an interval
without impurity scattering.  These probabilities are governed by the carrier
mean free path, which is allowed to have arbitrary magnitude.  By averaging the
(in general, spin-dependent) carrier currents in the individual ballistic
intervals over all ballistic configurations, a position-dependent total
(i.e., spin-summed) thermoballistic current as well as a thermoballistic 
spin-polarized current are derived.  These currents, in conjunction with the
associated ballistic densities, represent the key elements of the
thermoballistic approach.  They form the point of departure for the calculation
of the spin-resolved equilibrium chemical potentials.  From the latter
functions, in turn, transport quantities, like the current-voltage
characteristic, the magnetoresistance, and (position-dependent) current and
density spin polarizations, are obtained in terms of the potential energy
profile, the mean free path, the ballistic spin relaxation length, and other
parameters characterizing the sample.

The present article is devoted to a comprehensive exposition of the physical
concept underlying the thermoballistic approach and its detailed 
implementation.  We begin by describing a prototype model of thermoballistic 
transport (the ``prototype model'', for short).  This model introduces 
averaging over ballistic configurations as the constitutive element of the 
thermoballistic concept. It is based on the simplifying assumption that the 
carrier current is conserved across the points of local thermodynamic 
equilibrium linking the ballistic transport intervals. This assumption allows 
the derivation of an explicit, transparent expression for the current-voltage 
characteristic, as well as the construction of a local chemical 
potential in a heuristic form.  After describing the prototype model,
we present a general formulation of the thermoballistic concept proper, which 
makes use of a local chemical potential as the essential dynamical 
quantity, and takes into account arbitrarily shaped, spin-split potential 
energy profiles as well as spin relaxation during ballistic carrier motion.

Throughout this article, the formulation deviates in many respects from that in
our original publications.  Apart from correcting flaws and inconsistencies, we
introduce  modifications and extensions of the formalism that make it
more transparent and more comprehensible.  Aiming at a self-contained, unified 
presentation, we will develop the elements of the thermoballistic concept in 
considerable detail.  This includes a recollection of early attempts to 
describe charge carrier transport in semiconductors and, at one place or  
another, the coherent recapitulation of background material which otherwise can
be found only scattered over textbooks.

While the thermoballistic approach, in its most general form, would allow a
consistent treatment of three-dimensional, bipolar carrier transport in 
semiconducting systems, we confine ourselves here to the case of 
one-dimensional, unipolar transport throughout.  To be specific, we consider 
electron transport in the conduction band of n-doped systems, with the 
understanding that all results for hole transport in the valence band of 
p-doped systems can be obtained by transcribing the results for electron 
transport in an obvious way. In implementing the thermoballistic approach, we 
aim at carrying the development to the point where we obtain explicit equations
from which the relevant physical quantities can be calculated.
We do not enter here into applications requiring numerical calculations. 
Pertinent results presented in our previous publications will be  briefly
mentioned at various places in this article.     

The organization of this article is as follows.  In the next section, we
present an account of Drude's transport model and the standard
drift-diffusion and ballistic models.  In Sec.~\ref{sec:gendrift}, we begin
with an outline of the probabilistic approach to carrier transport, which is
prerequisite to the formulation of the thermoballistic model.  This is followed
by the description of the prototype model.  In Sec.~\ref{sec:therconc}, the
thermoballistic concept proper is developed at length, where particular 
emphasis is placed on the inclusion of spin degrees of freedom.  
The spin-resolved ballistic currents and associated
densities are introduced in terms of an ``average chemical potential'' and a
``spin accumulation function'' connected with the splitting of the
spin-resolved chemical potentials.  Total and spin-polarized thermoballistic
currents and densities are constructed by averaging the corresponding ballistic
quantities over the ballistic configurations, and  
are evaluated in the drift-diffusion regime and in the ballistic limit.
Energy dissipation (``heat production'') is analyzed within the thermoballistic
approach. In Sec.~\ref{sec:thersyno}, in particular,
a synopsis of the thermoballistic concept is presented, with emphasis on its 
physical content as well as its merits and weaknesses. The procedures devised 
for the implementation of the thermoballistic concept, i.e., for the explicit
calculation of the average chemical potential and of the spin accumulation
function as functions of intrinsic and external physical parameters, are 
described in Sec.~\ref{sec:therimpl}.  Expressions are derived for
the current and density spin polarizations and the magnetoresistance in terms
of the values of the spin accumulation function at the boundaries of a
semiconducting sample. As specific examples  of current interest in 
semiconductor and spintronics research, we  treat  spin-polarized transport
in heterostructures formed of a nonmagnetic semiconductor and two ferromagnetic
contacts, and spin-polarized transport in heterostructures involving diluted
magnetic semiconductors in their paramagnetic phase.  Finally,
in Sec.~\ref{sec:summconc}, we summarize the contents of this article and give
an outlook towards future developments.

\section{Drift-diffusion and ballistic transport}

\label{sec:dibatran}

As the thermoballistic approach is devised to bridge the gap between the
drift-diffusion and ballistic descriptions of transport, we survey, in this
section, the standard formulations of these two limiting cases.

\subsection{Drift-diffusion transport}

The drift-diffusion model is an extension of the transport model of Drude, so
that we begin here with an account of the latter. Drude's model is
far from being able to describe transport properties of semiconductors 
quantitatively;\ nevertheless, its exposition provides us with the opportunity 
to introduce and discuss the basic notions on which classical and semiclassical
transport theories rely, and which will appear ubiquitously throughout this 
article.

\subsubsection{Drude's model}

\label{sec:maxdrude}

In the transport model of Drude,\cite{dru00,ash76,sap95} one considers a 
(three-dimen\-sion\-al) homogeneous classical gas of non-inter\-acting 
electrons in thermodynamic equilibrium at temperature $T$, which is subjected 
to an externally applied, constant electric field ${\bf E}$.  The 
electrons collide with randomly
distributed, spatially fixed scattering centers (``impurities'').  The
collisions are assumed to cause the electrons to return instantaneously to
complete thermodynamic equilibrium.  The average vectorial velocity of the
electrons emerging from this equilibration is equal to zero.  Their mean
speed $u$, i.e., the thermal average of the magnitude of the electron
velocity as derived from the three-dimensional Maxwell-Boltzmann velocity
distribution function, is given by
\begin{equation}
u = \left( \frac{8}{\pi m^{\ast} \beta } \right) ^{1/2} ,
\label{eq:drude001}
\end{equation}
where $m^{\ast}$ is the effective electron mass, and $\beta = 1/k_{B} T$, with
$k_{B}$ the Boltzmann constant. [Note that in Drude's original
paper,\cite{dru00} the mean electron speed was derived from Boltzmann's
equipartition theorem;\ the Maxwell-Boltzmann velocity distribution was
introduced into the description of electron transport in metals by
Lorentz.\cite{lor09}]

Defining the {\em mean free path} (or {\em momentum relaxation length}) $l$ as
the average distance (measured along the transport direction) travelled by the 
electrons between two collisions, one finds for the {\em collision time} (or 
{\em relaxation time}) $\tau$, defined as the average time-of-flight between 
successive collisions,
\begin{equation}
\tau = \frac{l}{u} .
\label{eq:drude002}
\end{equation}
The mean free path $l$ characterizes a set of collision-free {\em ballistic
intervals} of average length $l$, across which electrons move ballistically
under the influence of the external field ${\bf E}$.  On average, they acquire
a velocity, the {\em drift velocity} ${\bf v}_{dr}$, given by the acceleration 
$- e {\bf E}/m^{\ast}$ times their average time-of-flight across the ballistic
intervals, i.e., times the collision time $\tau$,
\begin{equation}
{\bf v}_{dr} = - \frac{e {\bf E}}{m^{\ast}} \tau \equiv - \nu {\bf E} ,
\label{eq:drude003}
\end{equation}
where
\begin{equation}
\nu \equiv \frac{e}{m^{\ast}} \tau = \frac{e}{m^{\ast}} \frac{l}{u} = e \left(
\frac{\pi \beta}{8 m^{\ast}} \right) ^{1/2} l
\label{eq:drude004}
\end{equation}
is the electron mobility.

In the picture of Drude, the charge current density ${\bf{j}}$ (current, for
short) is a {\em drift current}, ${\bf{j}}_{dr}$,
driven by the external electric field ${\bf E}$,
\begin{equation}
{\bf{j}} \equiv {\bf{j}}_{dr} = - e n {\bf v}_{dr} = e n \nu {\bf E} ,
\label{eq:drude004a}
\end{equation}
where $n$ is the electron density.  Since, in Drude's model, $n$ as well as the
field ${\bf E}$ are independent of position, ${\bf j}$ is constant.  In terms
of the electron conductivity
\begin{equation}
\sigma = e \nu n ,
\label{eq:drude006}
\end{equation}
the current ${\bf{j}}$ is expressed as
\begin{equation}
{\bf{j}} =  \sigma {\bf E} ,
\label{eq:drude005}
\end{equation}
i.e., in the form of Ohm's law.

The transport mechanism in Drude's model can be elucidated\cite{sap95} by 
noting that Eq.~(\ref{eq:drude003}), when written as
\begin{equation}
- e {\bf E} - \frac{m^{\ast}}{\tau} {\bf v}_{dr} = 0 ,
\label{eq:drude009}
\end{equation}
states that the effect of the external force $- e {\bf E}$ is balanced, on
average, by that of a {\em friction force} $- \gamma {\bf v}_{dr}$, with
the friction coefficient $\gamma$ given by
\begin{equation}
\gamma = \frac{m^{\ast}}{\tau} = \frac{e}{\nu} .
\label{eq:drude010}
\end{equation}
The friction force reflects the reset to zero of their individual velocities
when, subsequent to their acceleration through the external field, the
electrons are thermally equilibrated due to impurity scattering.

Limits on the range of validity of Drude's model are set by the requirement
that the perturbation of the electron gas due to the external field is
sufficiently small so that the gas stays close to thermodynamic equilibrium.
This condition can be met by requiring the magnitude of the drift velocity to
be very small as compared with the mean speed of the electrons,
\begin{equation}
|{\bf v}_{dr}| \ll u .
\label{eq:drude004c}
\end{equation}
Using Eqs.~(\ref{eq:drude001}), (\ref{eq:drude003}), and (\ref{eq:drude004}),
condition (\ref{eq:drude004c}) can be re-expressed in a form that restricts,
for given mean free path $l$, the magnitude of the electric-field vector 
${\bf E}$,
\begin{equation}
|{\bf E}| \ll \frac{8}{\pi \beta e l} .
\label{eq:drude004b}
\end{equation}
For a homogeneous sample of length $S$ subjected to an external voltage bias 
$V$, when $|{\bf E}| = |V|/S$, one has
\begin{equation}
|V| \ll \frac{8 S}{\pi \beta e l} .
\label{eq:drude004d}
\end{equation}
Reversely, for given voltage bias $V$, the ratio of mean free path to sample
length is restricted by the condition
\begin{equation}
\frac{l}{S} \ll \frac{8}{\pi \beta e |V| } .
\label{eq:drude004e}
\end{equation}
For room temperature, i.e., $1/\beta \approx 0.025$ eV, and values of $e|V|$ of
a few tens of meV, which are typical for semiconducting devices, the right-hand
side of condition (\ref{eq:drude004e}) is of order unity.  Hence, one may use
the condition
\begin{equation}
\frac{l}{S} \ll 1
\label{eq:drude004f}
\end{equation}
as a rough criterion for delimiting the range of validity of the Drude model.
The value of $l$ remains small if the density of impurities is
sufficiently high.

\subsubsection{Drift-diffusion model}

\label{sec:diffmech}

Electron transport in inhomogeneous semiconductors is outside of the
scope of Drude's model.  The nonuniformity of the donor density in
inhomogeneous samples entails  a position dependence of the electron density, 
$n = n(x)$, and gives rise to an $x$-dependent internal (``built-in'')
electrostatic poten\-tial.\cite{ash76,boe90} [Throughout this article, we
consider {\em one-dimensional} transport in {\em three-dimensional}, 
``plane-parallel'' semiconducting samples, i.e., samples whose parameters do 
not vary in the directions perpendicular to the transport direction, which is 
taken as the $x$-direction (for a discussion of this assumption, see 
Sec.~\ref{sec:thersyno}). Further, the temperature $T$ is assumed to be 
constant across the sample.]

Transport in inhomogeneous systems near thermodynamic equilibrium can be 
described in terms of a {\em local}
thermodynamic equilibrium\cite{ash76} characterized by a {\em local chemical
potential} $\mu(x)$.\cite{chempo} Disregarding spin degrees of freedom and
adopting the effective-mass approximation,\cite{ash76,boe90} we have for
the equilibrium electron density $n(x)$ in a nondegenerate system 
\begin{eqnarray}
n(x) &=&  4 \pi \left( \frac{m^{\ast}}{h} \right)^{3} \int_{- \infty}^{\infty}
d v_{x} \int_{0}^{\infty} dw w \nonumber \\ &\ & \hspace{1.4cm} \times f^{MB}
\bm{(} E(v; x) - \mu (x) \bm{)} .
\label{eq:drude102}
\end{eqnarray}
Here, $h$ is Planck's constant, and we have introduc\-ed cylindrical
coordinates in three-dimensional velocity space (with Cartesian coordinates
$v_{x}, v_{y}, v_{z}$) such that
\begin{equation}
v = (v_{x}^{2} + w^{2})^{1/2} ,
\label{eq:drude090}
\end{equation}
with $w^{2} = v_{y}^{2} + v_{z}^{2}$. The function
\begin{equation}
f^{MB}(E) = e^{- \beta E}
\label{eq:drude101}
\end{equation}
is the Maxwell-Boltzmann energy distribution function, and
\begin{equation}
E(v; x) = \epsilon (v) + E_{c}(x)
\label{eq:drude103}
\end{equation}
is the total electronic energy at the equilibrium point $x$, with the kinetic
energy
\begin{equation}
\epsilon (v) = \frac{m^{\ast}}{2} v^{2} .
\label{eq:drude203}
\end{equation}
The potential energy profile $E_{c}(x)$ comprises the conduction band edge
potential (which includes the position-dependent internal potential) and the
external electrostatic potential.  In general, $E_{c}(x)$ exhibits a
``multiple-barrier'' shape associated with local maxima of the profile.
Evaluation of the integrals in Eq.~(\ref{eq:drude102}) results in the standard
form for the density,
\begin{equation}
n(x) = N_{c} e^{-\beta[E_{c}(x) - \mu(x)]} ,
\label{eq:drude014}
\end{equation}
where
\begin{equation}
N_{c} = 2 \left( \frac{2 \pi m^{\ast}}{ \beta h^{2}} \right) ^{3/2}
\label{eq:diff001a}
\end{equation}
is the effective density of states at the conduction band edge.\cite{ash76}

The effects of the spatial variation of the electron density and of the
associated occurrence of an internal potential in inhomogeneous semiconductors
can be described within an extension of Drude's model, the {\em
drift-diffusion model}.  In this model, a generalized, $x$-dependent drift
current $j_{dr}(x)$ is supplemented\cite{wag33,fre35,bar67,ash76} by a {\em
diffusion current} $j_{di}(x)$ proportional to the density gradient along the
$x$-axis, so that the total (conserved) current $j$ is given by
\begin{equation}
j = j_{dr}(x) + j_{di}(x) .
\label{eq:drude008f}
\end{equation}
The generalized drift current is obtained from expression (\ref{eq:drude005})
by replacing (i) the conductivity $\sigma$ with the {\em local conductivity}
$\sigma (x)$ given by Eq.~(\ref{eq:drude006}), with $n(x)$ in lieu of $n$, and
(ii) the constant external electric field ${\bf E}$ with the field $ [d
E_{c}(x)/dx]/e$ associated with the potential energy profile $E_{c}(x)$, so
that one obtains
\begin{equation}
j_{dr}(x) = \frac{1}{e} \sigma (x) \frac{d E_{c}(x)}{dx} .
\label{eq:drude007a}
\end{equation}
The diffusion current is written as
\begin{equation}
j_{di}(x) = e D \frac{d n(x)}{dx} ,
\label{eq:drude007}
\end{equation}
where $D$ is the diffusion coefficient, which is related to the
electron mobility $\nu$ via the Einstein relation\cite{ash76,sap95}
\begin{equation}
D = \frac{\nu}{\beta e} = \frac{\tau}{m^{\ast} \beta} ,
\label{eq:drude011}
\end{equation}
and $\nu$ is now assumed to have the form corresponding to one-dimensional
transport,
\begin{equation}
\nu  = e \left( \frac{2 \beta}{\pi m^{\ast}} \right)^{1/2} l
\label{eq:drude004hg}
\end{equation}
[see Eq.~(\ref{eq:drude004}) for the three-dimensional form of $\nu$]. Then,
generalizing Eq.~(\ref{eq:drude006}), one has
\begin{equation}
\sigma (x) \equiv e \nu n(x) = \beta e^{2} D n(x) .
\label{eq:drude006f}
\end{equation}
For the total current $j$, one now finds from
Eqs.~(\ref{eq:drude008f})--(\ref{eq:drude007}), using Eq.~(\ref{eq:drude006f}),
\begin{equation}
j = \frac{1}{e} \sigma (x) \left[ \frac{d E_{c}(x)}{dx} + \frac{1}{\beta n(x)}
\frac{d n(x)}{dx} \right] ,
\label{eq:drude008}
\end{equation}
which is the standard drift-diffusion expression for the total charge current.
For the derivation of this expression from Boltzmann's transport equation and
for its application in device simulation, see, e.g.,  Ref.~\onlinecite{jac10}.
In Ref.~\onlinecite{fab07}, the expression is derived within a time-dependent
tutorial treatment of diffusion in the presence of an external electric field
(``biased-random-walk model'').

Substituting expression (\ref{eq:drude014}) for the equilibrium density $n(x)$ 
in Eq.~(\ref{eq:drude008}), we obtain the current $j$ in the form
\begin{equation}
j = \frac{1}{e} \sigma (x) \frac{d \mu(x)}{dx} ,
\label{eq:diff001}
\end{equation}
which shows that, in the drift-diffusion model, it is the local chemical
potential which provides the driving force for electron transport. Using
Eqs.~(\ref{eq:drude014}) and (\ref{eq:drude006f}), we can rewrite 
Eq.~(\ref{eq:diff001}) as
\begin{equation}
\frac{\beta j}{\nu N_{c}} e^{\beta E_{c}(x)} = \frac{d}{dx} e^{\beta \mu (x)}
.
\label{eq:diff002}
\end{equation}
Considering a sample extending from $x_{1}$ to $x_{2}$ and integrating
Eq.~(\ref{eq:diff002}) over the interval $[x_{1}, x]$, we can express the local
chemical potential $\mu(x)$ in the form
\begin{equation}
e^{\beta \mu (x)} = e^{\beta \mu (x_{1})} + \frac{\beta j}{\nu N_{c}}
\int_{x_{1}}^{x} dx' e^{\beta E_{c}(x')} .
\label{eq:diff003}
\end{equation}
An equivalent representation of $\mu(x)$ is obtained by integrating
Eq.~(\ref{eq:diff002}) over the interval $[x, x_{2}]$.  Now, setting $x =
x_{2}$ in Eq.~(\ref{eq:diff003}), identifying the chemical potentials at the
end-points $x_{1,2}$ with the potentials $\mu_{1,2}$ in the contacts connected
to the semiconducting sample,
\begin{equation}
\mu(x_{1,2}) = \mu_{1,2} ,
\label{eq:diff005}
\end{equation}
and solving the resulting equation for the current $j$, we then obtain, using
Eq.~(\ref{eq:drude014}), the current-voltage characteristic of the
drift-diffusion model in the form
\begin{equation}
j = - n(x_{1}) e^{-\beta E_{b}^{l}(x_{1}, x_{2})} \frac{\nu}{\beta \tilde{S}}
(1 - e^{- \beta e V}) .
\label{eq:diff006}
\end{equation}
Here,
\begin{equation}
V = \frac{\mu_{1} - \mu_{2}}{e}
\label{eq:diff004}
\end{equation}
is the voltage bias, and
\begin{equation}
E_{b}^{l}(x_{1}, x_{2}) \equiv E_{c}^{m}(x_{1}, x_{2}) - E_{c}(x_{1}) \geq 0
\label{eq:diff007}
\end{equation}
is the maximum barrier height of the potential energy profile relative to its
value $E_{c}(x_{1})$ at the left end of the sample, where
$E_{c}^{m}(x_{1}, x_{2})$ is the overall maximum of $E_{c}(x)$ in the interval
$[x_{1}, x_{2}]$.  Finally, the quantity
\begin{equation}
\tilde{S} \equiv \int_{x_{1}}^{x_{2}} dx e^{- \beta [ E_{c}^{m}(x_{1}, x_{2}) -
E_{c}(x)]}
\label{eq:diff008}
\end{equation}
is the ``effective sample length''.  It has the appealing property of
becoming equal to the sample length $S = x_{2} - x_{1}$ for a flat profile,
i.e., if $E_{c}(x) = {\rm const.}$, and otherwise satisfies $\tilde{S} < S$.
Writing the current-volt\-age characteristic in the particular form
(\ref{eq:diff006}) facilitates comparison with analogous expressions given
below.

Expression (\ref{eq:diff006}) shows that the current-voltage characteristic of
the drift-diffusion model is controlled (i) by the ``barrier factor''
$e^{-\beta E_{b}^{l}(x_{1}, x_{2})}$, which involves the overall maximum of the
profile $E_{c}(x)$, and (ii) by the ratio $\nu/\tilde{S}$ or, owing to
Eq.~(\ref{eq:drude004hg}), by the ratio $l/\tilde{S}$, in which the effective
sample length reflects, in an integral way, the shape of $E_{c}(x)$.
[In the ballistic description of transport, the barrier factor $e^{-\beta
E_{b}^{l}(x_{1}, x_{2})}$ re-appears as the thermally averaged probability for
electron transmission from $x_{1}$ to $x_{2}$; see Eq.~(\ref{eq:ball005})
below.]

The drift-diffusion model is based on the assumption of a continuously varying
equilibrium chemical potential $\mu(x)$, which implies that the points of local
thermodynamic equilibrium lie arbitrarily dense.  Then, strictly speaking, the
mean free path $l$ must be confined to arbitrarily small values.  On the other 
hand, $l$ (or the mobility $\nu$) must be finite and large enough to give rise 
to a non-vanishing conductivity of a magnitude in the range of typical 
experimental values, which calls for a relaxation of the former condition.

To obtain a practical criterion for the range of validity of the
drift-diffusion model, one may require $l$ to be so small that the effective
number of points of local thermodynamic equilibrium along the length of the
sample is so large that the spatial variations in the potential energy profile
$E_{c}(x)$, and hence in the electron density $n(x)$, are ``resolved'' with
sufficient accuracy.  In terms of a local, $x$-dependent mean free path $l(x)$,
this requirement may be expressed as
\begin{equation}
l(x) \ll \Delta x ,
\label{eq:diff008a}
\end{equation}
where $\Delta x$ is the length of an interval, centered at the point $x$, over
which the relative variation of $E_{c}(x)$ is very small compared to unity,
i.e.,
\begin{equation}
\frac{1}{E_{c}(x)} \left| \frac{d E_{c}(x)}{dx} \right| \Delta x  \ll 1 .
\label{eq:diff008b}
\end{equation}
For constant mean free path $l$, one may fulfil the above requirement
in an overall way by adopting the condition
\begin{equation}
\frac{l}{\tilde{S}} \ll 1 .
\label{eq:diff009}
\end{equation}
The effective sample length $\tilde{S}$ defined by the integral
(\ref{eq:diff008}) tends to decrease exponentially when strong variations in
$E_{c}(x)$ are ``switched on''.  Condition (\ref{eq:diff009}) tightens
condition (\ref{eq:drude004f}) so as to permit the ``resolution'' of the
details of the profile $E_{c}(x)$.

\subsection{Ballistic (thermionic) transport}

\label{sec:ballmech}

In contrast to the drift-diffusion model, in which the points of local
thermodynamic equilibrium are assumed to lie arbitrarily dense, the ballistic
(thermionic) transport model presupposes the complete absence of such points
inside the sample.  Then, the electrons in the sample perform a 
collision-free {\em ballistic motion} in the field associated with the 
potential energy $E_{c}(x)$.  Without thermal equilibration, it is meaningless 
to speak of a local chemical potential. Only at the sample ends at $x_{1,2}$, 
the electrons are forced into equilibrium, with  densities
\begin{equation}
n(x_{1,2}) = N_{c} e^{-\beta [E_{c}(x_{1,2}) - \mu_{1,2}]}
\label{eq:ball000}
\end{equation}
determined by the boundary values of the potential energy profile,
$E_{c}(x_{1,2})$, and by the chemical potentials $\mu_{1,2}$  in
the contacts.  Here, the mean free path $l$, which has been of central 
importance in
Drude's model and in the drift-diffusion model, is effectively of infinite
length and does not appear in the formalism of the ballistic model.  Roughly
speaking, one may use the condition
\begin{equation}
\frac{l}{S} \gg 1
\label{eq:ball000a}
\end{equation}
to delimit the range of validity of the ballistic model.

\subsubsection{Nondegenerate case}

\label{sec:ballnonc}

In the ballistic model, the end-points $x_{1,2}$ of the sample are fixed points
of local thermodynamic equilibrium with chemical potentials $\mu(x_{1,2}) =
\mu_{1,2}$, out of which thermal electron currents are symmetrically emitted
towards the left and right, so that only {\em one half} of each of these 
currents are emitted towards the inner region of the sample.

For a nondegenerate system, the classical {\em electron current} $J^{l}(x_{1})$
emitted at the left end-point $x_{1}$ towards the right, say, is
expressed, in extension of the electron density $n(x)$ given by
Eq.~(\ref{eq:drude102}), in the form
\begin{eqnarray}
J^{l}(x_{1}) &=& 4 \pi \left( \frac{m^{\ast}}{h} \right)^{3} \int_{0}^{\infty}
d v_{x} v_{x} \int_{0}^{\infty} dw w \nonumber \\ &\ & \hspace{2.65cm} \times
f^{MB} \bm{(} E(v; x_{1}) - \mu_{1} \bm{)} \nonumber \\ &=& \frac{4 \pi m^{\ast
2}}{\beta h^{3}} \int_{0}^{\infty} d v_{x} v_{x} f^{MB} \bm{(} E(v_{x}; x_{1})
- \mu_{1} \bm{)} , \nonumber \\
\label{eq:ball001h}
\end{eqnarray}
from which we find, using Eq.~(\ref{eq:ball000}),
\begin{equation}
J^{l}(x_{1}) = v_{e} N_{c} e^{-\beta [E_{c}(x_{1}) - \mu_{1}]} =  v_{e}
n(x_{1}) .
\label{eq:ball001}
\end{equation}
Here,
\begin{eqnarray}
v_{e} &=& \left( \frac{m^{\ast} \beta}{2 \pi} \right)^{1/2} \int_{0}^{\infty}
dv_{x} v_{x} f^{MB}(m^{\ast} v_{x}^{2}/2) \nonumber \\ &=& \left( \frac{1}{2
\pi m^{\ast} \beta} \right) ^{1/2} 
\label{eq:ball002}
\end{eqnarray}
% t
is the {\em emission velocity}, which is actually equal to one half of the 
one-dimensional mean electron speed
\begin{equation}
u = \left( \frac{2}{\pi m^{\ast} \beta} \right) ^{1/2} ,
\label{eq:ball002hvl}
\end{equation}
the three-dimensional analogue of which is given by Eq.~(\ref{eq:drude001}).

The electrons emitted at $x_{1}$ with velocity component $v_{x}^{(1)}$ move
along ballistic trajectories, thereby conserving their total energy,
\begin{equation}
E(v_{x}; x) = E \bm{(} v_{x}^{(1)}; x_{1} \bm{)} ,
\label{eq:ball003}
\end{equation}
where $x$ is any point inside the interval $[x_{1}, x_{2}]$ (and is
not a point of local thermodynamic equilibrium).  The energy distribution
at $x$ is therefore equal to that at $x_{1}$.  However, only electrons with
total energy larger than the overall maximum $E_{c}^{m}(x_{1}, x_{2})$ of
$E_{c}(x)$ [see Eq.~(\ref{eq:diff007})] are classically able to reach the right
end-point of $[x_{1}, x_{2}]$ at $x_{2}$.  Thus, part of the current
$J^{l}(x_{1})$ emitted at $x_{1}$ will be reflected, and the electrons forming
it are absorbed when they return to their origin at $x_{1}$.  The other,
transmitted part $J^{l}(x_{1},x_{2}; x)$, called the (left) ``ballistic 
current'', is absorbed into the contact connected to the sample at $x_{2}$.  
Modifying expression (\ref{eq:ball001h}), we have for this part
\begin{eqnarray}
J^{l}(x_{1},x_{2}; x) &=& \frac{4 \pi m^{\ast 2}}{\beta h^{3}}
\int_{0}^{\infty} dv_{x} v_{x} f^{MB} \bm{(} E(v_{x}; x) - \mu_{1} \bm{)}
\nonumber \\ &\ & \hspace{1.6cm} \times \Theta \bm{(} E(v_{x}; x) - E_{c}^{m}
(x_{1}, x_{2}) \bm{)} . \nonumber \\
\label{eq:ball004}
\end{eqnarray}
In the integration, the potential energy profile $E_{c}(x)$ contained in the 
function $E(v_{x};x)$ drops out, and we obtain for the (left) ballistic current
the $x$-independent expression
\begin{equation}
J^{l}(x_{1},x_{2}) = v_{e} N_{c} e^{- \beta [E_{c}^{m}(x_{1},x_{2}) - \mu_{1}]}
,
\label{eq:ball004f}
\end{equation}
i.e., the ballistic current is conserved (independent of $x$), as expected.  
The (right) ballistic current $J^{r}(x_{1},x_{2})$ transmitted from the 
{\em right} end-point $x_{2}$ of the interval $[x_{1}, x_{2}]$ is given by
\begin{equation}
J^{r}(x_{1},x_{2}) = - v_{e} N_{c} e^{- \beta [E_{c}^{m}(x_{1},x_{2}) -
\mu_{2}]} ,
\label{eq:ball006}
\end{equation}
in analogy to Eq.~(\ref{eq:ball004f}).

The ballistic currents (\ref{eq:ball004f}) and (\ref{eq:ball006}) bear a close
analogy to the ``thermionic emission current'' associated with the evaporation
of electrons from a heated metal (``Richardson
effect'').\cite{ric03,ric29,som28,som33} In semiconductor physics, ``thermionic
emission'' was introduced as a mechanism of carrier transport by
Bethe\cite{bet42,sze81} in his treatment of electron transport across a
Schottky barrier.

Associated with the ballistic currents $J^{l,r}(x_{1},x_{2})$ are the
``ballistic densities'' $n^{l,r}(x_{1},x_{2}; x)$ of the electrons making up
the currents inside the ballistic interval $[x_{1}, x_{2}]$.  These
densities will turn out to be instrumental in establishing the spin-dependent
thermoballistic scheme (see Sec.~\ref{sec:balltrans}).

The density $n^{l}(x_{1},x_{2}; x)$ associated with the current
$J^{l}(x_{1},x_{2}; x)$ of Eq.~(\ref{eq:ball004}) is given by
\begin{eqnarray}
n^{l}(x_{1},x_{2}; x) &=& \frac{4 \pi m^{\ast 2}}{\beta h^{3}}
\int_{0}^{\infty} dv_{x} f^{MB} \bm{(} E(v_{x}; x) - \mu_{1}
\bm{)} \nonumber \\ &\ & \hspace{1.6cm} \times \Theta \bm{(} E(v_{x};
x) - E_{c}^{m} (x_{1}, x_{2}) \bm{)} , \nonumber \\
\label{eq:ball004k}
\end{eqnarray}
which is evaluated to yield
\begin{eqnarray}
n^{l}(x_{1},x_{2}; x) &=& \frac{N_{c}}{2} C^{m}(x_{1}, x_{2}; x)
e^{- \beta [E_{c}^{m} (x_{1}, x_{2}) - \mu_{1}]} . \nonumber \\
\label{eq:ball004ka}
\end{eqnarray}
Here,
\begin{eqnarray}
C^{m}(x_{1},x_{2}; x) &=& e^{\beta [E_{c}^{m} (x_{1}, x_{2}) - E_{c}(x)]}
\nonumber \\ &\ & \times {\rm erfc} \bm{(} \{ \beta [E_{c}^{m}(x_{1}, x_{2}) -
E_{c}(x)] \}^{1/2} \bm{)} , \nonumber \\
\label{eq:ball004kaa}
\end{eqnarray}
where the funct\-ion ${\rm erfc}(x)$ is the complementary error
function.\cite{abr65} The ballistic density is position-dependent via the
$x$-dependence of the function $C^{m}(x_{1},x_{2}; x)$, i.e., of the potential
energy profile $E_{c}(x)$.  The ballistic density $n^{r}(x_{1},x_{2}; x)$
associated with the current $J^{r}(x_{1},x_{2})$ is obtained by replacing
$\mu_{1}$ with $\mu_{2}$ in expression (\ref{eq:ball004ka}).

The function  $C^{m}(x_{1},x_{2}; x)$ determines the ``ballistic velocities''
\begin{eqnarray}
v^{l,r}(x_{1},x_{2}; x) &\equiv&
\frac{J^{l,r}(x_{1},x_{2})}{n^{l,r}(x_{1},x_{2}; x)} \nonumber \\ &=& \pm 
\frac{2 v_{e}}{C^{m}(x_{1}, x_{2}; x)} ,
\label{eq:ball004velo}
\end{eqnarray}
which have the same magnitude for the currents transmitted from the left and
right.  For constant potential energy profile, one has $C^{m}(x_{1}, x_{2}; x)
= 1$, and the electrons move with speed $2 v_{e}$, i.e, with the mean 
electron speed $u$ given by Eq.~(\ref{eq:ball002hvl}).  For position-dependent 
profiles, when $C^{m}(x_{1}, x_{2}; x) < 1$, the magnitude of the ballistic 
velocities is larger than $u$.

The {\em net ballistic current} $J(x_{1},x_{2})$ in the interval $[x_{1},
x_{2}]$,
\begin{equation}
J(x_{1},x_{2}) = J^{l}(x_{1},x_{2}) + J^{r}(x_{1},x_{2}) \equiv J
\label{eq:ball007}
\end{equation}
equals the (conserved) total current $J$, which we can express, using
Eqs.~(\ref{eq:ball004f}) and  (\ref{eq:ball006}),  as
\begin{equation}
J = v_{e} N_{c} e^{- \beta E_{c}^{m}(x_{1},x_{2})} ( e^{\beta \mu_{1}} -
e^{\beta \mu_{2}} ) .
\label{eq:ball008}
\end{equation}
This can be rewritten, using  Eqs.~(\ref{eq:diff004}) and  (\ref{eq:ball000}),
in the form 
\begin{equation}
J = v_{e} n(x_{1}) \bar{T}^{l}(x_{1},x_{2}) ( 1 - e^{- \beta eV} )
.
\label{eq:ball009}
\end{equation}
Here,
\begin{eqnarray}
\bar{T}^{l}(x_{1},x_{2}) &\equiv& \beta \int_{0}^{\infty} d \epsilon e^{-
\beta \epsilon} T \bm{(} x_{1},x_{2}; \epsilon + E_{c}(x_{1}) \bm{)}
\nonumber \\ &=& e^{- \beta E_{b}^{l}(x_{1},x_{2})} ,
\label{eq:ball005}
\end{eqnarray}
with $E_{b}^{l}(x_{1},x_{2})$ given by Eq.~(\ref{eq:diff007}), is the thermal
average of the classical transmission probability
\begin{equation}
T(x_{1},x_{2}; E) = \Theta \bm{(} E - E_{c}^{m}(x_{1},x_{2})
\bm{)}
\label{eq:ball005a}
\end{equation}
for electrons emitted at $x_{1}$ with total energy $E = \epsilon +
E_{c}(x_{1})$ to be transmitted to the point $x_{2}$.  If, in particular, the
potent\-ial energy profile is constant across the interval $[x_{1}, x_{2}]$, or
if its maximum lies at the emission point $x_{1}$ itself, then
$E_{c}^{m}(x_{1},x_{2}) = E_{c}(x_{1})$ in Eq.~(\ref{eq:diff007}), and hence
$\bar{T}^{l}(x_{1},x_{2}) = 1$.

Relation (\ref{eq:ball009}) is the current-voltage characteristic of the
(classical) ballistic transport model.  In contrast to the characteristic
(\ref{eq:diff006}) of the drift-diffusion model, which involves the mean free
path $l$ and the potential energy profile $E_{c}(x)$ [via the effective sample 
length $\tilde{S}$], the characteristic (\ref{eq:ball009}) is controlled by one
``material parameter'' only, {\em viz.}, the thermally averaged transmission
probability $\bar{T}^{l}(x_{1},x_{2})$.

For the {\em joint ballistic density} $n(x_{1},x_{2}; x)$,
\begin{equation}
n(x_{1},x_{2}; x) = n^{l}(x_{1},x_{2}; x ) + n^{r}(x_{1},x_{2}; x) ,
\label{eq:ball007z}
\end{equation}
we have
\begin{eqnarray}
n(x_{1},x_{2}; x) &=& \frac{N_{c}}{2} C^{m}(x_{1}, x_{2}; x) e^{- \beta
E_{c}^{m}(x_{1}, x_{2})} \nonumber \\ &\ & \times ( e^{\beta \mu_{1}} +
e^{\beta \mu_{2}} ) ,
\label{eq:ball007za}
\end{eqnarray}
in analogy to expression (\ref{eq:ball008}) for the net ballistic current.

\subsubsection{Electron tunneling}

\label{sec:balltunn}

The ballistic transport model is straightforwardly extended so as to include
{\em electron tunneling} by replacing the classical transmission probability
$T(x_{1}, x_{2}; E)$ of Eq.~(\ref{eq:ball005a}) with the corresponding {\em
quan\-tal} probability ${\cal T} ( x_{1}, x_{2}; E)$.  The thermally 
averaged quantal transmission probability is then, in generalization of 
expression (\ref{eq:ball005}), given by
\begin{eqnarray}
\bar{\cal T}(x_{1}, x_{2}) &=& \beta \int_{0}^{\infty}
d \epsilon e^{-\beta \epsilon} {\cal T} \bm{(} x_{1}, x_{2}; \epsilon
+ E_{c}^{>}(x_{1}, x_{2}) \bm{)} , \nonumber \\
\label{eq:ball010}
\end{eqnarray}
where
\begin{equation}
E_{c}^{>}(x_{1},x_{2}) \equiv \max \{ E_{c}(x_{1}), E_{c}(x_{2}) \} .
\label{eq:ball003b}
\end{equation}
The probability ${\cal T} ( x_{1}, x_{2}; E)$ is obtained by solving the
stationary Schr\"{o}\-dinger equation with the potential energy function
$E_{c}(x)$.  [Owing to time reversal invariance, the probability for
transmission from the left equals that for transmission from the right.] The
integration in Eq.~(\ref{eq:ball010}) starts at the total energy
$E_{c}^{>}(x_{1},x_{2})$, so that scattering boundary conditions can be imposed
on the wavefunction both in the ranges $x \leq x_{1}$ and $x \geq x_{2}$.

In WKB approximation,\cite{sch68,duk69} the transmission probability ${\cal
T}(x_{1}, x_{2}; E)$ to be used in Eq.~(\ref{eq:ball010}) is composed of the
classical (``over-barrier") part $T(x_{1}, x_{2}; E)$ given by
Eq.~(\ref{eq:ball005a}) and the remaining quantal (``sub-barrier'') part ${\cal
T}_{sb}(x_{1}, x_{2}; E)$,
\begin{eqnarray}
{\cal T}(x_{1}, x_{2}; E) &=& T(x_{1}, x_{2}; E) + {\cal T}_{sb}(x_{1}, x_{2};
E) . \nonumber \\
\label{eq:ball011}
\end{eqnarray}
The sub-barrier contribution has the form
\begin{eqnarray}
{\cal T}_{sb}(x_{1}, x_{2}; E) &=& \Theta \bm{(} E_{c}^{m}(x_{1}, x_{2}) - E
\bm{)} P_{c}(x_{1}, x_{2}; E) , \nonumber \\
\label{eq:ball012}
\end{eqnarray}
where
\begin{equation}
P_{c}(x_{1}, x_{2}; E) =  \exp \left( - 2 \int_{x_{1}}^{x_{2}} dx \kappa_{c}(x)
\right) ,
\label{eq:ball012f}
\end{equation}
with
\begin{equation}
\kappa_{c}(x) = \frac{1}{\hbar} \{ 2 m^{\ast} [E_{c}(x) - E] \}^{1/2} \Theta
\bm{(} E_{c}(x) - E \bm{)} ,
\label{eq:ball012g}
\end{equation}
is the barrier penetration factor.

In writing ${\cal T}_{sb}(x_{1}, x_{2}; E)$ in the form (\ref{eq:ball012}), we
disregard resonance effects that may occur when $E_{c}(x)$ exhibits two or more
local maxima in the interval $[x_{1}, x_{2}]$, with a corresponding number of
one or more minima in between.  Then, when the energy $E$ is located below the
second-highest maximum and above the lowest minimum, there is at least one
``valley'' in $E_{c}(x)$, across which the eletron motion is classically
allowed, so that resonance formation due to quantum coherence becomes possible.
In semiconductor physics, a concrete realization of this situation occurs in
resonant tunneling in multiple-barrier quantum-well structures.\cite{cha91}
For this case, the full WKB tunneling probability for double-barrier and
triple-barrier structures, respectively, has been presented in
Ref.~\onlinecite{xum93}.

From Eq.~(\ref{eq:ball011}), we now find for the WKB form of the thermally
averaged transmission probability $\bar{\cal T}(x_{1}, x_{2})$ of
Eq.~(\ref{eq:ball010})
\begin{equation}
\bar{\cal T}(x_{1}, x_{2}) =  \bar{T}(x_{1}, x_{2}) + \bar{\cal
T}_{sb}(x_{1}, x_{2}) .
\label{eq:ball010xa}
\end{equation}
Here, we have
\begin{equation}
\bar{T}(x_{1},x_{2})  = e^{- \beta E_{b}(x_{1}, x_{2})} , \nonumber \\
\label{eq:ball005d}
\end{equation}
with
\begin{equation}
E_{b}(x_{1}, x_{2}) = E_{c}^{m}(x_{1},x_{2}) - E_{c}^{>}(x_{1},x_{2}) ,
\label{eq:ball003a}
\end{equation}
for the over-barrier contribution,
and
\begin{eqnarray}
\bar{\cal T}_{sb}(x_{1}, x_{2}) &=& \beta \int_{0}^{\infty} d\epsilon e^{-
\beta \epsilon} P_{c}\bm{(} x_{1}, x_{2}; \epsilon + E_{c}^{>}(x_{1}, x_{2})
\bm{)} \nonumber \\ &\ & \hspace{1.5cm} \times \Theta \bm{(} E_{b}(x_{1},
x_{2}) - \epsilon \bm{)}
\label{eq:ball010x}
\end{eqnarray}
for the sub-barrier contribution.

The thermally averaged quantal probability $\bar{\cal T}^{l}(x_{1}, x_{2})$
for transmission from the left end-point at $x_{1}$ can be expressed in terms
of $ \bar{\cal T}(x_{1},x_{2})$, using Eqs.~(\ref{eq:diff007}) and
(\ref{eq:ball003a}), as
\begin{equation}
\bar{\cal T}^{l}(x_{1},x_{2}) = \bar{\cal T}(x_{1},x_{2}) e^{- \beta [
E_{c}^{>}(x_{1},x_{2}) - E_{c}(x_{1})]} .
\label{eq:ball006b}
\end{equation}
In analogy to Eq.~(\ref{eq:ball009}) for the classical case, we have
\begin{equation}
J = v_{e} n(x_{1}) \bar{\cal T}^{l}(x_{1}, x_{2}) ( 1 - e^{- \beta
eV} )
\label{eq:ball009b}
\end{equation}
for the current-voltage characteristic of tunneling-en\-han\-ced ballistic
transport.

\subsubsection{Degenerate case}

\label{sec:degener}

In the degenerate case, when the electron system obeys Fermi-Dirac statistics,
we write the ballistic current $J^{l}(x_{1}, x_{2})$ in the form
[see Eqs.~(\ref{eq:ball001h}) and (\ref{eq:ball004}) for the nondegenerate
case]
\begin{eqnarray}
J^{l}(x_{1}, x_{2};x) &=& 4 \pi  \left( \frac{m^{\ast}}{h}
\right) ^{3} \int_{0}^{\infty} dv_{x} v_{x} \int_{0}^{\infty} dw w
\nonumber \\ &\ & \times f^{FD} \bm{(} E(v; x) - \mu_{1} \bm{)} \nonumber
\\ &\ & \times \Theta \bm{(} E(v;x) - E_{c}^{m} (x_{1}, x_{2}) \bm{)} ,
\label{eq:degen040}
\end{eqnarray}
where $f^{FD}(E)$ is the Fermi-Dirac energy distribution function,
\begin{equation}
f^{FD}(E) = \frac{1}{1 + e^{\beta E}} .
\label{eq:degen011}
\end{equation}
Expression (\ref{eq:degen040}) for the current $J^{l}(x_{1}, x_{2};x)$ formally
agrees with the expression for the current of evaporated electrons encountered
in the degenerate treatment of the Richardson effect.\cite{som28,som33} Then,
following the procedure of Ref.~\onlinecite{som28}, we can reduce the threefold
integration in Eq.~(\ref{eq:degen040}) to a single integration over the kinetic
energy $\epsilon = m^{\ast} w^{2}/2$, obtaining
\begin{eqnarray}
J^{l}(x_{1},x_{2}) &=& \frac{4 \pi m^{\ast}}{\beta h^{3}} \int_{0}^{\infty}
d \epsilon \ln \bm{(} 1 + e^{- \beta (\epsilon - \mu_{1})}
\bm{)} \nonumber \\ &\ & \hspace{1.2cm} \times \Theta \bm{(} \epsilon -
E_{c}^{m} (x_{1}, x_{2}) \bm{)} .
\label{eq:degen010}
\end{eqnarray}
The ballistic current $J^{r}(x_{1},x_{2})$ transmitted from the right end-point
of the sample at $x_{2}$ is the negative of expression (\ref{eq:degen010}),
with $\mu_{2}$ substituted for $\mu_{1}$. The total current $J$ [see
Eq.~(\ref{eq:ball007})] is thus obtained as
\begin{eqnarray}
J &=& v_{e} N_{c} \beta \nonumber \\ &\ & \times \int_{0}^{\infty} d \epsilon
[ \ln \bm{(} 1 + e^{- \beta (\epsilon - \mu_{1} )} \bm{)} - \ln \bm{(} 1 +
e^{- \beta (\epsilon - \mu_{2} )} \bm{)} ] \nonumber \\ &\ &
\hspace{3.0cm} \times \Theta \bm{(} \epsilon - E_{c}^{m} (x_{1}, x_{2}) \bm{)}
.
\label{eq:degen011e}
\end{eqnarray}
Here, we have expressed the factor preceding the integral in
Eq.~(\ref{eq:degen010}) in terms of the emission velocity, $v_{e}$, and the
effective density of states, $N_{c}$, which are nondegenerate quantities given
by Eqs.~(\ref{eq:ball002}) and (\ref{eq:diff001a}), respectively. Expression
(\ref{eq:degen011e})  for the total currrent $J$ is the degenerate analogue to
the nondegenerate current-voltage characteristic (\ref{eq:ball008}).

For zero bias, the chemical potentials at the sample ends, $\mu_{1}$ and
$\mu_{2}$, differ only by an infinitesimal $\delta$,
\begin{equation}
\mu_{2} = \mu_{1} - \delta .
\label{eq:degen002}
\end{equation}
Then, expanding the right-hand side of Eq.~(\ref{eq:degen011e}) to first order
in $\delta$, we find
\begin{eqnarray}
J &=& v_{e} N_{c} \beta \delta \nonumber \\ &\ & \hspace{-0.8cm} \times \left[
\frac{\partial}{\partial \mu} \int_{0}^{\infty} d \epsilon \ln \bm{(} 1+
e^{-\beta (\epsilon - \mu)} \bm{)} \Theta \bm{(} \epsilon - E_{c}^{m} (x_{1},
x_{2}) \bm{)} \right]_{\mu = \mu_{1}} \nonumber \\ &=& v_{e} N_{c} \beta \delta
\ln \bm{(} 1+ e^{-\beta[E_{c}^{m}(x_{1}, x_{2}) - \mu_{1}]} \bm{)} .
\label{eq:degen001}
\end{eqnarray}
Setting $\delta \equiv \mu_{1} - \mu_{2} = eV$, we now have for the zero-bias
conductance per unit area in the ballistic transport model
\begin{eqnarray}
g &\equiv& \left( \frac{eJ}{V} \right)_{V \rightarrow 0} \nonumber \\ &=&
 \beta e^{2} v_{e} N_{c} \ln \bm{(} 1 + e^{- \beta [E_{c}^{m}(x{_1}, x_{2}) 
- \mu_{1}] } \bm{)} .
\label{eq:degen009}
\end{eqnarray}
In highly doped, degenerate semiconductors, we may have 
\begin{equation}
E_{c}^{m}(x{_1}, x_{2}) - \mu_{1} < 0 
\label{eq:degen009uhb}
\end{equation}
(see, e.g., Ref.~\onlinecite{pri98}, where grain-boundary-limited transport in 
polycrystalline materials is considered). Then, if 
\begin{equation}
\beta [\mu_{1} - E_{c}^{m}(x{_1}, x_{2})] \gg 1 , 
\label{eq:degen009ofd}
\end{equation}
we find from Eq.~(\ref{eq:degen009})
\begin{equation}
g  = \beta^{2} e^{2} v_{e} N_{c} \left[ \mu_{1} -  E_{c}^{m}(x{_1}, x_{2})
\right] ,
\label{eq:degen009ze}
\end{equation}
which equals the conductance of a ballistic point con\-tact.\cite{sha65,bee91}

In closing this subsection, we note that the ballistic transport model does not
provide information on where the resistance causing the voltage drop is located
along the sample.  Evidently, it cannot be inside the collision-free sample. In
the quantal description of ballistic electron transport in mesoscopic
systems as formulated by Landauer,\cite{lan87,dat95,imr99} the resistance is
made up solely of the {\em interface resistances} arising from the abrupt
change in the density of states (``transverse modes'') that the electrons
encounter when they move across the interfaces separating the contacts (with
infinitely many modes) from the sample (with a few modes only).  The voltage
drop is located, therefore, in the immediate vicinity of the interfaces, so
that, when a chemical potential is introduced {\em ad hoc}, this must be 
constant inside the sample and discontinuous at the interfaces.  Prior to the 
work of Landauer, the importance of interface resistances in ballistic 
transport had been emphasized by Sharvin.\cite{sha65}

Anticipating the later development, we remark at this point that in the
thermoballistic approach, i.e., for {\em finite} magnitude of the mean free 
path, the (local) equilibrium chemical potential is a constitutive element of 
the transport mechanism;\ it is defined, and can be explicitly calculated, 
all along a semiconducting sample.  This potential has 
discontinuities  at the contact-sample interfaces, whose magnitude  
increases from near-zero in the small-$l$,  drift-diffusion regime to the 
Sharvin value in the large-$l$, ballistic regime. For more details, see 
Sec.~\ref{sec:sharvin}.

\section{Prototype thermoballistic model}

\label{sec:gendrift}

In the drift-diffusion and ballistic transport models, the parameter of central
importance is the electron mean free path $l$ or, equivalently, the collision
time $\tau$ originally introduced in Drude's model.  The collision time
represents the average time-of-flight that elapses between successive electron
collisions with the randomly distributed scattering centers in the sample.  The
drift-diffusion and ballistic transport mechanisms are limiting cases
associated with very small and very large (effectively, infinite) collision
times, respectively.  In the present section, we proceed to the ``prototype
thermoballistic model'',\cite{lip01} which represents an attempt to unify the
drift-diffusion and ballistic models within a stationary (time-independent)
description in which the mean free path is not limited in magnitude and enters
as a parameter that controls collision {\em probabilities}.  [In our original
paper,\cite{lip01} we have called this model the ``generalized Drude model'';
the new name appears to give a better description of its features.]
We begin this section by introducing the probabilistic definition of $l$ (or 
$\tau$).

\subsection{Probabilistic approach to carrier transport}

\label{sec:probappr}

In the time-dependent probabilistic approach, one introduces the probability
$1/\tau$ that an electron undergoes a collision, i.e., is equilibrized with its
surroundings, within unit time.  That is, $dt/ \tau$ is the probability for the
collision to occur within an infinitesimally short time interval $dt$.  Then,
assuming the collision time $\tau$ to be constant, the probability $p(t)$ that
an electron moves without collision over a finite time interval $t$ is 
given\cite{sap95,fey64} by
\begin{equation}
p(t) = e^{- t/ \tau} ,
\label{eq:prob001}
\end{equation}
and the conditional probability  $P(t) dt$ that an electron undergoes a
collision within the time interval $dt$ after a collision-free flight over a
time interval $t$, by
\begin{equation}
P(t) dt = p(t) \frac{dt}{\tau} .
\label{eq:prob002}
\end{equation}
The mean time-of-flight between successive collisions now becomes
\begin{equation}
\bar {t} \equiv \int_{0}^{\infty} dt t P(t)  = \tau  .
\label{eq:prob003}
\end{equation}
Identifying $\bar{t}$ with the collision time $\tau$ of the Drude
model, one sees that the inverse of this quantity is just equal to the
collision probability per unit time, $1 / \tau$, of the probabilistic approach.

We recall that the term ``collision'' used here is that proffered by Drude, 
meaning a collision leading to instantaneous, complete equilibration of the 
momentum of an electron with its surroundings. In that picture, $\tau$ is the 
average time of collision-free flight between points of equilibration, hence 
the appellation. However, $1/\tau$ is, more generally, the probability for a 
{\em complete equilibration} to occur in unit time;\ this may happen as a 
consequence of a ``complete collision'' (in the Drude sense) after the average
time $\tau$, or by ``incomplete collisions'' (leading to incomplete 
equilibration) in a sequence of shorter collision times, which add up, on 
average, to the time $\tau$. 
 
Within the time-dependent probabilistic picture, one can set up\cite{sap95} an 
expression for the net electron current $J(x,t)$ at position $x$ and time $t$, 
which embodies features that are relevant for devising the thermoballistic 
concept (see Sec.~\ref{sec:thermcurr}). Assuming a sample with 
position-dependent electron density $n(x)$, one writes
\begin{equation}
J(x,t) = J^{l}(x,t) + J^{r}(x,t) .
\label{eq:prob003e}
\end{equation}
Considering for the moment electrons with {\em arbitrary, constant} velocity
$v_{x}$, the currents $J^{l,r}(x, t)$ are expressed, using 
Eqs.~(\ref{eq:prob001}) and (\ref{eq:prob002}), as
\begin{eqnarray}
J^{l,r}(x, t) &=& \pm  v_{x} \int_{- \infty}^{t} \frac{dt'}{\tau} e^{- (t
- t')/\tau} n \bm{(} x \mp v_{x}(t - t') \bm{)} . \nonumber \\
\label{eq:prob004}
\end{eqnarray}
The contributions to these currents for fixed time $t$ arise from electrons 
that are transmitted, subsequent to their
emission at the points $x_{\mp} \equiv x \mp v_{x}(t - t')$ [lying to the left
(right) of the point $x$], without collision from $x_{-}$ $(x_{+})$ to the 
point $x$.  The intervals $[x_{-}, x]$ and $[x, x_{+}]$ define ``ballistic
intervals'';\ the currents transmitted across these intervals contribute to the
total current with weight $e^{- (t - t')/\tau}$. Note that, in contrast to
the situation considered above, where ballistic motion occurs {\em before} 
equilibration in the time interval $dt$ (with probability $dt/\tau$), here the
ballistic motion occurs {\em after} the electrons have been equilibrated in 
the time interval $dt'$ (with probability $dt'/\tau$). 

When the collision time $\tau$ is sufficiently small, such that $v_{x} \tau$
is much smaller than the average length over which the density $n(x)$ changes 
appreciably, one has, by expanding in expressions (\ref{eq:prob004}) 
$n(x)$ to first order, 
\begin{eqnarray}
J(x,t) &=&  - 2 v_{x}^{2} \frac{d n(x)}{dx} \int_{- \infty}^{t}
\frac{dt'}{\tau} e^{- (t - t')/\tau} (t- t') \nonumber \\ &=&
 - 2 v_{x}^{2} \tau \frac{d n (x)}{dx}  \equiv J(x) .
\label{eq:prob005}
\end{eqnarray}
Taking into account that only electrons with $v_{x} > 0$ ($v_{x}< 0$)
contribute to the current arriving at the point $x$ from the left (right),
{\em thermal averaging} of $v_{x}^{2}$ now yields
\begin{equation}
\langle v_{x}^{2} \rangle
\label{eq:prob005b} = \frac{1}{2 m^{\ast} \beta} ,
\end{equation}
so that one obtains the local expression
\begin{equation}
J(x) = - \frac{\tau}{m^{\ast} \beta} \frac{d n (x)}{dx}
\label{eq:prob006}
\end{equation}
for the thermally averaged electron current.  The charge current $-e J(x)$ is
then found, in view of expression (\ref{eq:drude011}) for the diffusion
coefficient $D$, to agree with relation (\ref{eq:drude007}) for the diffusion
(charge) current $j_{di}(x)$ of the drift-diffusion model.  This result may
thus be regarded as a proof of the Einstein relation (\ref{eq:drude011}).

When $\tau \rightarrow \infty$, only electrons emitted at the left and right
end, respectively, of the sample  contribute to the currents $J^{l,r}(x, t)$. 
Then, the (thermally averaged) net electron current $J(x,t)$ essentially 
reduces to the ballistic 
current given by expression (\ref{eq:ball008}).  In the intermediate regime, 
i.e., for nonzero, finite values of $\tau$, $J(x,t)$ combines elements of both 
drift-diffusion and ballistic transport.

In analogy to the time-dependent probability $p(t)$ given by
Eq.~(\ref{eq:prob001}), one can introduce the probability $p(x)$ for an
electron to travel without collision over a finite distance $x$,
\begin{equation}
p(x) = e^{- x/ l} ,
\label{eq:prob001a}
\end{equation}
with a constant electron mean free path $l$. The conditional probability for 
an electron to undergo a collision in the infinitesimal collision interval
$dx$ after a collision-free flight over the distance $x$ reads
\begin{equation}
P(x) dx = p(x) \frac{dx}{l} .
\label{eq:prob002a}
\end{equation}
Then one finds
\begin{equation}
\bar {x} \equiv \int_{0}^{\infty} dx x P(x)  = l
\label{eq:prob003a}
\end{equation}
for the mean distance $\bar{x}$ between successive electron collisions.

In the following, the mean free path $l$ will be assumed to take on any finite
value, while, as before, the collisions are assumed to lead instantaneously to
a complete equilibration of the electron momenta, and are to be interpreted as 
a simulation of the real situation where the distances between successive
collisions are short (of atomic dimensions), while the equilibration during a
collision is (usually) far less than complete. The quantity $l$, in parallel to
the collision time $\tau$, is simply to be regarded as a parameter which 
determines the probability (\ref{eq:prob001a}) of collision-free (complete or
not) traversal by an electron of the distance $x$.

It is noted here that when treating inhomogeneous systems, one should, strictly
speaking, allow for a position dependence of the collision time, $\tau = \tau
(x)$, and of the mean free path, $l = l(x)$.  In that case, expressions
(\ref{eq:prob001}) and (\ref{eq:prob001a}) for the probabilities $p(t)$ and
$p(x)$, respectively, must be replaced with the more general forms
\begin{equation}
p(t) = \exp \left(- \int_{0}^{\infty} \frac{dt'}{\tau \bm{(} x(t') \bm{)}}
\right)
\label{eq:prob006a}
\end{equation}
and
\begin{equation}
p(x) = \exp \left(- \int_{0}^{\infty} \frac{dx'}{l (x')} \right) ,
\label{eq:prob007}
\end{equation}
respectively.  Then, of course, the relation $l = u \tau$ [see
Eq.~(\ref{eq:drude002})] ceases to be valid.

\subsection{Prototype thermoballistic model:\ Concept and implementation}

\label{sec:gendrude}

The unification of the drift-diffusion and ballistic transport models within 
the prototype thermoballistic model\cite{lip01} is based on the introduction of
configurations of ballistic transport intervals, which cover the length of the
sample. The individual ballistic intervals in a configuration are linked by 
points of local thermodynamic equilibrium, and their lengths  are stochastic 
variables occurring with probabilities determined by the probabilities for 
collision-free electron motion across the intervals, where the electron mean
free path is allowed to have {\em arbitrary} magnitude.  The electron current
is, of course, conserved in each ballistic interval.  In addition, it is 
assumed that 
{\em the current is conserved also from one interval to the next}.  Averaging 
over the ballistic configurations then leads to a transparent and intuitively 
appealing expression for the current voltage-characteristic in terms of an 
effective transport length, in which the detailed shape of the potential energy
profile is taken into account.  Moreover, a local chemical potential can be 
constructed in a heuristic way.

\subsubsection{Ballistic configurations}

\label{sec:ballconf}

Here, as well as in Sec.~\ref{sec:averball}, we confine ourselves to 
(one-dimensional) classical transport in nondegenerate systems.  The effects of
tunneling and degeneracy will be considered in Secs.~\ref{sec:quantum} and 
\ref{sec:degener1}, respectively.

\begin{figure}[t]
\includegraphics[width=0.45\textwidth]{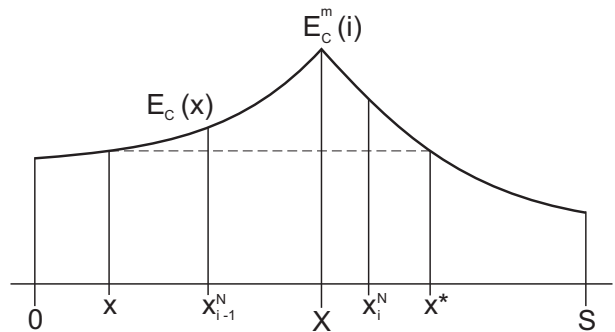}
\caption{Schematic diagram of a potential energy profile $E_{c}(x)$ exhibiting 
a single barrier with maximum at $x=X$.  The partitioning of the sample length 
$S$ into $N$ ballistic intervals $[x_{i-1}^{N},x_{i}^{N}]$ 
($i = 1, \ldots , N$) is indicated, where $E_{c}^{m}(i)$ is the maximum of 
$E_{c}(x)$ in interval $i$. For the definition of the mirror point $x^{\ast}$
associated with a point $x$, see Eq.~(\ref{eq:gendr009uvy}) below. 
}
\label{fig:1}
\end{figure}

For given $N$ ($N = 1, \ldots , \infty$), the length $S$ of the sample is
randomly partitioned into $N$ ``ballistic intervals'' labeled $i$ ($i = 1,
\ldots, N$), in which the electrons move ballistically across the potential
energy profile $E_{c}(x)$ [see Fig.~\ref{fig:1}].  These intervals are linked
by points of local thermodynamic equilibrium $x_{i-1}^{N}$ and $x_{i}^{N}$
($x_{i-1}^{N} < x_{i}^{N}$) which, in addition, include the two (true) 
equilibrium points at the interfaces with the left and right contacts at 
$x^{N}_{0} = 0$ and $x^{N}_{N} = S$, respectively. [In this section, in order 
to avoid confusion with the labeling of the equilibrium points inside, we 
denote the end-point coordinates of the sample by $0$ and $S$, instead of 
$x_{1}$ and $x_{2}$, as used elsewhere.] At $x_{i-1}^{N}$ and $x_{i}^{N}$, 
electrons are thermally emitted towards the right and left, respectively, into 
the interval $[x_{i-1}^{N},x_{i}^{N}]$.  Any set
\begin{equation}
\{ x_{j}^{N} \} \equiv \{ x_{0}^{N}, x_{1}^{N}, \ldots , x_{j}^{N}, \ldots ,
x_{N}^{N} \}
\label{eq:gendr000a}
\end{equation}
of $N+1$ equilibrium points linking $N$ ballistic intervals characterizes a
{\em ballistic configuration}.

Introducing an equilibrium chemical potential $\mu(x)$ all across the sample,
in terms of which the values $\mu (x_{i-1}^{N})$ and $\mu (x_{i}^{N})$ are
defined, the net ballistic current $J^{N}(i) \equiv
J^{N}(x_{i-1}^{N},x_{i}^{N})$ transmitted across the interval $i$ in a
partition with $N$ intervals is given, in analogy to expression
(\ref{eq:ball008}) for the current in the ballistic transport model, by
\begin{equation}
J^{N}(i) = v_{e} N_{c} e^{- \beta E_{c}^{m}(i)} [ e^{\beta \mu
(x_{i-1}^{N})} - e^{\beta \mu (x_{i}^{N})} ] ,
\label{eq:gendr001}
\end{equation}
where
\begin{equation}
E_{c}^{m}(i) \equiv E_{c}^{m}(x_{i-1}^{N},x_{i}^{N})
\label{eq:gendr001m}
\end{equation}
denotes the absolute maximum of the potential energy profile $E_{c}(x)$ in the
interval $i$. In terms of the (classical) thermally averaged transmission
probability
\begin{equation}
\bar{T}(i) = e^{- \beta [E_{c}^{m}(i) - E_{c}^{>}(i)]}
\label{eq:gendr001x}
\end{equation}
defined in analogy to expression (\ref{eq:ball005d}), we can write $J^{N}(i)$
in the form
\begin{eqnarray}
J^{N}(i) &=& v_{e} N_{c} \bar{T}(i) e^{- \beta E_{c}^{>}(i)} [
e^{\beta \mu (x_{i-1}^{N})} - e^{\beta \mu (x_{i}^{N})} ] , \nonumber \\
\label{eq:gendr001y}
\end{eqnarray}
where
\begin{equation}
E_{c}^{>}(i) \equiv \max\{ E_{c}(x_{i-1}^{N}),  E_{c}(x_{i}^{N})\} .
\label{eq:gendr001k}
\end{equation}
The current $J^{N}(i)$ is, of course, conserved across each interval $i$.

In its general form, expression (\ref{eq:gendr001y}) obviously does not lend
itself for the calculation of transport properties in terms of external 
physical quantities. Therefore, we make the simplifying assumption that 
the electron current is conserved also across the points of local thermodynamic
equilibrium linking one ballistic interval to its adjacent intervals (and hence
is constant all along the length of the sample), so that in 
Eq.~(\ref{eq:gendr001y}) 
\begin{equation}
J^{N}(i) = J = {\rm const.}
\label{eq:gendr001qal}
\end{equation}
for all $i$ and $N$, where $J$ is the {\em physical current}. We can then 
iterate Eq.~(\ref{eq:gendr001y}) with respect to $i$ (for fixed $N$), with 
the result
\begin{equation}
e^{\beta \mu^{N}(0)} - e^{\beta \mu^{N}(S)} = \frac{J}{v_{e} N_{c}} e^{\beta
E_{c}^{m}(0,S)} {\cal Z}^{N} .
\label{eq:gendr002}
\end{equation}
Here, $E_{c}^{m}(0,S)$ denotes the overall maximum of the profile $E_{c}(x)$ in
the interval $[0,S]$, and the function ${\cal Z}^{N}$, defined as
\begin{eqnarray}
{\cal Z}^{N} \equiv {\cal Z}^{N}(\{ x_{j}^{N} \}) &=& \sum_{i=1}^{N}  
\frac{1}{\bar{T}(i)}  e^{-\beta [E_{c}^{m}(0,S) - E_{c}^{>}(i)]} \nonumber 
\\ &\equiv& \sum_{i=1}^{N} e^{-\beta [E_{c}^{m}(0,S) - E_{c}^{m}(i)]} ,
\label{eq:gendr002a}
\end{eqnarray}
can be formally viewed as the partition function corresponding to the ``energy
spectrum'' $\{ E_{c}^{m}(0,S) - E_{c}^{m}(i) \}$. It depends on the equilibrium
points $x^{N}_{j}$ in the ballistic configuration  $\{ x^{N}_{j}\}$ via
Eq.~(\ref{eq:gendr001m}).

\subsubsection{Averaging over ballistic configurations}

\label{sec:averball}

In order to arrive at results in terms of physical quantities, we must average
Eq.~(\ref{eq:gendr002}) over the complete set of  ballistic configurations 
$\{ x_{j}^{N} \}$ $(j = 0, 1, \ldots, N;\ N = 1, \ldots , \infty)$.
In doing so, we identify the average of the left-hand side of
Eq.~(\ref{eq:gendr002}) with the ``bias term'' $e^{\beta \mu_{0}} - e^{\beta
\mu_{S}}$, where $\mu_{0}$ and $\mu_{S}$ are the values of the
chemical potential at the contact side of the left and right 
contact-semiconductor interface, respectively, and obtain
\begin{equation}
e^{\beta \mu_{0}} - e^{\beta \mu_{S}} = \frac{J}{v_{e} N_{c}} e^{\beta
E_{c}^{m}(0,S)} {\cal R} .
\label{eq:gendr004}
\end{equation}
Here,
\begin{equation}
{\cal R} = \left\langle  {\cal Z}^{N}( \{ x_{j}^{N} \} )  \right\rangle
\label{eq:gendr003}
\end{equation}
denotes the average of the partition functions ${\cal Z}^{N}$ over all 
ballistic configurations $\{ x_{j}^{N} \}$.  From Eq.~(\ref{eq:gendr004}), the
current-voltage characteristic of the prototype thermoballistic model is now
obtained in the general form
\begin{equation}
J = v_{e} N_{c} e^{- \beta [E_{c}^{m}(0,S) - \mu_{0}]} \frac{1}{\cal R} 
( 1 - e^{-\beta e V} ) ,
\label{eq:gendr004a}
\end{equation}
with the voltage bias $V$ given by
\begin{equation}
V = \frac{\mu_{0} - \mu_{S}}{e} .
\label{eq:gendr004pd}
\end{equation}
Aside from the barrier factor involving the absolute maximum, $E_{c}^{m}(0,S)$,
of the potential energy profile, the current $J$ is controlled here by the
parameter ${\cal R}$, which alone comprises the detailed dependence of the
current-voltage characteristic on the shape of the profile
and on the material parameters characterizing the
sample.  As it appears in Eq.~(\ref{eq:gendr004a}), ${\cal R}$ can be viewed as
a (dimensionless) {\em reduced resistance}.

In the averaging process implied in Eq.~(\ref{eq:gendr003}), the contribution 
of each ballistic
interval $i$ in a partition with $N$ intervals is to be weighted with the
conditional probability $P_{i}^{N} dx_{i}^{N}$ for an electron to make a
collision in the infinitesimal interval $dx_{i}^{N}$ after having traveled
freely across the interval $i$ of length $x_{i}^{N} - x_{i-1}^{N}$, into which 
it was emitted after a collision in the interval $dx_{i-1}^{N}$. The collision
probability is connected with the {\em two} ends of each ballistic interval, 
and the collision interval $dx_{i}^{N}$ is the {\em equilibration interval} 
for the ballistic interval $i$, but it  is simultaneously the {\em emission 
interval} for the ballistic interval $i+1$. This ``sharing'' of collision 
intervals $dx_{i}^{N}$ between the ballistic intervals $i$ and $i+1$ implies 
that effectively 
the density of points of equilibration is only {\em one-half} of that given  
originally. Therefore, the conditional probability $P_{i}^{N} dx_{i}^{N}$ is, 
in conformance with Eqs.~(\ref{eq:prob001a}) and (\ref{eq:prob002a}), expressed
as
\begin{equation}
P_{i}^{N} dx_{i}^{N} = e^{- (x_{i}^{N} -x_{i-1}^{N}) / \ell}
 \frac{dx_{i}^{N}}{\ell} ,
\label{eq:gendr005}
\end{equation}
with the {\em effective mean free path} $\ell$ given by 
\begin{equation}
\ell = 2l .
\label{eq:gendr05asf}
\end{equation}
Here, the meaning of the mean free path $l$ is the original one introduced
within the Drude model. 

The conditional probability $d P^{N}$ for an electron
emitted at $x_{0}^{N} = 0$ to undergo $N-1$ collisions at the equilibrium 
points $x_{i}^{N}$ $(i=1, \ldots \ N-1)$, and finally be absorbed with unit 
probability at $x_{N}^{N} = S$, is then found as
\begin{eqnarray}
d P^{N} &=& \left[ \prod_{i = 1}^{N-1} \frac{dx_{i}^{N}}{\ell} e^{- (x_{i}^{N}
-x_{i-1}^{N}) / \ell} \Theta(x_{i}^{N} -x_{i-1}^{N}) \right] \nonumber \\ &\ &
\times e^{-(S - x^{N}_{N-1})/\ell}
\label{eq:gendr007}
\end{eqnarray}
for $N \geq 2$, while
\begin{equation}
d P^{1} = e^{-S/\ell} .
\label{eq:gendr006}
\end{equation}
In the product in Eq.~(\ref{eq:gendr007}), the exponentials cancel out except
for the factor $e^{- S/\ell}$, so that
\begin{equation}
d P^{N} =  e^{- S/\ell} \prod_{i = 1}^{N-1} \frac{dx_{i}^{N}}{\ell}
\Theta(x_{i}^{N} -x_{i-1}^{N})
\label{eq:gendr008}
\end{equation}
and
\begin{equation}
\sum_{N=1}^{\infty} \int_{0}^{S} d P^{N} = 1
\label{eq:gendr008x}
\end{equation}
[the symbol $\int_{0}^{S}$ is meant to imply $(N-1)$-fold integration over
$x_{1}^{N}, \ldots , x_{N-1}^{N}$ from $0$ to $S$], i.e., the total probability
is unity.

The reduced resistance ${\cal R}$ defined by Eq.~(\ref{eq:gendr003}) can now 
be expressed as
\begin{eqnarray}
{\cal R} &=& \sum_{N=1}^{\infty} \int_{0}^{S} d P^{N} {\cal Z}^{N} \nonumber \\
&\equiv& e^{- S/\ell} \sum_{N=1}^{\infty} {\cal O}^{N} {\cal Z}^{N}
( \{ x_{j}^{N} \} ) , 
\label{eq:gendr009}
\end{eqnarray}
where the operators ${\cal O}^{N}$ acting on the partition functions 
${\cal Z}^{N}$ are given by ${\cal O}^{1} = 1$ and
\begin{equation}
{\cal O}^{N} = \int_{0}^{S} \frac{dx_{1}^{N}}{\ell} \int_{x_{1}^{N}}^{S} 
\frac{dx_{2}^{N}}{\ell} \ldots \int_{x_{N-2}^{N}}^{S} \frac{dx_{N-1}^{N}}{\ell}
\label{eq:gendr009ya}
\end{equation}
for $N \geq 2$.
It can be shown that, owing to the separable form of ${\cal Z} ^{N}$ [see
Eqs.~(\ref{eq:gendr002a}) and (\ref{eq:gendr002c})], the multi-dimensional
integrals in Eq.~(\ref{eq:gendr009}) can always be reduced to one-dimensional
and two-dimensional integrals.  The remaining infinite series can be summed up
in terms of exponentials.  For the classical partition function
${\cal Z}^{N}$ of Eq.~(\ref{eq:gendr002a}), the reduced resistance 
${\cal R}$ is thus obtained in the form
\begin{eqnarray}
{\cal R} &=& e^{- S/\ell} \bar{R}(0,S) \nonumber \\ &+& \hspace{-0.1cm} 
\int_{0}^{S}
\frac{dx'}{\ell} [ e^{-x'/\ell} \bar{R}(0,x') + e^{-(S-x')/\ell} \bar{R}(x',S)]
\nonumber \\ &+& \hspace{-0.1cm} \int_{0}^{S} \frac{dx'}{\ell} \int_{x'}^{S}
\frac{dx''}{\ell} e^{-(x''-x')/\ell} \bar{R}(x',x'') ,
\label{eq:gendr009a}
\end{eqnarray}
where the (dimensionless) function $\bar{R}(x',x'')$ is given by
\begin{eqnarray}
\bar{R}(x',x'') &\equiv& \bar{R}(x'',x') \nonumber \\ &=& 
\frac{1}{\bar{T}(x',x'')} e^{- \beta [E_{c}^{m}(0,S) -
E_{c}^{>}(x',x'')]} \nonumber \\ &=& e^{- \beta [E_{c}^{m}(0,S) -
E_{c}^{m}(x',x'')]} .
\label{eq:gendr009b}
\end{eqnarray}
Here, the thermally averaged classical transmission probability
$\bar{T}(x',x'')$ is defined in analogy to Eqs.~(\ref{eq:ball005d}) and
(\ref{eq:gendr001x}), $E_{c}^{m}(x',x'') \equiv E_{c}^{m}(x'',x') $ is the 
absolute maximum of the potential energy profile $E_{c}(x)$ in the interval 
$[x', x'']$, and
\begin{equation}
E_{c}^{>}(x',x'') \equiv \max \{E_{c}(x'), E_{c}(x'') \} .
\label{eq:gendr009xg}
\end{equation}
In general, we have
\begin{equation}
\bar{R}(x',x'') \leq 1
\label{eq:gendr009uhd}
\end{equation}
and, in particular, $\bar{R}(0,S) = 1$.

For a flat potential energy profile, $E_{c}(x) \equiv {\rm const.}$, when
$\bar{R}(x',x'') \equiv 1$, expression (\ref{eq:gendr009a}) can be immediately
evaluated with the result
\begin{equation}
{\cal R} \equiv {\cal R}_{0} = 1 + \frac{S}{\ell} ,
\label{eq:gendr009x}
\end{equation}
so that in the general case, owing to Eq.~(\ref{eq:gendr009uhd}),
\begin{equation}
{\cal R} \leq {\cal R}_{0} ,
\label{eq:gendr009y}
\end{equation} 
i.e., ${\cal R}$ {\em decreases} when a position-dependent potential energy 
profile is introduced. In the current-voltage characteristic 
(\ref{eq:gendr004a}), this decrease of ${\cal R}$ is overcompensated by the 
effect of the barrier factor $e^{- \beta [E_{c}^{m}(0,S) - \mu_{0}]}$, so that
the current $J$ decreases as well.

By reducing expression (\ref{eq:gendr009}) for ${\cal R}$ to the form
(\ref{eq:gendr009a}), we have transcribed the prototype thermoballistic
model, which, according to its original concept, relies on a discrete 
partitioning of the
sample length into ballistic intervals, into a pure continuum model.
Expression (\ref{eq:gendr009a}) is composed of three contributions.  The first
corresponds to ballistic transport all across the sample length $S$, with
associated probability $e^{-S/\ell}$.  The second reflects the integrated effect
of ballistic transport across intervals $[0,x']$ and $[x',S]$, respectively,
with probabilities $e^{-x'/\ell}$ and $e^{-(S-x')/\ell}$.  The third, finally,
corresponds to the integrated effect of ballistic transport across intervals
$[x',x'']$, with probabilities $e^{-(x''-x')/\ell}$.

The central task in implementing the prototype thermoballistic model is to
evaluate the reduced resistance ${\cal R}$ from Eq.~(\ref{eq:gendr009a}) [or 
from the analogous equation derived from the quantal partition function 
(\ref{eq:gendr002c})] for given potential energy 
profile $E_{c}(x)$.  In general, the  computation of $E_{c}(x)$ is accomplished
(see, e.g., Ref.~\onlinecite{jac10}) by solving, for given distribution of the
space-fixed charges in the sample, a nonlinear Poisson equation.

\subsubsection{Electron tunneling}

\label{sec:quantum}

In the ballistic transport model (see Sec.~\ref{sec:balltunn}), we have taken
into account the effect of electron tunneling by introducing quantal 
probabilities to describe electron transmission across the whole sample length.
In the prototype model, we can include tunneling effects by using quantal
transmission probabilities in expression (\ref{eq:gendr001y}) for the current 
$J^{N}(i)$ in the individual ballistic intervals $[x^{N}_{i-1}, x^{N}_{i}]$.
 
Replacing in Eq.~(\ref{eq:gendr001y}) the probability $\bar{T}(i)$ with the 
thermally averaged quantal transmission probability $\bar{\cal T}^{l}(i) 
\equiv\bar{\cal T}^{l}(x^{N}_{i-1}, x^{N}_{i})$ defined in analogy to 
expression (\ref{eq:ball010}), with $x_{1,2}$ replaced with $x^{N}_{i-1,i}$,
the partition function ${\cal Z}^{N}$ in Eq.~(\ref{eq:gendr002}) 
acquires the form
\begin{eqnarray}
{\cal Z}^{N}_{q} &=& \sum_{i=1}^{N} \frac{1}{\bar{\cal T}(i)}
e^{-\beta [E_{c}^{m}(0,S) - E_{c}^{>}(i)]} \nonumber \\ &\equiv& \sum_{i=1}^{N}
\frac{\bar{T}(i)}{\bar{\cal T}(i)} e^{-\beta [E_{c}^{m}(0,S) - E_{c}^{m}(i)]} ,
\label{eq:gendr002c}
\end{eqnarray}
where Eq.~(\ref{eq:gendr001x}) for  $\bar{T}(i)$ has been used..

In generalization of expression (\ref{eq:gendr009a}), the reduced resistance
${\cal R}_{q}$ including tunneling effects is  obtained from the quantal 
partition function ${\cal Z}^{N}_{q}$ as
\begin{eqnarray}
{\cal R}_{q} &=& e^{- S/\ell} \bar{\cal R}(0,S) \nonumber \\ &+& \int_{0}^{S}
\frac{dx'}{\ell} [ e^{-x'/\ell} \bar{\cal R}(0,x') + e^{-(S-x')/\ell} \bar{\cal
R}(x',S) ] \nonumber \\ &+& \int_{0}^{S} \frac{dx'}{\ell} \int_{x'}^{S}
\frac{dx''}{\ell} e^{-(x''-x')/\ell} \bar{\cal R}(x',x'') ,
\label{eq:gendr009e}
\end{eqnarray}
where
\begin{eqnarray}
\bar{\cal R}(x',x'') &=&  \frac{1}{\bar{\cal T}(x',x'')} e^{- \beta 
[E_{c}^{m}(0,S) - E_{c}^{>}(x',x'')]} \nonumber \\ &\equiv& 
\frac{\bar{T}(x',x'')}{\bar{\cal T}(x',x'')} \bar{R}(x', x'') .
\label{eq:ball003y}
\end{eqnarray}
Here, the thermally averaged quantal transmission probability $\bar{\cal
T}(x',x'')$ is defined in analogy to expression (\ref{eq:ball010}), and
$\bar{R}(x', x'')$ is given by Eq.~(\ref{eq:gendr009b}). In view of
Eq.~(\ref{eq:ball003y}), the inequality (\ref{eq:gendr009y}) providing an
upper limit on the values of the classical function ${\cal R}$ holds also for 
the quantal function ${\cal R}_{q}$.

We now adopt the WKB approximation, for which we have, using
Eq.~(\ref{eq:ball010xa}),
\begin{equation}
\bar{\cal R}(x',x'') = \bar{R}(x', x'') + \bar{\cal R}_{sb}(x',x'') ,
\label{eq:ball003za}
\end{equation}
with
\begin{eqnarray}
\bar{\cal R}_{sb}(x',x'') &=& - \frac{\bar{\cal
T}_{sb}(x',x'')}{\bar{T}(x',x'') + \bar{\cal T}_{sb}(x',x'')}
\bar{R}(x', x'') . \nonumber \\
\label{eq:ball003zy}
\end{eqnarray}
Inserting expression (\ref{eq:ball003za}) in Eq.~(\ref{eq:gendr009e}), we find
that ${\cal R}_{q}$ separates in the form
\begin{equation}
{\cal R}_{q} = {\cal R}_{cl} + {\cal R}_{sb} ,
\label{eq:ball003p}
\end{equation}
where the classical contribution ${\cal R}_{cl}$ is given by
Eq.~(\ref{eq:gendr009a}), and the tunneling (sub-barrier) contribution
${\cal R}_{sb}$ by Eq.~(\ref{eq:gendr009e}), with $\bar{\cal R}(x',x'')$
replaced with $\bar{\cal R}_{sb}(x',x'')$. Evidently, ${\cal R}_{sb} < 0$ and
$|{\cal R}_{sb}| < {\cal R}_{cl}$.

According to its definition in analogy to Eq.~(\ref{eq:ball010x}), the
thermally averaged sub-barrier transmission probability $\bar{\cal
T}_{sb}(x',x'')$ is nonzero only if the potential energy profile
$E_{c}(x)$ exhibits at least one local maximum in the range $x' < x < x''$ such
that $E_{b}(x', x'') > 0$.  In the evaluation of the sub-barrier function
${\cal R}_{sb}$ from Eq.~(\ref{eq:gendr009e}), ballistic intervals for which
$E_{c}(x)$ does not meet this requirement can, therefore, be excluded from the
outset.

\subsubsection{Degenerate case}

\label{sec:degener1}

Recalling the procedure developed in the nondegenerate case (see
Secs.~\ref{sec:ballconf} and \ref{sec:averball}), in which expression
(\ref{eq:gendr001}) for the net ballistic current $J^{N}(i)$ forms the starting
point of the derivation of the current-voltage characteristic
(\ref{eq:gendr004a}), we find that an analogous procedure cannot be set up, in
general, in the degenerate case.  The reason is that for the argument leading
to Eq.~(\ref{eq:gendr004a}), $J^{N}(i)$ must factorize into a term depending on
the potential energy profile $E_{c}(x)$, and terms depending on the chemical
potential $\mu (x)$.  In the degenerate case, the current analogous to
$J^{N}(i)$ is the current $J(x_{1}, x_{2})$ given by Eqs.~(\ref{eq:ball007})
and (\ref{eq:degen011e}), with $x_{1}$ and $x_{2}$, respectively, replaced with
$x^{N}_{i-1}$ and $x^{N}_{i}$.  This current does not factorize if the bias is
nonzero.

In the zero-bias case, however,  the chemical potentials at the sample ends, 
$\mu(0)$ and $\mu(S)$, differ only infinitesimally, and the increment of 
$\mu(x)$ across a ballistic interval $i$,
\begin{equation}
\delta_{i} = \mu(x_{i-1}^{N}) - \mu(x_{i}^{N}) ,
\label{eq:degen002z}
\end{equation}
is infinitesimally small as well.  Then, we find in analogy to 
Eq.~(\ref{eq:degen001})
\begin{eqnarray}
J^{N}(i) &=& v_{e} N_{c} \beta \delta_{i} \nonumber \\ &\ & \hspace{-1.65cm}
\times \left[ \frac{\partial}{\partial \mu} \int_{0}^{\infty} d \epsilon \ln
\bm{(} 1+ e^{-\beta (\epsilon - \mu)} \bm{)} \Theta \bm{(} \epsilon -
E_{c}^{m}(i) \bm{)} \right]_{\mu = \mu(x_{i-1}^{N})} \nonumber \\ &=& v_{e}
N_{c} \beta \delta_{i} \ln \bm{(} 1+ e^{-\beta[E_{c}^{m}(i) - \mu(0)]} \bm{)}
,
\label{eq:degen001y}
\end{eqnarray}
where we have replaced, in the second equation, $\mu(x_{i-1}^{N})$ with $\mu
(0)$.

Expression (\ref{eq:degen001y}) has the factorization property alluded to
above, so that, following the procedure leading to Eq.~(\ref{eq:gendr002}) in
the nondegenerate case, we here find
\begin{equation}
\beta [\mu(0) - \mu(S)] =  \frac{J}{v_{e} N_{c}} 
\frac{{\cal Z}^{N}_{d} }{\ln \bm{(} 1+ e^{- \beta [E_{c}^{m}(0,S) - \mu(0)]} 
\bm{)} }  ,
\label{eq:degen003}
\end{equation}
where the function
\begin{equation}
{\cal Z}^{N}_{d} = \sum_{i=1}^{N}\frac{\ln \bm{(} 1+ e^{- \beta
[E_{c}^{m}(0,S) - \mu(0)]} \bm{)}}{\ln \bm{(} 1+ e^{- \beta [E_{c}^{m}(i) - 
\mu(0)]} \bm{)}} \nonumber \\
\label{eq:degen003a}
\end{equation}
generalizes the partition function ${\cal Z}^{N}$ given by
Eq.~(\ref{eq:gendr002a}) to the degenerate case.  Averaging
Eq.~(\ref{eq:degen003}) over the ballistic configurations $\{
x_{j}^{N}\}$ yields, in analogy to Eq.~(\ref{eq:gendr004}),
\begin{equation}
\beta (\mu_{0} - \mu_{S}) =  \frac{J}{v_{e} N_{c}} 
\frac{{\cal R}_{\rm d} }{\ln \bm{(}1 + e^{- \beta [E_{c}^{m}(0,S) - \mu_{0}]} 
\bm{)}}  ,
\label{eq:degen005}
\end{equation}
where
\begin{equation}
{\cal R}_{d} = \left\langle {\cal Z}^{N}_{d}(\{ x_{j}^{N} \}) \right\rangle
\label{eq:degen004}
\end{equation}
is the reduced resistance for the degenerate case.

\subsubsection{Chemical potential}

\label{sec:prochem}

Within the prototype model, we can construct, in a heuristic way, a unique 
chemical potential $\mu(x)$ all along the sample length
by modifying relation (\ref{eq:gendr004}) that connects the values 
of the chemical potential at the contact-semiconductor interfaces, $\mu_{0}$ 
and $\mu_{S}$. We proceed as follows.

On the one hand, by replacing the fixed position $S$ with a 
variable position $x$ inside the sample, we turn $\mu_{S}$ into a 
chemical-potential function $\mu_{1}(x)$ and, simultaneously, reduce the range
of averaging over the ballistic configurations in Eq.~(\ref{eq:gendr003}) for 
the reduced resistance ${\cal R}$ from $[0, S]$ to $[0, x]$.  We then obtain
\begin{equation}
e^{\beta \mu_{1}(x)} = e^{\beta \mu_{0}} - \frac{J}{v_{e} N_{c}} 
e^{\beta E_{c}^{m}(0,x)} {\cal R}_{1}(x) 
\label{eq:prochem01}
\end{equation}
$(0 < x < S)$, where we have introduced the ``resistance function'' 
${\cal R}_{1}(x)$  as
\begin{equation}
{\cal R}_{1}(x) \equiv \left\langle {\cal Z}^{N} ( \{ x_{j}^{N} \} ) 
\right\rangle_{[0, x]} , 
\label{eq:prochem03}
\end{equation}
which, in turn, can be expressed as 
\begin{equation}
{\cal R}_{1}(x) = \frac{v_{e} N_{c}}{J} e^{- \beta E_{c}^{m}(0,x)} [
e^{\beta \mu_{0}} - e^{\beta \mu_{1}(x)} ] .
\label{eq:prochem03gcx}
\end{equation}
On the other hand, by replacing in Eq.~(\ref{eq:gendr004}) the fixed position 
$0$ with the variable $x$, so that $\mu_{0}$ turns into a chemical-potential
function $\mu_{2}(x)$ and the range of averaging is reduced to $[x, S]$, we 
have 
\begin{equation}
e^{\beta \mu_{2}(x)} =  e^{\beta \mu_{S}}  + \frac{J}{v_{e} N_{c}}
e^{\beta E_{c}^{m}(x,S)} {\cal R}_{2}(x) 
\label{eq:prochem02}
\end{equation}
$(0 < x < S)$, with the resistance function ${\cal R}_{2}(x)$ defined by
\begin{equation}
{\cal R}_{2}(x) \equiv  \left\langle {\cal Z}^{N} (\{ x_{j}^{N} \} ) 
\right\rangle _{[x, S]} , 
\label{eq:prochem04}
\end{equation}
so that
\begin{equation}
{\cal R}_{2}(x) = \frac{v_{e} N_{c}}{J} e^{- \beta E_{c}^{m}(x,S)} 
[e^{\beta \mu_{2}(x)} - e^{\beta \mu_{S}}] . 
\label{eq:prochem04ldp}
\end{equation}
From the definitions (\ref{eq:gendr003}), (\ref{eq:prochem03}), and 
(\ref{eq:prochem04}), it follows that
\begin{equation}
{\cal R}_{1}(S) =  {\cal R}_{2}(0) = {\cal R} .
\label{eq:prochem04zwe}
\end{equation}
The resistance functions ${\cal R}_{1}(x)$ and ${\cal R}_{2}(x)$ are 
obtained in explicit form from expression (\ref{eq:gendr009a}) 
[or from expression (\ref{eq:gendr009e}), which includes tunneling 
effects] by transcribing it so as to correspond to samples
with end-point coordinates $0,x$ and $x,S$, respectively. 

From the potentials $\mu_{1,2}(x)$, we now construct a unique chemical 
potential $\mu (x)$  by setting 
\begin{equation}
e^{\beta \mu (x)} = \smfrac{1}{2} [e^{\beta \mu_{1} (x)} + 
e^{\beta \mu_{2} (x)} ] 
\label{eq:prochem05}
\end{equation}
for $0 < x < S$, while
\begin{equation}
\mu (0) = \mu_{0} , \; \mu (S) = \mu_{S} .
\label{eq:prochem05jcs}
\end{equation}
Using Eqs.~(\ref{eq:prochem01}) and (\ref{eq:prochem02}), along with
Eq.~(\ref{eq:gendr004}) in Eq.~(\ref{eq:prochem05}), we obtain
\begin{equation}
e^{\beta \mu (x)} = \smfrac{1}{2} (e^{\beta \mu_{0}} + e^{\beta \mu_{S}}) 
- \frac{{\cal R}_{-}(x)}{2 {\cal R}}(e^{\beta \mu_{0}} - e^{\beta \mu_{S}})  ,  
\label{eq:prochem06}
\end{equation}
where the function ${\cal R}_{-}(x)$ is defined as
\begin{eqnarray} 
{\cal R}_{-}(x) &=& e^{- \beta E_{c}^{m}(0,S)} \nonumber \\
&\ & \times [e^{\beta E_{c}^{m}(0,x)} {\cal R}_{1}(x) - 
e^{\beta E_{c}^{m}(x,S)} {\cal R}_{2}(x)] .  \nonumber \\
\label{eq:prochem06gcl}
\end{eqnarray}
With the resistance functions ${\cal R}_{1,2}(x)$  and the reduced resistance 
${\cal R}$ calculated from Eq.~(\ref{eq:gendr009a}) [or 
Eq.~(\ref{eq:gendr009e})], the equilibrium chemical potential in the 
prototype model can now be explicitly evaluated in terms of the potential 
energy profile $E_{c}(x)$ and the momentum relaxation length $l$.

A noteworthy feature of the chemical potential $\mu(x)$ is the occurrence of 
discontinuities at $x = 0$ and $x = S$. Using Eqs.~(\ref{eq:gendr004}), 
(\ref{eq:gendr009a}), (\ref{eq:gendr009b}), and (\ref{eq:prochem04zwe}), we 
find from Eq.~(\ref{eq:prochem06})
\begin{equation}
e^{\beta [\mu (0^{+}) - \mu_{0}]} - 1 = - \frac{J}{2 v_{e} N_{c}} e^{\beta 
[E_{c}(0) - \mu_{0}]} 
\label{eq:prochem06amf}
\end{equation}
for $x = 0$, and 
\begin{equation}
e^{\beta [\mu (S^{-}) - \mu_{S}]} - 1 =  \frac{J}{2 v_{e} N_{c}} e^{\beta 
[E_{c}(S) - \mu_{S}]}  
\label{eq:prochem06pky}
\end{equation}
for $x = S$. 

In the fully thermoballistic approach (see Sec.~\ref{sec:equichem} below), we 
will derive an expression for the average chemical potential which closely 
resembles expression (\ref{eq:prochem06}) and which exhibits discontinuities 
analogous to those expressed by Eqs.~(\ref{eq:prochem06amf}) and 
(\ref{eq:prochem06pky}), but relies on different assumptions.

\subsection{Prototype thermoballistic model:\ Examples}

\label{sec:gendimpl}

The present subsection is devoted to the application of the prototype 
thermoballistic 
model to specific examples. With a view to inhomogeneous semiconductors and 
heterostructures, in which barriers in the potential energy profile arise, 
e.g., at heterojunctions, grain boundaries, and metal-semiconductor contacts 
(``Schottky barriers''),  we deal with the evaluation of the reduced 
resistance ${\cal R}$ for profiles $E_{c}(x)$ exhibiting an arbitrary number of
barriers and interjacent valleys. As a prelude to the general case, we 
treat the cases of a single barrier and of two barriers enclosing a 
valley separately. Another example considered here is that of field-driven 
transport in homogeneous semoconductors, for which we evaluate the position 
dependence of the chemical potential.      

\subsubsection{Single potential energy barrier}

\label{sec:onepeak}

Here, we consider the case of a single barrier in the profile $E_{c}(x)$, with
its maximum located at some position $X \in [0,S]$ (see Fig.~\ref{fig:1}).  
Then, we have $E_{c}^{m}(0,S) = E_{c}(X)$.  If, in particular, $X = 0$ or 
$X = S$, the profile is monotonic.

In the calculation of the corresponding reduced resistance ${\cal R}$ from 
expression (\ref{eq:gendr009a}), one must distinguish three cases.  If the 
ballistic interval $[x', x'']$ contains $X$, we have $E_{c}^{m}(x', x'') = 
E_{c}(X)$;\ if it lies to the left or right of $X$, we have $E_{c}^{m}(x',x'')
= E_{c}(x'')$ or $E_{c}^{m}(x',x'') = E_{c}(x')$, respectively.  We then find
\begin{equation}
{\cal R} = 1 + \frac{\tilde{S}}{\ell} ,
\label{eq:gendr011}
\end{equation}
where $\tilde{S}$ is the effective sample length,
\begin{equation}
\tilde{S} = \int_{0}^{S} dx e^{- \beta [E_{c}(X) - E_{c}(x)]} .
\label{eq:gendr012}
\end{equation}
This expression is formally equal to the effective sample length
(\ref{eq:diff008}) introduced in the context of the drift-diffusion model, but
it differs in physical origin.  While expression (\ref{eq:gendr012}) results 
from averaging over ballistic configurations involving intervals of arbitrary, 
finite length, expression (\ref{eq:diff008}) arises from the assumption of 
arbitrarily short ballistic intervals, on which the drift-diffusion model is 
implicitly based.

In expression (\ref{eq:gendr011}) for ${\cal R}$, the unit term corresponds to
ballistic transmission all across the sample, governed by the barrier maximum
at $X$, while the effective sample length $\tilde{S}$ represents the effect of
the potential energy profile in an integral way.  The (classical)
current-voltage characteristic (\ref{eq:gendr004a}) for the case of a single
potential energy barrier now reads
\begin{equation}
J = v_{e} N_{c} e^{- \beta [E_{c}(X) - \mu_{0}]} \frac{\ell}{\ell +
\tilde{S}} ( 1 - e^{-\beta e V} ) .
\label{eq:gendr004c}
\end{equation}
If the effective mean free path is much longer than the effective sample 
length, $\ell  \gg \tilde{S}$, the characteristic (\ref{eq:gendr004c}) reduces 
to that of the ballistic transport model, Eq.~(\ref{eq:ball009}).  In the 
opposite case, $\ell \ll \tilde{S}$, using the relation
\begin{equation}
\nu = 2 \beta e v_{e} l =  \beta e v_{e} \ell 
\label{eq:ballim20a}
\end{equation}
for the electron mobility $\nu$ [see Eqs.~(\ref{eq:drude004hg}) and 
(\ref{eq:ball002})], the characteristic of the drift-diffusion model, 
Eq.~(\ref{eq:diff006}), is retrieved.  Expression
(\ref{eq:gendr004c}) exemplifies the unification of the ballistic and
drift-diffusion transport mechanisms in the prototype thermoballistic model.

We note that a heuristic attempt to unify drift-diffusion and ballistic
transport has been made, for the special case of transport across a Schottky
barrier, by Crowell and Sze\cite{cro66} (see also Ref.~\onlinecite{sze81}), who
assumed ballistic transport to prevail in the vicinity of the barrier maximum,
and drift-diffusion transport elsewhere.  They obtained an expression for the
current-voltage characteristic equivalent to (\ref{eq:gendr004c}),
\begin{equation}
J = N_{c} e^{- \beta [E_{c}(X) - \mu_{0}]} \frac{v_{e} v_{di}}{v_{e} +
v_{di}} ( 1 - e^{-\beta e V} ) ,
\label{eq:gendr004f}
\end{equation}
where
\begin{equation}
v_{di} = \frac{\nu}{ \beta e \tilde{S}} =  \frac{v_{e} \ell}{ \tilde{S}}
\label{eq:gendr004d}
\end{equation}
is an effective diffusion velocity, with the electron mobility $\nu$ given by
Eq.~(\ref{eq:ballim20a}).  A picture similar to that of
Ref.~\onlinecite{cro66} has been developed and applied by Evans and
Nelson\cite{eva91} in a study of transport across a single grain boundary
barrier.  Other studies elucidating the transition region between
drift-diffusion and ballistic transport were presented by de Jong\cite{jon94}
and Prins {\em et al.}\cite{pri98}

Considering now, for the single barrier, the effect of electron tunneling, we 
obtain, following Sec.~\ref{sec:quantum},
\begin{eqnarray}
{\cal R}_{sb} &=& e^{- S/\ell} \bar{\cal R}_{sb}(0,S) + \int_{X}^{S}
\frac{dx''}{\ell} e^{-x''/\ell} \bar{\cal R}_{sb}(0,x'')
\nonumber \\ &\ & \hspace{3.5mm} +\int_{0}^{X} \frac{dx'}{\ell}
e^{-(S-x')/\ell} \bar{\cal R}_{sb}(x',S) \nonumber \\ &\ & \hspace{3.5mm} +
\int_{0}^{X} \frac{dx'}{\ell} \int_{X}^{S} \frac{dx''}{\ell} e^{-(x''-x')/\ell}
\bar{\cal R}_{sb}(x',x'')  \nonumber \\
\label{eq:gendr009s}
\end{eqnarray}
for the sub-barrier part ${\cal R}_{sb}$ of the reduced resistance 
${\cal R}_{q}$, while the classical part ${\cal R}_{cl}$ is given by 
Eq.~(\ref{eq:gendr011}). Since $| \bar{\cal R}_{sb}(x',x'')| < 1$, we have
from Eq.~(\ref{eq:gendr009s})
\begin{equation}
|{\cal R}_{sb}| < 1 .
\label{eq:gendr009sh}
\end{equation}
Since $E_{c}^{m}(x', x'') \equiv E_{c}(X)$, the transmission probabilities
$\bar{T}(x',x'')$ and $\bar{\cal T}_{sb}(x',x'')$, and consequently the
function $\bar{\cal R}_{sb}(x',x'')$, depend on $x'$ and $x''$ only via the
function $E_{c}^{>}(x', x'')$ [see Eqs.~(\ref{eq:ball005d}) and
(\ref{eq:ball010x})]. Then, $\bar{\cal R}_{sb}(x',x'')$ is a function of
one coordinate only, determined by the potential energy profile $E_{c}(x)$,
\begin{equation}
\bar{\cal R}_{sb}(x',x'') \equiv \left\{ \begin{array}{ll} \bar{\cal
R}_{sb}(x') & \hspace{-1.5mm} ; E_{c}^{>}(x', x'') = E_{c}(x') , \\
\bar{\cal R}_{sb}(x'') & \hspace{-1.5mm} ; E_{c}^{>}(x', x'') = E_{c}(x'')
. \end{array} \right.  \nonumber \\
\label{eq:gendr009glp}
\end{equation}
As $x''$ increases from $X$ to $S$, for fixed $x' \in [0, X]$, $\bar{\cal
R}_{sb}(x'')$ changes to $\bar{\cal R}_{sb}(x')$ when $x'' = x'^{\ast}$ or,
equivalently, $x' = x''^{\ast}$;\ here, the ``mirror point'' $x^{\ast}$ is
defined as that position to the right (left) of a maximum or minimum of the 
potential energy profile where it has the same value as at the point $x$ to 
the left (right), i.e., 
\begin{equation}
E_{c}(x^{\ast}) = E_{c}(x)
\label{eq:gendr009uvy}
\end{equation}
(see Figs.~\ref{fig:1} and \ref{fig:2}).   The double integral in 
expression (\ref{eq:gendr009s}) then reduces to two single integrals which, 
together with the single integrals preceding it, combine to just one single 
integral, so that ${\cal R}_{sb}$ attains the simple form
\begin{eqnarray}
{\cal R}_{sb} &\equiv& {\cal R}_{sb}(\ell) \nonumber \\ &=& e^{- 0^{\ast}/\ell}
\bar{\cal R}_{sb}(0) + \int_{0}^{0^{\ast}} \frac{dx}{\ell} 
e^{-|x - x^{\ast}|/\ell} \bar{\cal R}_{sb}(x) . \nonumber \\
\label{eq:gendr009st}
\end{eqnarray}
This expression bears a close formal similarity to the shape term appearing in
the classical treatment of the double-barrier case 
[see Eq.~(\ref{eq:vall002a}) below].

\subsubsection{Double barrier}

\label{sec:twopeak}

\begin{figure}[t]
\includegraphics[width=0.45\textwidth]{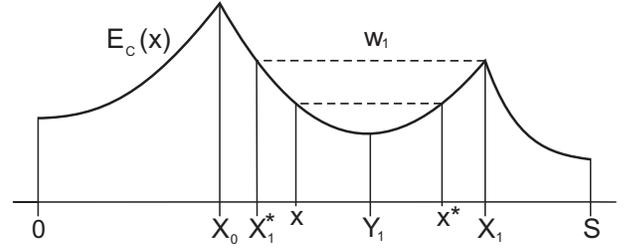} 
\caption{Schematic diagram of a potential energy profile $E_{c}(x)$ exhibiting 
two barriers with maxima at $x = X_{0}$ and $x = X_{1}$, respectively, which 
enclose a valley with minimum at $x = Y_{1}$ and width $w_{1}$. For the 
definition of the mirror point $x^{\ast}$ associated with a point $x$, see 
Eq.~(\ref{eq:gendr009uvy}).  
}
\label{fig:2}
\end{figure}

Next, we consider a potential energy profile of the type shown in
Fig.~\ref{fig:2}, exhibiting two barriers with maxima at $x = X_{0}$ and $x =
X_{1}$, respectively (without loss of generality, the maximum at $X_{0}$ is
assumed to be the higher one), and a valley with minimum at $x = Y_{1}$ in
between.

In the evaluation of the reduced resistance ${\cal R}$, 
a number of cases are to be distinguished when 
calculating the function $E_{c}^{m}(x', x'')$ in dependence on the location of 
the ballistic interval $[x', x'']$ with respect to the different ranges 
delimited by the points $x = 0$, $X_{0}$, $X_{1}^{\ast}$, $Y_{1}$, $X_{1}$, and
$S$.  From expression (\ref{eq:gendr009a}), we obtain 
\begin{eqnarray}
{\cal R} = 1 + \frac{\tilde{S} + \tilde{\Lambda}}{\ell} ,
\label{eq:vall002}
\end{eqnarray}
where the effective sample length $\tilde{S}$ is given by
Eq.~(\ref{eq:gendr012}) with $X = X_{0}$.  The ``shape term'' $\tilde{\Lambda}$
has the form
\begin{eqnarray}
\tilde{\Lambda} \equiv \tilde{\Lambda} (\ell) &=&\int_{X_{1}^{\ast}}^{X_{1}} dx
e^{-|x-x^{\ast}|/\ell} e^{- \beta E_{c}(X_0)} \nonumber \\ &\ & \hspace{0.7cm}
\times [ e^{\beta E_{c}(X_{1})} - e^{\beta E_{c}(x)} ] .
\label{eq:vall002a}
\end{eqnarray}
In writing $\tilde{\Lambda}$ in this compact form, we have used the identity
\begin{equation}
1 - e^{-w_{1}/\ell} \equiv \int_{X_{1}^{\ast}}^{X_{1}} \frac{dx}{\ell}
e^{-|x-x^{\ast}|/\ell} ,
\label{eq:vall002b}
\end{equation}
where $w_{1} = X_{1} - X_{1}^{*}$ is the width of the valley (see
Fig.~\ref{fig:2}).  The position of the minimum of the valley,
$Y_{1}$, enters ${\cal R}$ implicitly via the shape of $E_{c}(x)$ in the 
interval $[X_{1}^{\ast}, X_{1}]$.

Comparing in Eq.~(\ref{eq:vall002a}) the magnitude of the integral involving
the term $e^{\beta E_{c}(X_{1})}$ in the brackets to that of the integral
involving $e^{\beta E_{c}(x)}$, it is seen that for $b_{0,1} \ll l \ll
w_{1}$, where $b_{0,1}$ are the widths of the two barriers centered about
$X_{0,1}$, the second integral can be neglected. Hence,
\begin{equation}
{\cal R} = 1 + \frac{\tilde{S}}{\ell} + e^{-\beta [ E_{\rm c}(X_{0}) - 
E_{\rm c}(X_{1})]} ,
\label{eq:vall004}
\end{equation}
i.e., the two barriers contribute independently to the reduced resistance
(first and third term).

When $\ell \gg S$ (ballistic limit), we have
\begin{equation}
\tilde{\Lambda} < w_{1} e^{-\beta [E_{c}(X_{0}) - E_{c}(X_{1})]} < S ,
\label{eq:vall005}
\end{equation}
so that both the terms $\tilde{S}/\ell$ and $\tilde{\Lambda}/\ell$ in
Eq.~(\ref{eq:vall002}) can be neglected.  Then, ${\cal R}$ reduces to the unit 
term, which reflects ballistic transmission across the higher barrier maximum 
at $X_{0}$;\ this maximum ``eclipses'' the lower maximum at $X_{1}$.

\subsubsection{Arbitrary number of barriers}

\label{sec:peaval}

In the foregoing cases of a single barrier and of a valley in-between two
barriers, we have obtained explicit expressions for the reduced resistance
${\cal R}$ in terms of the potential energy profile $E_{c}(x)$ and the 
effective momentum relaxation length $\ell$.  If the
profile contains an arbitrary combination of barriers and valleys, ${\cal R}$ 
must be evaluated, in general, for each case anew, starting from the basic 
formula (\ref{eq:gendr009a}).  However, an explicit expression for ${\cal R}$ 
can still be obtained in the special case where the {\em height} of the 
barriers decreases or increases {\em monotonically} along the sample length.

\begin{figure}[t]
\includegraphics[width=0.45\textwidth]{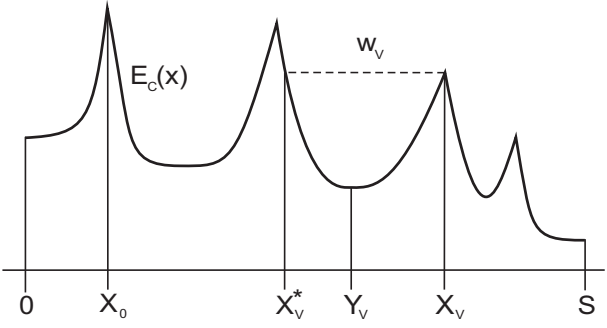}
\caption{Schematic diagram of a potential energy profile $E_{c}(x)$ exhibiting 
$M+1$ barriers with maxima at $x = X_{v}$ ($v=0,1, \ldots ,M$), and
heights decreasing monotonically with increasing $x$. The valleys
enclosed by two adjacent barriers have minima located at $x = Y_{v}$ 
($v=1, \ldots ,M$) and widths $w_{v}$.
}
\label{fig:3}
\end{figure}

Assuming, without loss of generality, the barrier height to decrease
monotonically when $x$ increases from 0 to $S$ (which includes the case of equal
height of all barriers), we consider $M+1$ barriers, with maxima at 
$x = X_{v}$ $(v = 0, 1, \ldots, M)$, enclosing $M$ valleys, with minima at 
$x = Y_{v}$ ($v=1, \ldots ,M$) and 
widths $w_{v}$ (see Fig.~\ref{fig:3}).  Then, using Eq.~(\ref{eq:gendr009a})
and proceeding as in the case of a single valley, we obtain ${\cal R}$ in the
general form given by Eq.~(\ref{eq:vall002}), with $\tilde{S}$ again given by
Eq.~(\ref{eq:gendr012}) with $X = X_{0}$, where $X_{0}$ is now the position of
the maximum of the heighest (left-most) barrier of the $M+1$ barriers
considered.  The shape term $\tilde{\Lambda}$ entering expression
(\ref{eq:vall002}),
\begin{eqnarray}
\tilde{\Lambda} \equiv \tilde{\Lambda} (\ell) &=& \sum_{v=1}^{M}
\int_{X_{v}^{\ast}}^{X_{v}}dx e^{-|x-x^{\ast}|/\ell}  \nonumber \\ &\ & \times
e^{-\beta E_{c}(X_{0})} [ e^{\beta E_{c}(X_{v})} - e^{\beta E_{c}(x)}] , 
\nonumber \\
\label{eq:vall006a}
\end{eqnarray}
generalizes expression (\ref{eq:vall002a}) to the case of $M$ valleys;\ it
appears as a sum over separate contributions from the different valleys.  The
contribution of valley $v$, with the maximum of the adjoining lower barrier
located at $X_{v}$, consists of an integral which extends over the width of the
valley from $X_{v}^{*}$ to $X_{v}$.

When the barrier heights in the potential energy profile do not behave
monotonically, inspection of formula (\ref{eq:gendr009a}) shows that the
reduced resistance ${\cal R}$ can always be expressed as a sum of a term
formally identical to expression (\ref{eq:vall002}) [with $E_{c}(X_{0})$ in
$\tilde{S}$ and $\tilde{\Lambda}$ replaced with $E_{c}(X_{m})$, where $X_{m}$
is the position of the maximum of the heighest barrier] plus additional terms
arising from the combined effect of barriers and valleys on the electron
transport in the ballistic intervals.

Introducing now the effective transport length as
\begin{equation}
L  \equiv {\cal R} \ell ,
\label{eq:vall007a}
\end{equation}
we have from
Eqs.~(\ref{eq:gendr009x}), (\ref{eq:gendr009y}), and (\ref{eq:vall002})
\begin{equation}
L  = \ell + \tilde{S} + \tilde{\Lambda}(\ell) \leq \ell + S .
\label{eq:vall007}
\end{equation}
From Eqs.~(\ref{eq:gendr004a}) and (\ref{eq:vall007a}), we then obtain the
current-voltage characteristic for a potential energy profile of the type
depicted in Fig.~\ref{fig:3} in the form
\begin{equation}
J = v_{e} N_{c} e^{-\beta [E_{c}(X_{0}) - \mu_{0}]} \frac{\ell}{L} ( 1-
e^{-\beta eV} ) .
\label{eq:vall008}
\end{equation}
This expression is the principal result of the prototype thermoballistic
model.  It has been derived here in the framework of classical transport in
nondegenerate systems, where its interpretation is most transparent.

The properties of the characteristic (\ref{eq:vall008}) are determined by the
barrier factor $e^{- \beta [E_{c}(X_{0}) - \mu_{0}]}$ and by the ratio $\ell/L$.
In the effective transport length $L$, the effective mean free path $\ell$ 
represents the
ballistic contribution to the current, which is associated with the maximum
$E_{c}(X_{0})$ of the highest barrier in the potential energy profile.  The
remaining terms give a quantitative measure of the influence of that part of
the electron motion which is not purely ballistic.  Their contribution amounts
to at most the sample length $S$.  The effective sample length $\tilde{S}$
given by Eq.~(\ref{eq:gendr012}) with $X = X_{0}$ represents a contribution
that characterizes the potential energy profile $E_{\rm c}(x)$ in an integral
way; it does not manifestly depend on $\ell$, only implicitly so via the 
profile
(an indirect relationship between profile and mean free path arises from their
common dependence on the donor density).  The shape term $\tilde{\Lambda}$
given by Eq.\ (\ref{eq:vall006a}), on the other hand, depends on the detailed
structure of the profile as well as explicitly on the mean free path, and thus
represents the interplay of ballistic and drift-diffusion transport.  This term
is a distinctive feature of expression (\ref{eq:vall008}), and therefore of the
prototype thermoballistic model.

Tunneling effects in the current-voltage characteristic (\ref{eq:vall008}) can 
be taken into account by generalizing the effective transport length $L$ of 
Eq.~(\ref{eq:vall007a}) so as to include the sub-barrier contribution 
${\cal R}_{sb}$ to the reduced resistance. Using Eq.~(\ref{eq:ball003p}), we 
have
\begin{eqnarray}
L_{q} &\equiv& {\cal R}_{q} \ell  = ({\cal R}_{cl} + {\cal R}_{sb})\ell 
\nonumber \\ &=& ( 1 - |{\cal R}_{sb}(\ell)| ) \ell + \tilde{S}  +
 \tilde{\Lambda}(\ell) .
\label{eq:gendr009r}
\end{eqnarray}
Here, as exemplified by expression (\ref{eq:gendr009st}) for the single-barrier
case, ${\cal R}_{sb}$ depends explicitly on the effective mean free path 
$\ell$.  From expression (\ref{eq:gendr009r}), the inclusion of tunneling is
seen to {\em lower} the {\em ballistic} contribution to the effective transport
length [see Eq.~(\ref{eq:vall007a})].  This behavior is in line with the role
of ${\cal R}_{q}$ as a resistance, which ought to decrease when 
the barriers become ``transparent'' to ballistic electron transport.

The (classical) reduced resistance in the degenerate case, ${\cal R}_{d}$ [see 
Eqs.~(\ref{eq:degen003a}) and (\ref{eq:degen004})], is expressed, in analogy to
Eqs.~(\ref{eq:vall007a}) and (\ref{eq:vall007}) for the nondegenerate case, 
in terms of a (classical) effective transport length $L_{d}$ given by
\begin{equation}
L_{d} \equiv {\cal R}_{d} \ell = \ell + \tilde{S}_{d} + 
\tilde{\Lambda}_{d}(\ell) ,
\label{eq:degen006}
\end{equation}
where, in generalization of Eq.~(\ref{eq:gendr012}),
\begin{equation}
\tilde{S}_{\rm d} = \int_{0}^{S} dx \frac{\ln \bm{(} 1+ e^{- \beta [
E_{c}(X_{0}) - \mu_{0}]} \bm{)}}{\ln \bm{(} 1+ e^{- \beta [E_{c}(x) - \mu_{0}]}
\bm{)}}
\label{eq:degen007}
\end{equation}
is the effective sample length for the degenerate case, and
\begin{eqnarray}
\tilde{\Lambda}_{d} &\equiv& \tilde{\Lambda}_{d}(l) = \sum_{v=1}^{M}
\int_{X_{v}^{\ast}}^{X_{v}} dx e^{-|x-x^{\ast}|/\ell} \nonumber \\ &\ &
\hspace{-0.8cm} \times \left\{ \frac{\ln \bm{(} 1 + e^{- \beta [E_{c}(X_{0}) -
\mu_{0}]} \bm{)} }{\ln \bm{(} 1 + e^{- \beta [E_{c}(X_{v}) - \mu_{0}]} \bm{)} }
- \frac {\ln \bm{(} 1 + e^{- \beta [E_{c}(X_{0}) - \mu_{0}]} \bm{)} }{\ln
\bm{(} 1 + e^{ -\beta [ E_{c}(x) - \mu_{0}]}\bm{)} } \right\} \nonumber \\
\label{eq:degen008}
\end{eqnarray}
generalizes the shape term (\ref{eq:vall006a}).  The inclusion of tunneling
effects is, in the degenerate case, a highly intricate task and will not be
considered here.

Setting $\mu_{0} - \mu_{S} = eV$ in Eq.~(\ref{eq:degen005}),  we now have for
the zero-bias conductance per unit area in the prototype thermoballistic model
\begin{eqnarray}
g &\equiv& \left( \frac{eJ}{V} \right)_{V \rightarrow 0} \nonumber \\
&=& \beta e^{2} v_{e} N_{c} \ln \bm{(} 1 + e^{- \beta
(E_{c}^{m} - \mu_{0}) } \bm{)} \frac{\ell}{L_{d}} , 
\label{eq:degen009n}
\end{eqnarray}
in generalization of expression (\ref{eq:degen009}) for the conductance in the
ballistic transport model.

Previously, we have applied\cite{lip01,wei98,wei99,lip01a,wei02} the prototype
model in calculations of transport properties of poly-\ and microcrystalline 
semiconducting materials, in particular, of materials relevant to 
photovoltaics. The occurence of grain boundaries in this 
kind of materials gives rise to a multi-barrier structure of the band edge 
profile. Adopting the trapping model\cite{kam71,set75,wei98,wei99} to describe 
the grain boundaries, we have solved\cite{wei99} the corresponding nonlinear 
Poisson equation to obtain zero-bias potential energy profiles for chains of 
grains. Then, taking into account tunneling corrections and using a 
phenomenological relation\cite{aro82} to express the momentum relaxation 
length $l$ in terms of the donor density, we have calculated zero-bias 
conductivities and electron mobilities as a function of $l$, and of the number 
and lengths of the grains. It turns out that neither the ballistic 
(thermionic) model nor the drift-diffusion model can provide an adequate
description of electron transport in poly-\ and microcrystalline materials.
The application of the prototype model in the analysis of experimental 
data\cite{che05,mug05,che06,opr07,bik12} has led to promising results.

\subsubsection{Chemical potential for field-driven transport}

\label{sec:potfi}

The chemical potential $\mu (x)$ given by Eq.~(\ref{eq:prochem06}) can be 
expressed in closed form for the case of electron transport in a homogeneous 
semiconductor (no space charges), driven by a constant external electric field
of magnitude ${\cal E}$. 

With the field assumed to be directed antiparallel to the $x$-axis, the 
corresponding potential energy profile reads 
\begin{equation}
E_{c}(x) = E_{c}(0) - \frac{\epsilon}{\beta} x ,
\label{eq:potfi01}
\end{equation} 
where $\epsilon = \beta e {\cal E}$. Using this in the properly transcribed
expression (\ref{eq:gendr009a}) [or, alternatively, in  Eq.~(\ref{eq:gendr011}),
considering the profile (\ref{eq:potfi01}) a particular case of a barrier], we 
obtain the resistance functions ${\cal R}_{1,2}(x)$ in the form
\begin{equation}
{\cal R}_{1}(x)  = 1 + \frac{1}{\epsilon \ell} (1 - e^{- \epsilon x}) 
\label{eq:potfi02}
\end{equation} 
and
\begin{equation}
{\cal R}_{2}(x) = 1 + \frac{1}{\epsilon \ell} [1 - e^{- \epsilon (S - x)}]
\equiv {\cal R}_{1}(S-x) .
\label{eq:potfi03}
\end{equation} 
From Eq.~(\ref{eq:prochem04zwe}), we then have 
\begin{equation}
{\cal R} = {\cal R}_{1}(S)  = 1 + \frac{1}{\epsilon \ell} 
(1 - e^{- \epsilon S}) , 
\label{eq:potfi04a}
\end{equation} 
and from Eq.~(\ref{eq:prochem06gcl}),
\begin{eqnarray}
{\cal R}_{-}(x) &=& {\cal R}_{1}(x) - e^{- \epsilon x} 
{\cal R}_{2}(x) \nonumber \\ &=&  1 - e^{- \epsilon x} 
+  \frac{1}{\epsilon \ell} (1 - 2 e^{- \epsilon x} +   e^{- \epsilon S}) .
\label{eq:potfi04}
\end{eqnarray} 
Inserting expressions (\ref{eq:potfi04a}) and (\ref{eq:potfi04}) in 
Eq.~(\ref{eq:prochem06}) gives $\mu(x)$ in closed form. 
 
For zero bias, when $ \epsilon x \leq \epsilon S \ll 1$, we have
\begin{equation}
{\cal R} = 1 + \frac{S}{\ell} 
\label{eq:potfi05a}
\end{equation} 
and
\begin{equation}
{\cal R}_{-}(x) = \frac{2 x - S}{\ell} ,
\label{eq:potfi05}
\end{equation} 
i.e., $e^{\beta \mu(x)}$ varies linearly with position.

\section{Thermoballistic approach:\ Concept}

\label{sec:therconc}

The prototype thermoballistic model developed in the preceding section has been
based on the random partitioning of the length of a semiconducting sample into
ballistic transport intervals linked by points of local thermodynamic
equilibrium, which make up a ballistic configuration.  Electrons thermally
emitted at either end-point of a ballistic interval, and subsequently
transmitted across the potential energy profile in the sample, form the net
ballistic electron current in the interval.  This current is conserved across
an individual ballistic interval.  A distinctive assumption of the prototype
model has been that the current is conserved also across the 
points of local thermodynamic equilibrium linking the ballistic intervals, and 
equals the physical current.  By averaging over all ballistic configurations, 
the current-voltage characteristic can then be expressed, without requiring 
knowledge of the equilibrium chemical potential inside the sample, essentially 
in terms of a reduced resistance that comprises the effect of the sample 
parameters. The position dependence of the chemical potential has been 
constructed in a heuristic way only.

While adhering to the idea of introducing ballistic configurations and
averaging thereover, the thermoballistic concept proper refines the prototype
model in that it abandons the assumption of current conservation across the 
points of local thermodynamic equilibrium. [A simple example contradicting this 
assumption is provided by the case of the ballistic currents in a homogeneous
semiconductor at zero bias considered in Sec.~\ref{sec:homsem} below.]   
Position-dependent total and spin-polarized thermoballistic currents as well as
the associated densities are defined in terms of an average chemical potential
and a spin accumulation function related to the splitting of the spin-resolved
chemical potentials. By imposing appropriately chosen physical conditions on
these dynamical functions, procedures for their explicit determination are
devised. 

We have developed the thermoballistic approach to semiclassical carrier
transport\cite{semcl} in a series of papers.  In Ref.~\onlinecite{lip03}, the 
concept of a thermoballistic current was introduced and implemented without 
regard to spin degrees of freedom.  The extension of this concept to 
spin-polarized electron transport across a spin-degenerate potential energy 
profile was presented in Ref.~\onlinecite{lip05}, in which spin injection out 
of ferromagnetic contacts into a nonmagnetic semiconducting sample was treated 
in detail.  In Ref.~\onlinecite{lip09}, we have generalized the thermoballistic
concept to the case of arbitrary spin splitting of the profile, thereby 
covering, in particular, spin-polarized transport in diluted magnetic 
semiconductors in their paramagnetic phase.  In the present section, we 
formulate the thermoballistic
concept within the frame set by Ref.~\onlinecite{lip09}.  Classical transport
in nondegenerate systems is considered throughout.  Effects of electron
tunneling and degeneracy can be included, in principle, by resorting to the 
corresponding developments in Secs.~\ref{sec:ballmech} and \ref{sec:gendrude}.

For comprehensive surveys of the fundamentals of spin physics in semiconductors
and their application in the field of spintronics, we refer the reader to
Refs.~\onlinecite{fab07} and \onlinecite{aws02,zut04,dya08}.

\subsection{Electron densities at local thermodynamic equilibrium}

\label{sec:eqeldens}

\begin{figure}[t]
\includegraphics[width=0.45\textwidth]{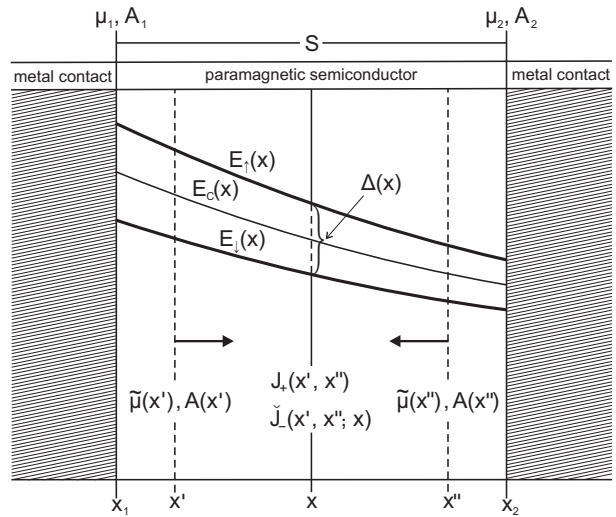}
\caption{Schematic diagram showing, in a one-dimensional representation, a
paramagnetic semiconducting sample of length $S$ enclosed between two
metal contacts.  The spin splitting $\Delta (x)$ of the
electrostatic potential energy profile $E_{c}(x)$ gives rise to the
spin-dependent potential energy profiles $E_{\uparrow , \downarrow}(x)$.  The
coordinates $x'$ and $x''$ denote points of local thermodynamic equilibrium
which delimit a ballistic transport interval $[x',x'']$.  Electrons thermally 
emitted at $x' (x'')$ toward the right (left) move ballistically across the 
profiles $E_{\uparrow , \downarrow}(x)$ to reach the end-point $x'' (x')$ of 
the ballistic interval, where they are absorbed (equilibrated).  
Two net electron currents are indicated:\ The
(conserved) net total ballistic current $J_{+}(x',x'')$ determined by
the values $\tilde{\mu} (x')$ and $\tilde{\mu} (x'')$ of the average chemical
potential $\tilde{\mu}(x)$ and the values $A(x')$ and $A(x'')$  of the spin
accumulation function $A(x)$ according to Eq.~(\ref{eq:ball004jvx}) [with the 
mean spin function $\tilde{A}(x)$ expressed in terms of $\tilde{\mu}(x)$ via 
Eq.~(\ref{eq:eqden03zk})], and the
net relaxing ballistic spin-polarized current $\check{J}_{-}(x',x'';x)$
determined by $A(x')$ and $A(x'')$ according to Eq.~(\ref{eq:netbal02}).  The 
quantities $\mu_{1,2}$ and $A_{1,2}$ are the values of the equilibrium chemical
potential and the spin accumulation function, respectively, at the contact side
of the contact-semiconductor interfaces [see Eqs.~(\ref{eq:ballim05}) and 
(\ref{eq:ballim05s})].
}
\label{fig:4}
\end{figure}

In Sec.~\ref{sec:gendrude}, we have introduced ballistic transport intervals 
with end-point coordinates $x^{N}_{i-1}, x^{N}_{i}$ characterized by discrete 
labels $N, i$.  When the ballistic configurations made up of these intervals are
averaged over, the description in terms of discrete coordinates turns into one
in terms of the {\em continuous} coordinates $x', x''$ in expressions
(\ref{eq:gendr009a}) and (\ref{eq:gendr009e}) for the reduced resistance
${\cal R}$. Transferring this feature into the formulation of the
thermoballistic concept, we work here with ballistic intervals $[x', x'']$ with
continuous end-point coordinates $x'$ and $x''$  representing points of local 
thermodynamic equilibrium (see also Sec.~\ref{sec:probappr}). For notational 
convenience, we will henceforth label
the end-points of a semiconducting sample by $x_{1}$ and $x_{2}$, respectively,
so that we have
\begin{equation}
x_{1} \leq x' < x'' \leq x_{2} ,
\label{eq:eqden01wla}
\end{equation}
and the sample length $S$ is given by  $S = x_{2} - x_{1}$.

In Fig.~\ref{fig:4}, a paramagnetic semiconducting sample enclosed between two 
metal contacts is depicted in a schematic diagram. Various physical
quantities appearing in the thermoballistic description are indicated.

\subsubsection{Spin-resolved densities}

\label{sec:spinres}

We write the spin-dependent potential energy profiles $E_{\uparrow ,
\downarrow}(x')$ corresponding to spin-up $(\uparrow)$ and spin-down
$(\downarrow)$ conduction band electrons, respectively, in the form
\begin{equation}
E_{\uparrow , \downarrow}(x') = E_{c}(x') \pm \smfrac{1}{2} \Delta (x') .
\label{eq:eqden01}
\end{equation}
Here, the spin-independent part $E_{c}(x')$ is assumed to com\-prise the
conduction band edge potential and the external electrostatic potential, and
$\Delta(x')$ is the spin splitting of the conduction band.  Having in mind
electron transport in diluted magnetic semiconductors in their paramagnetic
phase, we identify this splitting with the (giant) Zeeman splitting due to an
external magnetic field \cite{fur88,mos94,die94,cib08,bha11,gaj11} [we restrict
ourselves to considering a
single Landau level whose energy is assumed to be included in $E_{c}(x')$].  In
presenting the general formalism, we assume both $E_{c}(x')$ and $\Delta(x')$,
and hence $E_{\uparrow , \downarrow}(x')$, to be continuous functions of $x'$
in the interval $[x_{1}, x_{2}]$.  The case of abrupt changes in one or the
other of these functions, which occur at the interfaces in heterostructures,
may be described, in a simplified picture, in terms of discontinuous functions
(see Sec.~\ref{sec:dmsnmshe} below).

In terms of the Boltzmann factors $e^{-\beta E_{\uparrow, \downarrow}(x')}$,
the (static) local spin polarization of the conduction band electrons, $P(x')$,
is given by
\begin{equation}
P(x') = \frac{B_{-}(x')}{B_{+}(x')} ,
\label{eq:eqden06}
\end{equation}
where
\begin{equation}
B_{\pm}(x') = e^{-\beta E_{\uparrow}(x')} \pm e^{-\beta E_{\downarrow}(x')} ,
\label{eq:eqden60}
\end{equation}
so that we have
\begin{equation}
P(x') = - \tanh \bm{(} \beta \Delta(x')/2 \bm{)} .
\label{eq:eqden61}
\end{equation}
The function $Q(x')$ defined by
\begin{eqnarray}
Q(x') &\equiv&  \left\{ 1 - [P(x')]^{2} \right\}^{1/2} \nonumber \\ 
&=& \frac{1}{\cosh \bm{(} \beta \Delta(x')/2 \bm{)}}  
\label{eq:spfra17c}
\end{eqnarray}
will be frequently used below.

For the spin-resolved equilibrium electron densities $n_{\uparrow ,
\downarrow}(x')$, we have, in generalization of expression (\ref{eq:drude014}),
\begin{equation}
n_{\uparrow , \downarrow}(x') = \frac{N_{c}}{2} e^{- \beta [E_{\uparrow ,
\downarrow}(x') - \mu_{\uparrow , \downarrow}(x')]} .
\label{eq:eqden02}
\end{equation}
Here, $\mu_{\uparrow , \downarrow}(x')$ are the spin-resolved chemical
potentials associated with the local thermodynamic equilibrium at $x'$, and
$N_{c}/2$, with $N_{c}$ given by Eq.~(\ref{eq:diff001a}), is the effective
density of states of either spin at the conduction band edge (for simplicity,
the effective electron mass $m^{\ast}$ entering $N_{c}$ is assumed here to be
independent of position and of the external magnetic field).

The spin-resolved chemical potentials $\mu_{\uparrow}(x')$ and
$\mu_{\downarrow}(x')$ are independent dynamical functions
whose position dependence is to be determined within the thermoballistic
approach and from which, subsequently, all transport properties are to be
derived.  However, in implementing the thermoballistic approach, we will work
not with $\mu_{\uparrow}(x')$ and $\mu_{\downarrow}(x')$, but with suitably
defined  combinations of these functions:\ (i) an ``average
chemical potential'', and (ii) a ``spin accumulation function'' related to the
splitting of the potentials $\mu_{\uparrow}(x')$ and $\mu_{\downarrow}(x')$.

\subsubsection{Average chemical potential and spin accumulation function}

\label{sec:spinequi}

We define the ``spin functions'' $A_{\uparrow, \downarrow}(x')$ as
\begin{equation}
A_{\uparrow, \downarrow} (x') = e^{\beta \mu_{\uparrow,\downarrow }(x')} ,
\label{eq:eqden03mm}
\end{equation}
and the ``mean spin function'' $\tilde{A}(x)$ as
\begin{equation}
\tilde{A}(x')  = \smfrac{1}{2}  A_{+}(x') ,
\label{eq:eqden03}
\end{equation}
where
\begin{equation}
A_{+}(x') \equiv A_{\uparrow}(x') + A_{\downarrow}(x') .
\label{eq:eqden03oh}
\end{equation}
We can express $\tilde{A}(x')$ in the form
\begin{equation}
\tilde{A}(x') = e^{\beta \bar{\mu}(x')} \cosh \bm{(} \beta \mu_{-}(x')/2
\bm{)} ,
\label{eq:eqden03nn}
\end{equation}
where
\begin{equation}
\bar{\mu}(x') = \smfrac{1}{2} [\mu_{\uparrow}(x') +\mu_{\downarrow}(x')]
\label{eq:spfra15xxx}
\end{equation}
is the mean value of $\mu_{\uparrow}(x')$ and $\mu_{\downarrow}(x')$ [mean
chemical potential], and
\begin{equation}
\mu_{-} (x') \equiv \mu_{\uparrow} (x') - \mu_{\downarrow} (x')
\label{eq:spfra15x}
\end{equation}
their splitting.  [In the following, the $``\pm''$ notation introduced in
Eqs.~(\ref{eq:eqden03oh}) and(\ref{eq:spfra15x}), respectively, will be
frequently used {\em mutatis mutandis} to denote spin-summed and spin-polarized
quantities.] Further, writing
\begin{equation}
\tilde{A}(x') = e^{\beta \tilde{\mu}(x')} ,
\label{eq:eqden03zk}
\end{equation}
we introduce the {\em ``average chemical potential''} $\tilde{\mu}(x')$.

The {\em ``spin accumulation function''} $A_{-}(x')$ is defined as
\begin{eqnarray}
A_{-}(x') &\equiv& A_{\uparrow}(x') - A_{\downarrow}(x') \nonumber \\&=& 
2e^{\beta \bar{\mu}(x')} \sinh \bm{(} \beta \mu_{-}(x')/2 \bm{)} . 
\label{eq:spfra01}
\end{eqnarray}
[This function agrees with the ``spin transport function'' $A(x')$ of
Ref.~\onlinecite{lip09}, but differs by a factor of two from the identically
named 
function introduced in Ref.~\onlinecite{lip05}.]  The name here chosen for
$A_{-}(x')$ derives from the fact that in the limit $|\beta \mu_{-}(x')| \ll
1$, this function becomes proportional to the splitting $\mu_{-}(x')$, which is
the dynamical quantity used in the drift-diffusion approach to spin-polarized
transport, and which is commonly\cite{zut04} referred to as the ``spin
accumulation`` there.

Using Eq.~(\ref{eq:eqden03nn}), we can rewrite $A_{-}(x')$ in the form
\begin{equation}
A_{-}(x') = 2 \tilde{A}(x') \tanh \bm{(} \beta \mu_{-}(x')/2 \bm{)} .
\label{eq:spfra01fv}
\end{equation}
Introducing the ``reduced'' spin accumulation function $\breve{A}(x')$ as
\begin{equation}
\breve{A}(x') \equiv \frac{A_{-}(x')}{\tilde{A}(x')}  = 2 \tanh \bm{(}
\beta \mu_{-}(x')/2 \bm{)} ,
\label{eq:spfra01hl}
\end{equation}
we then obtain, using Eqs.~(\ref{eq:eqden03zk})--(\ref{eq:spfra01fv}),
the relations
\begin{eqnarray}
\tilde{\mu}(x') &=& \bar{\mu}(x') + \frac{1}{\beta} \ln \bm{(} \cosh
\bm{(}\beta \mu_{-}(x')/2 \bm{)} \bm{)} \nonumber \\ &=& \bar{\mu}(x') -
\frac{1}{2 \beta} \ln \bm{(} 1 - \breve{A}^{2}(x')/4 \bm{)}
\label{eq:eqden08jfc}
\end{eqnarray}
expressing the difference between average and mean chemical potential in terms 
of the splitting $\mu_{-}(x')$ and of $\breve{A}(x')$, respectively.

The spin functions $A_{\uparrow, \downarrow}(x')$ are proportional to the
spin-resolved electron densities $n_{\uparrow , \downarrow}(x')$ [see
Eqs.~(\ref{eq:eqden02})], so that their use results in a formulation in terms
of {\em linear} equations, instead of the nonlinear description ensuing from
using the spin-resolved chemical potentials $\mu_{\uparrow , \downarrow}(x')$
themselves.  [This aspect has been emphasized previously \cite{yuf02,yuf02a}
within a study, based on the standard drift-diffusion approach, of
electric-field effects on spin-polarized transport in nondegenerate
semiconductors.]

From Eqs.~(\ref{eq:eqden02}), the total (i.e., spin-summed) equilibrium
density, $n_{+}(x')$, and the spin-polarized equilibrium density, $n_{-}(x')$,
are now obtained in terms of the functions $A_{\pm}(x')$ as
\begin{equation}
n_{\pm}(x') = \frac{N_{c}}{4} B_{+}(x') [A_{\pm}(x') + P(x') A_{\mp}(x')] ,
\label{eq:ball004jux}
\end{equation}
where Eqs.~(\ref{eq:eqden60}) and (\ref{eq:eqden61}) have been used.

\subsection{Ballistic spin-polarized transport}

\label{sec:balltrans}

In this subsection, the spin-resolved electron currents transmitted across a 
ballistic transport
interval as well as the associated densities (called ``ballistic currents''
and ``ballistic densities'', for short) are constructed  by closely
following the development in the ballistic (see Sec.~\ref{sec:ballmech}) and
the prototype thermoballistic (see Sec.~\ref{sec:ballconf}) transport models,
in which spin degrees of freedom have been disregarded.  Introducing spin 
relaxation during the ballistic electron motion, we obtain the ballistic 
spin-polarized currents and densities, whose dynamics are determined from a 
balance equation.

\subsubsection{Ballistic currents}

\label{sec:ballemis}

The left end-point, $x'$, as well as the right end-point, $x''$,  of the 
ballistic interval $[x', x'']$ are points of local thermodynamic equilibrium 
but are, in general, {\em not} points of spin equilibrium. 

We first assume that the electrons thermally emitted at $x'$ 
towards the right are {\em not} affected by spin relaxation during their motion
across the interval $[x', x'']$. The spin-resolved densities 
$n_{\uparrow , \downarrow}(x')$ [see Eq.~(\ref{eq:eqden02})] then give rise to 
conserved ballistic spin-resolved currents (i.e., currents independent of the
position $x \in [x',x'']$) in that interval, which have the form
\begin{equation}
J^{l}_{\uparrow ,\downarrow}(x',x'') = v_{e} n_{\uparrow ,\downarrow}(x')
\bar{T}^{l}_{\uparrow ,\downarrow}(x',x'')
\label{eq:ball004ii}
\end{equation}
(see Sec.~\ref{sec:ballnonc}), where the thermally averaged (classical)
transmission probabilities $ \bar{T}^{l}_{\uparrow ,\downarrow}(x',x'')$ are
given [see Eqs.~(\ref{eq:ball005}) and(\ref{eq:ball005a})] by
\begin{equation}
\bar{T}^{l}_{\uparrow ,\downarrow}(x',x'') = e^{- \beta [
E_{\uparrow, \downarrow}^{m}(x',x'') - E_{\uparrow, \downarrow}(x')]} ,
\label{eq:ball004iii}
\end{equation}
with $E_{\uparrow, \downarrow}^{m}(x',x'')$ the overall maxima of the potential
energy profiles $E_{\uparrow, \downarrow}(x)$ in $[x', x'']$.  In terms of the
spin functions $A_{\uparrow, \downarrow}(x')$ given by
Eqs.~(\ref{eq:eqden03mm}), we can rewrite the currents $J^{l}_{\uparrow,
\downarrow}(x',x'')$ in the form
\begin{equation}
J^{l}_{\uparrow ,\downarrow}(x',x'') = \frac{v_{e} N_{c}}{2} e^{- \beta
E_{\uparrow, \downarrow}^{m}(x',x'')} A_{\uparrow, \downarrow}(x') .
\nonumber \\
\label{eq:ball004i}
\end{equation}
We then have for the total (i.e., spin-summed) ballistic current,
$J^{l}_{+}(x',x'')$, and the ballistic  spin-polarized current,
$J^{l}_{-}(x',x'')$,
\begin{eqnarray}
J^{l}_{\pm}(x',x'') &=& \frac{v_{e} N_{c}}{4} B^{m}_{+}(x', x'') \nonumber \\
&\ & \times  [A_{\pm}(x') + P^{m}(x', x'')
A_{\mp}(x')] , \nonumber \\
\label{eq:ball004ju}
\end{eqnarray}
where
\begin{equation}
B^{m}_{\pm} (x', x'') = e^{- \beta E_{\uparrow}^{m} (x',x'')} \pm e^{- \beta
E_{\downarrow}^{m} (x',x'')}
\label{eq:ball004u}
\end{equation}
and
\begin{equation}
P^{m}(x',x'') = \frac{B^{m}_{-} (x', x'')}{B^{m}_{+} (x', x'')} ,
\label{eq:emtra09}
\end{equation}
which are nonlocal extensions of expressions (\ref{eq:eqden60}) and
(\ref{eq:eqden06}) for the functions $B_{\pm}(x')$ and the local polarization
$P(x')$, respectively.

We now consider the effect of spin relaxation on the ballistic currents 
inside the interval $[x', x'']$. We introduce functions $A^{l}_{\pm}(x',x'';x)$
that depend on the position $x \in [x', x'']$ and are required to satisfy the 
initial conditions
\begin{equation}
A^{l}_{\pm}(x',x'';x = x') = A_{\pm}(x') .
\label{eq:spirel05}
\end{equation}
Using these functions, we generalize the definitions of the currents
$J^{l}_{\pm}(x',x'')$ given by Eqs.~(\ref{eq:ball004ju}) to
\begin{eqnarray}
J^{l}_{\pm}(x',x'';x) &=& \frac{v_{e}N_{c}}{4} B^{m}_{+}(x',x'') \nonumber
\\ &\ & \hspace{-1.0cm} \times [ A_{\pm}^{l}(x',x'';x) + P^{m}(x',x'')
A_{\mp}^{l}(x',x'';x)] . \nonumber \\
\label{eq:spirel08}
\end{eqnarray}
Now, also in the presence of spin relaxation, the {\em total} ballistic current
must be conserved,
\begin{equation}
J^{l}_{+}(x',x'';x) = J^{l}_{+}(x',x'';x=x') \equiv J^{l}_{+}(x',x'') ,
\label{eq:spirel10}
\end{equation}
where, from Eq.~(\ref{eq:ball004ju}),
\begin{eqnarray}
J^{l}_{+}(x',x'') &=& \frac{v_{e} N_{c}}{4} B^{m}_{+}(x', x'') \nonumber \\ &\
& \times  [A_{+}(x') + P^{m}(x', x'') A_{-}(x')] . \nonumber \\
\label{eq:ball004juhu}
\end{eqnarray}
Equating this to expression (\ref{eq:spirel08}) for $J^{l}_{+}(x',x'';x)$, we
obtain the relation
\begin{eqnarray}
A_{+}^{l}(x',x'';x) &=& A_{+}(x') + P^{m}(x',x'') \nonumber \\  &\ & \times
[A_{-}(x') - A_{-}^{l}(x',x'';x)] . \nonumber \\
\label{eq:balden09}
\end{eqnarray}
Therefore, the spin-polarized current can be written as
\begin{eqnarray}
J^{l}_{-}(x',x'';x) &=& J^{l}_{+}(x',x'') P^{m}(x',x'') \nonumber \\ &\ & +
\frac{v_{e}N_{c}}{4} B^{m}(x',x'') A_{-}^{l}(x',x'';x) , \nonumber \\
\label{eq:spirel11}
\end{eqnarray}
where
\begin{equation}
B^{m}(x',x'') =  B_{+}^{m}(x',x'') [Q^{m}(x',x'')]^{2} ,
\label{eq:spirel15}
\end{equation}
and
\begin{equation}
Q^{m}(x',x'') \equiv \left\{ 1 - [P^{m}(x',x'')]^{2} \right\}^{1/2}
\label{eq:emtra14}
\end{equation}
is the nonlocal extension of expression (\ref{eq:spfra17c}) for the function
$Q(x')$.

The ballistic spin-polarized current of Eq.~(\ref{eq:spirel11}) has the form
\begin{equation}
J^{l}_{-}(x',x'';x) = \overstar{J}^{l}_{-}(x',x'') +
\check{J}^{l}_{-}(x',x'';x) ,
\label{eq:spirel11ij}
\end{equation}
where the first ($x$-independent, nonrelaxing) term is the ``persistent''
ballistic spin-polarized current,
\begin{equation}
\overstar{J}^{l}_{-}(x',x'') \equiv J^{l}_{+}(x',x'') P^{m}(x',x'') ,
\label{eq:spirel12}
\end{equation}
while the second ($x$-dependent) term is the ``relaxing'' ballistic
spin-polarized current,
\begin{equation}
\check{J}^{l}_{-}(x',x'';x)  = \frac{v_{e}N_{c}}{4} B^{m}(x',x'')
A_{-}^{l}(x',x'';x) . \nonumber \\
\label{eq:spirel14}
\end{equation}
The latter describes the spin dynamics in the current $J^{l}_{-}(x',x'';x)$ via
the ``spin relaxation function'' $A_{-}^{l}(x',x'';x)$.  The $x$-dependence of
$A_{-}^{l}(x',x'';x)$ will be determined explicitly in Sec.~\ref{sec:balance},
while the procedure for calculating the spin accumulation function
\begin{equation}
A_{-}(x') \equiv A_{-}^{l}(x',x'';x = x')
\label{eq:spirel14eka}
\end{equation}
will be described in Sec.~\ref{sec:accufunc}.

The persistent ballistic spin-polarized current $\overstar{J}^{l}_{-}(x',x'')$
depends, via the total ballistic current $J^{l}_{+}(x',x'')$, on the spin
accumulation function $A_{-}(x')$. When $A_{-}(x')$ is calculated within
the thermoballistic approach, spin relaxation in all ballistic intervals
(including the interval $[x', x'']$ under consideration) is taken into account.
Therefore, while being not {\em directly} affected by spin relaxation
{\em inside} the interval $[x', x'']$, the persistent current
$\overstar{J}^{l}_{-}(x',x'')$ depends, via the dependence of the factor 
$J_{+}^{l}(x', x'')$ on  $A_{-}(x')$ [see Eqs.~(\ref{eq:ball004juhu}) and 
(\ref{eq:spirel12})], in an {\em indirect} way on spin relaxation.

\subsubsection{Ballistic densities}

\label{sec:offden}

The ballistic densities $n^{l}_{\pm}(x',x''; x)$ associated with the ballistic 
currents $J^{l}_{\pm}(x',x''; x)$ of Eqs.~(\ref{eq:spirel08}) are given by
\begin{eqnarray}
n^{l}_{\pm}(x',x''; x) &=& \frac{N_{c}}{8} D^{m}_{+}(x', x''; x) \nonumber
\\  &\ & \hspace{-1.5cm} \times [ A_{\pm}^{l}(x',x'';x) + P^{m}_{C}(x',x'';x)
A_{\mp}^{l}(x',x'';x)] \nonumber \\
\label{eq:ball004m}
\end{eqnarray}
[see Eq.~(\ref{eq:ball004ka})]. Here, we have taken into account that the 
ballistic densities correspond to {\em one half} of the thermal currents 
emitted symmetrically at a point of local thermodynamic equilibrium [see, e.g.,
Eq.~(\ref{eq:ball001h})]. The functions $D_{\pm}^{m}(x', x''; x)$ and 
$P_{C}^{m}(x',x'';x)$ are defined as
\begin{eqnarray}
D_{\pm}^{m}(x', x''; x) &=& C^{m}_{\uparrow}(x',x''; x) e^{- \beta
E_{\uparrow}^{m} (x', x'')} \nonumber \\ &\ & \pm C^{m}_{\downarrow}(x',x''; x)
e^{- \beta E_{\downarrow}^{m} (x', x'')} , \nonumber \\
\label{eq:ball004p}
\end{eqnarray}
where
\begin{eqnarray}
C^{m}_{\uparrow ,\downarrow}(x',x''; x) &=& e^{\beta [E_{\uparrow ,
\downarrow}^{m} (x', x'') - E_{\uparrow , \downarrow}(x)]} \nonumber \\ &\ &
\times {\rm erfc} \bm{(} \{ \beta [E_{\uparrow , \downarrow}^{m}(x', x'') -
E_{\uparrow , \downarrow}(x)] \}^{1/2} \bm{)}  \nonumber \\
\label{eq:ball004n}
\end{eqnarray}
[see Eq.~(\ref{eq:ball004kaa})], and
\begin{equation}
P_{C}^{m}(x',x'';x) = \frac{D_{-}^{m}(x', x''; x)}{D_{+}^{m}(x', x''; x)} ,
\label{eq:emtra09z}
\end{equation}
in generalization of expression (\ref{eq:emtra09}) for the function $P^{m}(x',
x'')$.

Now, inserting expression (\ref{eq:balden09}) for $A_{+}^{l}(x',x'';x)$ in
Eqs.~(\ref{eq:ball004m}), we find that the total ballistic density,
$n^{l}_{+}(x',x''; x)$, and the ballistic spin-polarized density,
$n^{l}_{-}(x',x''; x)$, have the form
\begin{equation}
n^{l}_{\pm}(x',x'';x) = \overstar{n}^{l}_{\pm}(x',x'';x) +
\check{n}^{l}_{\pm}(x',x'';x) ,
\label{eq:spirel11kx}
\end{equation}
Here, the first ($x$-dependent, but nonrelaxing) term is the persistent part,
\begin{eqnarray}
\overstar{n}^{l}_{\pm}(x',x'';x) &=& \frac{N_{c}}{8} D^{m}_{\pm}(x', x''; x)
\nonumber \\ &\ & \times [A_{+}(x') + P^{m}(x',x'') A_{-}(x')]   \nonumber
\\
\label{eq:balden13}
\end{eqnarray}
and the second term is the relaxing part,
\begin{eqnarray}
\check{n}^{l}_{\pm}(x',x'';x) &=& \frac{N_{c}}{8} {\cal D}^{m}_{\pm}(x',x'';x)
A^{l}_{-}(x',x'';x) , \nonumber \\
\label{eq:balden1}
\end{eqnarray}
where
\begin{eqnarray}
{\cal D}^{m}_{\pm}(x',x'';x) &=& D^{m}_{\mp}(x',x'';x) \nonumber \\ &\ & -
P^{m}(x',x'') D^{m}_{\pm}(x',x'';x) . \nonumber \\
\label{eq:balden15}
\end{eqnarray}
In contrast to the $x$-independent currents $J^{l}_{+}(x',x'')$ and
$\overstar{J}^{l}_{-}(x',x'')$, the persistent parts of the ballistic
densities, $\overstar{n}^{l}_{\pm}(x',x'';x)$, depend on position via the
$x$-dependence of the potential energy profiles $E_{\uparrow, \downarrow}(x)$.
For zero spin splitting of the conduction band, $\Delta (x) \equiv 0$, when
$P^{m}(x',x'') =  D^{m}_{-}(x',x'';x) = {\cal D}^{m}_{+}(x',x'';x) = 0$, the 
persistent part of the spin-polarized density as well as the relaxing part of 
the total density vanish, $\overstar{n}^{l}_{-}(x',x'';x) = 
\check{n}^{l}_{+}(x',x'';x) = 0$.

\subsubsection{Spin balance equation}

\label{sec:balance}

In a time-dependent formulation, spin relaxation in a system described by
spin-resolved electron densities $n_{\uparrow, \downarrow} (x,t)$ and currents
$J_{\uparrow, \downarrow} (x,t)$ is governed by the local coupled spin balance
equations\cite{yuf02a}
\begin{equation}
\frac{\partial}{\partial t} n_{\uparrow} (x,t) + \frac{\partial}{\partial x}
J_{\uparrow} (x,t) = - \frac{n_{\uparrow} (x,t)}{\tau_{\uparrow \downarrow}} +
\frac{n_{\downarrow} (x,t)}{\tau_{\downarrow \uparrow}}
\label{eq:spirel01aa}
\end{equation}
and
\begin{equation}
\frac{\partial}{\partial t} n_{\downarrow} (x,t) + \frac{\partial}{\partial x}
J_{\downarrow} (x,t) = - \frac{n_{\downarrow} (x,t)}{\tau_{\downarrow
\uparrow}} + \frac{n_{\uparrow} (x,t)}{\tau_{\uparrow \downarrow}} ,
\label{eq:spirel01ab}
\end{equation}
where $1/\tau_{\uparrow \downarrow}$ $(1/\tau_{\downarrow \uparrow})$ is the
rate for spin-flip scattering from spin-up (spin-down) to spin-down (spin-up)
states. In the stationary case, when $\partial n_{\uparrow, \downarrow} (x,t) /
\partial t \equiv 0$, this leads to the balance equation
\begin{equation}
\frac{d}{dx} \check{J}_{-} (x) = - \frac{1}{\tau_{s}} \check{n}_{-} (x) +
\left( \frac{1}{\tau_{\downarrow \uparrow}} -
\frac{1}{\tau_{\uparrow \downarrow}} \right) \check{n}_{+}(x)
\label{eq:spirel01ac}
\end{equation}
connecting the relaxing part of the spin-polarized current, $\check{J}_{-}
(x)$, to the relaxing parts of the spin-polarized, $\check{n}_{-} (x)$, and of
the total density, $\check{n}_{+} (x)$.  Here, $\tau_{s}$, defined as
\begin{equation}
\frac{1}{\tau_{s}} = \frac{1}{\tau_{\uparrow \downarrow}} +
\frac{1}{\tau_{\downarrow \uparrow}} ,
\label{eq:spirel02}
\end{equation}
is the spin relaxation time.

We apply Eq.~(\ref{eq:spirel01ac}) to spin relaxation during ballistic
transport. In doing this, we will disregard the term involving 
$\check{n}_{+}(x)$ for two
reasons:\ (i) As it is preceded by the {\em difference} of the two relaxation 
rates $1/\tau_{\uparrow \downarrow}$ and $1/\tau_{\downarrow \uparrow}$, which
are estimated to be of comparable magnitude, the term may
generally be considered small in comparison with the term involving 
$\check{n}_{-}(x)$. (ii) Since both $\check{n}_{-}(x)$ and $\check{n}_{+}(x)$ 
are proportional to the spin relaxation function $A^{l}_{-}(x',x'';x)$ [see
Eq.~(\ref{eq:balden1})], we can account for $\check{n}_{+}(x)$ by combining its
prefactors with those of $\check{n}_{-}(x)$ in an {\em effective} spin
relaxation time (still denoted by $\tau_{s}$) depending, in general, on the
potential energy profiles. [Note that we must not assume $\check{n}_{+}(x) = 0$ 
from the outset by adopting the arguments leading to Eq.~(2.7) of 
Ref.~\onlinecite{yuf02a}. The density $n_{\uparrow} + n_{\downarrow}$ appearing
in that equation is the deviation of
the total density from its spin equilibrium value. By contrast, the relaxing 
total density $\check{n}_{+}(x)$, when used in the thermoballistic 
description, gives the deviation of the total density from the 
persistent total density. In the latter, a spin non-equilibrium part enters 
via the spin accumulation function $A_{-}(x)$ [see Eq.~(\ref{eq:balden13})].]

In the thermoballistic approach, it is assumed that the thermally emitted 
electrons spend only an infinitesimally short time span at the emission point, 
and it is only during their motion across the ballistic interval that they can 
undergo spin relaxation.  Spin relaxation in ballistic transport is commonly 
described\cite{yaf52,ell54,yaf63} in terms of a (ballistic) spin relaxation 
length $l_{s}$ given by
\begin{equation}
l_{s} = 2 v_{e} \tau_{s} .
\label{eq:spirel01}
\end{equation}
[As in the case of the effective electron mass $m^{\ast}$, we assume
$\tau_{s}$, and hence $l_{s}$, to be independent of position and of the
external magnetic field;\ we consider $l_{s}$ here an {\em effective} quantity,
in line with the interpretation of $\tau_{s}$.] In our description of
spin-polarized electron transport, spin relaxation is thus completely separated
from momentum relaxation at the points of local thermodynamic equilibrium and,
in this respect, is similar to the D'yakonov-Perel' relaxation
mechanism.\cite{aws02,zut04,fab07,dya08,dya71,dya08a,kor10}

In terms of the relaxing  ballistic spin-polarized current
$\check{J}^{l}_{-}(x',x'';x)$ and the corresponding density
$\check{n}^{l}_{-}(x',x'';x)$, the balance equation governing spin relaxation
during ballistic transport now reads
\begin{equation}
\frac{d}{dx} \check{J}^{l}_{-}(x',x'';x) +  \frac{2 v_{e}}{l_{s}}
\check{n}^{l}_{-}(x',x'';x) = 0 ,
\label{eq:baleq01}
\end{equation}
where the spin relaxation length $l_{s}$ is given by Eq.~(\ref{eq:spirel01}).

Inserting in Eq.~(\ref{eq:baleq01}) the expressions for
$\check{J}^{l}_{-}(x',x'';x)$ and $\check{n}^{l}_{-}(x',x'';x)$ from
Eqs.~(\ref{eq:spirel14}) and (\ref{eq:balden1}), respectively, we obtain a
first-order differential equation for the spin relaxation function
$A^{l}_{-}(x',x'';x)$,
\begin{eqnarray}
\frac{d}{dx} A^{l}_{-}(x',x'';x) + \frac{C^{m}(x',x'';x)}{l_{s}}
A^{l}_{-}(x',x'';x) &=& 0 , \nonumber \\
\label{eq:baleq02}
\end{eqnarray}
where
\begin{equation}
C^{m}(x',x'';x) = \frac{{\cal D}^{m}_{-}(x',x'';x)}{B^{m}(x',x'')} .
\label{eq:baleq03}
\end{equation}
The solution of Eq.~(\ref{eq:baleq02}) obeying the initial condition
(\ref{eq:spirel05}) is
\begin{equation}
A_{-}^{l}(x',x'';x) = A_{-}(x') e^{-{\cal C}^{m}(x',x'';x',x)/l_{s}} ,
\label{eq:baleq05}
\end{equation}
where
\begin{equation}
{\cal C}^{m}(x',x'';z_{1},z_{2}) = \int_{z_{<}}^{z_{>}} dz C^{m}(x',x'';z) ,
\label{eq:baleq06}
\end{equation}
with $z_{<} = \min (z_{1}, z_{2}), \; z_{>} = \max (z_{1}, z_{2})$.

Now, inserting expression (\ref{eq:baleq05}) in Eqs.~(\ref{eq:spirel14}) and
(\ref{eq:balden1}), respectively, we obtain the relaxing ballistic
spin-polarized current and density explicitly in terms of the spin accumulation
function $A_{-}(x')$,
\begin{eqnarray}
\check{J}^{l}_{-}(x',x'';x) &=& \frac{v_{e} N_{c}}{4} B^{m}(x',x'')  A_{-}(x')
\nonumber \\ &\ & \times e^{-{\cal C}^{m}(x',x'';x',x)/l_{s}}
\label{eq:baleq07}
\end{eqnarray}
and
\begin{eqnarray}
\check{n}^{l}_{-}(x',x'';x) &=& \frac{N_{c}}{8} {\cal D}^{m}_{-}(x',x'';x )
A_{-}(x') \nonumber \\ &\ & \times e^{-{\cal C}^{m}(x',x'';x',x)/l_{s}} .
\label{eq:baleq08}
\end{eqnarray}
The $x$-dependence of the relaxing spin-pol\-arized current and density in the
ballistic interval $[x',x'']$ is hence governed by the factor $e^{-{\cal
C}^{m}(x',x'';x',x)/l_{s}}$.  It departs from a purely exponential behavior
unless the potential energy profiles $E_{\uparrow , \downarrow}(x)$ are
constant over the interval, in which case ${\cal C}^{m}(x',x'';x',x) = x - x'$.

\subsubsection{Net ballistic currents and joint ballistic densities}

\label{sec:netball}

So far, we have only considered thermal emission at the {\em left} end-point, 
$x'$, of the ballistic interval $[x',x'']$, obtaining a variety of ballistic 
currents and densities summarily denoted here by $J^{l}(x', x''; x)$ and 
$n^{l}(x', x''; x)$, respectively. The analogous ballistic currents and 
densities $J^{r}(x', x''; x)$ and $n^{r}(x', x''; x)$  
corresponding to emission at the {\em right} end-point $x''$ can be expressed 
in terms of those emitted at $x'$ as
\begin{equation}
J^{r}(x', x''; x) = - J^{l}(x'', x'; x)
\label{eq:ball004rfd}
\end{equation}
and
\begin{equation}
n^{r}(x', x''; x) = n^{l}(x'', x'; x) .
\label{eq:ball004uhv}
\end{equation}
[Note that, owing to the symmetry of the functions $E^{m}_{\uparrow , 
\downarrow}(x',x'')$, the functions $B^{m}_{\pm}(x',x'')$, 
$B^{m}(x',x'')$, $D^{m}_{\pm}(x',x'';x)$, and ${\cal D}^{m}_{\pm}(x',x'';x)$ 
entering the expressions for $J^{l}(x', x''; x)$  and $n^{l}(x', x''; x)$ 
are symmetric under the exchange of $x'$ and $x''$.] We then have
\begin{eqnarray}
J(x', x''; x) &\equiv&  J^{l}(x', x''; x) +  J^{r}(x', x''; x) \nonumber \\
&=&  J^{l}(x', x''; x) -  J^{l}(x'', x'; x)
\label{eq:ball004zgx}
\end{eqnarray}
and
\begin{eqnarray}
n(x', x''; x) &\equiv& n^{l}(x', x''; x) + n^{r}(x', x''; x) \nonumber \\
 &=& n^{l}(x', x''; x) + n^{l}(x'', x'; x)
\label{eq:ball004pla}
\end{eqnarray}
for the {\em net} ballistic currents and {\em joint} ballistic densities 
summarily denoted by $J(x', x''; x)$ and $ n(x', x''; x)$, respectively.

For the (conserved) net {\em total} ballistic current $J_{+}(x',x'')$ inside 
the ballistic interval, we now find, using Eq.~(\ref{eq:ball004juhu}),
\begin{eqnarray}
J_{+}(x',x'') &=& \frac{v_{e} N_{c}}{2} B^{m}_{+}(x', x'') \{ [\tilde{A}(x') - 
\tilde{A}(x'')] \nonumber \\ &\ & + \smfrac{1}{2} P^{m}(x', x'')[A(x') -A(x'')] 
\} . \nonumber \\
\label{eq:ball004jvx}
\end{eqnarray}
Here, the function $A_{+}(x')$ has been replaced with $2 \tilde{A}(x')$ [see
Eq.~(\ref{eq:eqden03})], and the subscript attached to the spin accumulation
function $A_{-}(x')$ has been omitted, i.e., we have set
\begin{equation}
A(x') \equiv A_{-}(x') .
\label{eq:ball004jhm}
\end{equation}
The current $J_{+}(x',x'')$ is seen to be dynamically determined, in general,
both by the average chemical potential $\tilde{\mu}(x')$ [via the mean spin
function$\tilde{A}(x')$] and the spin accumulation function $A(x')$.  The same
then holds for the net {\em persistent} ballistic {\em spin-polarized} current
$\overstar{J}_{-}(x',x'')$, for which we have, using Eq.~(\ref{eq:spirel12}),
\begin{equation}
\overstar{J}_{-}(x',x'')  = J_{+}(x',x'') P^{m}(x',x'') .
\label{eq:netbal06}
\end{equation}
For zero spin splitting, when $\Delta(x) = 0$ and hence $P^{m}(x',x'') = 0$,
the dependence of $J_{+}(x',x'')$ on $A(x)$ drops out and, further,
$\overstar{J}_{-}(x',x'') = 0$.

For the net {\em relaxing}  ballistic {\em spin-polarized} current, we have,
using Eq.~(\ref{eq:baleq07}),
\begin{eqnarray}
\check{J}_{-}(x',x'';x)  &=& \frac{v_{e} N_{c}}{4} B^{m}(x',x'') [ A(x') 
e^{-{\cal C}^{m}(x',x'';x',x)/l_{s}} \nonumber \\ &\ & - A(x'') 
e^{-{\cal C}^{m}(x',x'';x,x'')/l_{s}}] ,
\label{eq:netbal02}
\end{eqnarray}
which is dynamically determined by the spin accumulation function $A(x)$ alone.

For the joint {\em total} ballistic density, $n_{+}(x',x'';x)$, and the joint 
ballistic {\em spin-polarized} density, $n_{-}(x',x'';x)$, respectively,
 we have from Eq.~(\ref{eq:spirel11kx})
\begin{equation}
n_{\pm}(x',x'';x)  = \overstar{n}_{\pm}(x',x'';x) + \check{n}_{\pm}(x',x'';x) ,
\label{eq:netbal03ax}
\end{equation}
with {\em persistent} parts $\overstar{n}_{\pm}(x',x'';x)$ and {\em relaxing}
parts $\check{n}_{\pm}(x',x'';x)$ given by
\begin{eqnarray}
\overstar{n}_{\pm}(x',x'';x) &=& \frac{N_{c}}{4} D^{m}_{\pm}(x', x'';x) \{
[\tilde{A}(x') + \tilde{A}(x'')] \nonumber \\ &\ & + \smfrac{1}{2} P^{m}(x',
x'')[A(x') + A(x'')] \} \nonumber \\
\label{eq:ball004juy}
\end{eqnarray}
[see Eq.~(\ref{eq:balden13})] and
\begin{eqnarray}
\check{n}_{\pm}(x',x'';x) &=& \frac{N_{c}}{8} {\cal D}^{m}_{\pm}(x',x'';x)
\nonumber \\ &\ & \times [ A(x') e^{-{\cal C}^{m}(x',x'';x',x)/l_{s}} \nonumber
\\ &\ & \hspace{0.5cm} + A(x'') e^{-{\cal C}^{m}(x',x'';x,x'')/l_{s}}]
\nonumber \\
\label{eq:netbal02xy}
\end{eqnarray}
[see Eqs.~(\ref{eq:balden1}) and (\ref{eq:baleq08})], in analogy to
Eqs.~(\ref{eq:ball004jvx}), (\ref{eq:netbal06}), and (\ref{eq:netbal02}) for
the net currents.

From expressions (\ref{eq:netbal02}) and (\ref{eq:netbal02xy}), we now derive,
using Eqs.~(\ref{eq:baleq03}) and (\ref{eq:baleq06}), the balance equation
\begin{equation}
\frac{d}{dx} \check{J}_{-}(x',x'';x) +  \frac{2 v_{e}}{l_{s}}
\check{n}_{-}(x',x'';x) = 0
\label{eq:baleq01zs}
\end{equation}
connecting the net relaxing ballistic spin-polarized current
$\check{J}_{-}(x',x'';x)$ and the associated joint density
$\check{n}_{-}(x',x'';x)$.  This equation can be obtained, of course,
simply by adding the balance equation Eq.~(\ref{eq:baleq01}) and the 
corresponding equation for emission at $x''$.

\subsection{Thermoballistic spin-polarized transport}

\label{sec:thermtrans}

In a significant advance over its prototype presented in 
Sec.~\ref{sec:gendrift}, where overall current conservation was introduced via
the condition (\ref{eq:gendr001qal}), the thermoballistic concept 
proper rests on the introduction of a ``reference coordinate'' $x$ that
characterizes an arbitrary point inside the semiconducting
sample extending from $x_{1}$ to $x_{2}$, as shown in Fig.~\ref{fig:4}. 
Singling out
this coordinate, we consider the net ballistic currents and joint ballistic
densities within the ensemble of all ballistic intervals $[x',x'']$ enclosing
$x$. These currents and densities form the building blocks for establishing, at
$x$, the corresponding thermoballistic quantities.
The point labeled by the reference coordinate $x$ is {\em not} a point of local
thermodynamic equilibrium (we may call it a ``ballistic point''). However, the
``equilibrium points'' $x', x''$ may come infinitesimally close to $x$.

\subsubsection{Thermoballistic currents and densities}

\label{sec:thermcurr}

The thermoballistic currents and densities at the point $x$ are constructed by
performing weighted summations of the corresponding net ballistic currents and
joint ballistic densities over all ballistic intervals  $[x',x'']$ subjected to
the condition
\begin{equation}
x_{1} \leq x' < x < x'' \leq x_{2} .
\label{eq:thebal00}
\end{equation}
Just as in the prototype thermoballistic model (see Sec.~\ref{sec:averball}), 
we adopt here the  probabilistic picture outlined in Sec.~\ref{sec:probappr}:\ 
the contributions from the interval $[x',x'']$ are
weighted with the probability $e^{-(x''-x')/l}$ (one-dimensional transport is
assumed here;\ see the remark at the beginning of Sec.~\ref{sec:diffmech}) that
the electrons traverse the interval without collisions, multiplied by the
probability $dx'/l$ ($dx''/l$) that they are absorbed or emitted in an
interval $dx'$ ($dx''$) around the end-point $x'$ ($x''$).  At the ends of the
semiconducting sample at $x_{1,2}$, absorption and emission occur with unit
probability.  Like the effective electron mass $m^{\ast}$ and the spin
relaxation length $l_{s}$, the momentum relaxation length $l$ is assumed here
to be independent of position and of the external magnetic field.

Representing the net ballistic currents and joint ballistic
densities of Eqs.~(\ref{eq:ball004jvx})--(\ref{eq:netbal02xy}) summarily by a
function $F(x',x'';x)$, and the corresponding thermoballistic currents
and densities by $\mathfrak{F}(x)$, we write $\mathfrak{F} (x)$ in the form
\begin{eqnarray}
\mathfrak{F}(x) &\equiv& \mathfrak{F}(x_{1},x_{2}; x;l) \nonumber \\ &=&
e^{-(x_{2}-x_{1})/l} F(x_{1},x_{2};x) \nonumber \\ &\ & +
\int_{x_{1}}^{x^{-}} \frac{dx'}{l} e^{-(x_{2}-x')/l} F(x',x_{2};x)
\nonumber \\ &\ & + \int_{x^{+}}^{x_{2}} \frac{dx''}{l} e^{-(x''-x_{1})/l}
F(x_1, x'';x) \nonumber \\ &\ & + \int_{x_{1}}^{x^{-}} \frac{dx'}{l}
\int_{x^{+}}^{x_2} \frac{dx''}{l} e^{-(x'' - x')/l} F (x',x'';x) ,
\nonumber \\
\label{eq:thebal01}
\end{eqnarray}
where $x^{\pm} = x \pm \delta$, and the infinitesimal $\delta$  has been
introduced in accordance with condition (\ref{eq:thebal00}). Further, we set
\begin{equation}
\mathfrak{F}(x_{1,2}) = \left. \mathfrak{F}(x_{1,2} \pm \eta) \right|_{\eta 
\rightarrow 0} .
\label{eq:thebal01ocm}
\end{equation}
In parallel to Eq.~(\ref{eq:gendr008x}) for the total probability in the 
prototype model, we have $\mathfrak{F}(x) = 1$ for $F(x', x''; x) = 1$.

\begin{figure}[t]
\includegraphics[width=0.45\textwidth]{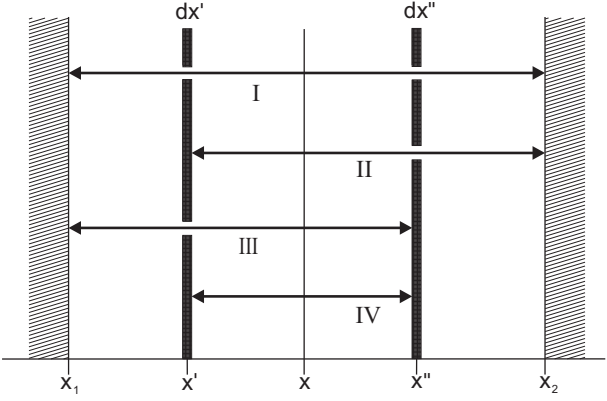}
\caption{Schematic diagram illustrating the four types of ballistic current 
contributing to the total thermoballistic current $\mathfrak{J}_{+}(x)$
according to Eq.~(\ref{eq:thebal01}).
}
\label{fig:5}
\end{figure}

Specifically, the total thermoballistic current $\mathfrak{J}_{+}(x)$ is given 
by expression (\ref{eq:thebal01}) with $J_{+}(x',x'')$ of
Eq.~(\ref{eq:ball004jvx}) substituted for $F (x',x'';x)$.  The persistent
thermoballistic spin-pola\-ri\-zed current $\overstar{\mathfrak{J}}_{-}(x)$
follows by identifying $F(x',x'';x)$ with $\overstar{J}_{-}(x',x'')$ of
Eq.~(\ref{eq:netbal06}).  The relaxing thermoballistic spin-polarized
current $\check{\mathfrak{J}}_{-}(x)$ is obtained by replacing $F(x',x'';x)$
with $\check{J}_{-}(x',x''; x)$ of Eq.~(\ref{eq:netbal02}).  Further, the
thermoballistic densities $\mathfrak{n}_{+}(x)$,
$\overstar{\mathfrak{n}}_{-}(x)$, and $\check{\mathfrak{n}}_{-}(x)$
corresponding to the currents $\mathfrak{J}_{+}(x)$,
$\overstar{\mathfrak{J}}_{-}(x)$, and $\check{\mathfrak{J}}_{-}(x)$ follow by
substituting the respective joint ballistic densities [see
Eqs.~(\ref{eq:netbal03ax})--(\ref{eq:netbal02xy})] for $F(x',x'';x)$ in 
Eq.~(\ref{eq:thebal01}).

In expression (\ref{eq:thebal01}) when specialized to the total thermoballistic 
current $\mathfrak{J}_{+}(x)$, the first term on the right-hand side represents
the net electron current passing through the point $x$
while being ballistically transmitted 
between $x_{1}$ and $x_{2}$ [which occurs with probability $e^{-(x_{2} - 
x_{1})/l}$], the second term refers to ballistic electron motion between any
point $x'$ $(x_{1} \leq x' \leq x^{-})$ and $x_{2}$ [probability $e^{-(x_{2} - 
x')/l} dx'/l$], the third term to ballistic electron motion between $x_{1}$ and
any point $x''$ $ (x^{+} \leq x'' \leq x_{2})$ [probability $e^{-(x'' - 
x_{1})/l} dx''/l$], etc. The four types of ballistic current appearing in 
expression (\ref{eq:thebal01}) are illustrated by the double arrows labeled I 
to IV in Fig.~\ref{fig:5}.
 
The derivatives with respect to $x$ of the various thermoballistic quantities 
can be written in the general form
\begin{equation}
\frac{d}{dx} \mathfrak{F}(x) = \left. \frac{d}{dx} \mathfrak{F}(x) \right|_{F}
+ \mathfrak{D}(x) ,
\label{eq:thebal01dz}
\end{equation}
where the first term on the right-hand side comprises the contributions arising
from differentiating the functions $F(x', x''; x)$ in the integrands of
expression (\ref{eq:thebal01}), and the second, those from differentiating the 
limits of integration:\
\begin{eqnarray}
\mathfrak{D}(x) &\equiv& \mathfrak{D}(x_{1}, x_{2}; x; l) \nonumber \\ &\equiv&
\left( \frac{\partial}{\partial x^{+}} + \frac{\partial}{\partial x^{-}}
\right) \mathfrak{F}(x_{1}, x_{2}; x; l) \nonumber \\ &=& - \frac{1}{l} \{
\mathbb{F}_{1}(x; [F]) - \mathbb{F}_{2}(x; [F]) \} . 
\label{eq:thebal06jb}
\end{eqnarray}
Here, the functionals $\mathbb{F}_{1}(x; [F])$ and $\mathbb{F}_{2}(x; [F])$,
defined by
\begin{eqnarray}
\mathbb{F}_{1}(x; [F]) &=& e^{- (x - x_{1})/l} F(x_{1}, x^{+}; x) \nonumber \\
&\ &  + \int_{x_{1}}^{x^{-}} \frac{dx'}{l} e^{- (x - x')/l} F(x', x^{+}; x)
\nonumber \\
\label{eq:thec00ohx}
\end{eqnarray}
and 
\begin{eqnarray}
\mathbb{F}_{2}(x; [F]) &=& e^{- (x_{2} - x)/l} F(x^{-}, x_{2}; x)  \nonumber \\
&\ &  + \int_{x^{+}}^{x_{2}} \frac{dx''}{l} e^{- (x'' - x)/l} F(x^{-}, x''; x) 
, \nonumber \\
\label{eq:thec00gma}
\end{eqnarray}
respectively, represent the contributions of the function $F(x', x''; x)$ 
arising from the ranges to the left and right of the point $x$.

\begin{figure}[t]
\includegraphics[width=0.45\textwidth]{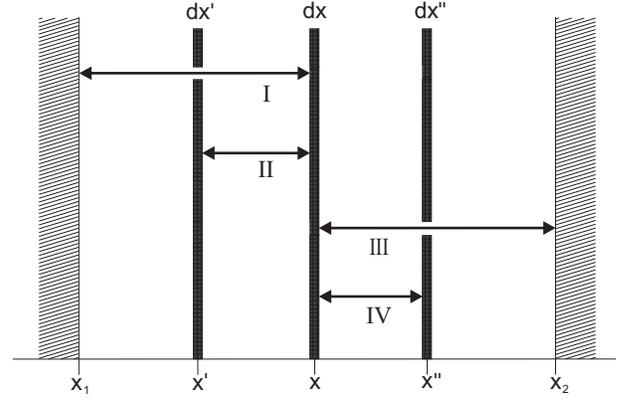}
\caption{Schematic diagram illustrating the four types of ballistic current 
contributing to the infinitesimal increment $\mathfrak{D}_{+}(x) dx$ of 
the total thermoballistic current $\mathfrak{J}_{+}(x)$ according to
Eq.~(\ref{eq:thebal06jb}). 
}
\label{fig:6}
\end{figure}

For the total thermoballistic current $\mathfrak{J}_{+}(x)$ 
constructed from the $x$-independent ballistic current $J_{+}(x',x'')$, the
first term on the right-hand side of Eq.~(\ref{eq:thebal01dz}) vanishes, and we
have
\begin{equation}
\frac{d}{dx} \mathfrak{J}_{+}(x) = \mathfrak{D}_{+}(x) ,
\label{eq:thebal01ek}
\end{equation}
with $\mathfrak{D}_{+}(x)$ given by expression (\ref{eq:thebal06jb}) for $
F(x', x''; x) \equiv J_{+}(x',x'')$.  We note that $ \mathfrak{D}_{+}(x)$ is 
not, in general, equal to zero, so that the total thermoballistic current 
$\mathfrak{J}_{+}(x)$ is not conserved. The quantity $\mathfrak{D}_{+}(x) dx $ 
is the increment of $\mathfrak{J}_{+}(x)$  across the infinitesimal 
interval $dx$ at position $x$ of the sample. The four types of ballistic 
current appearing in expression (\ref{eq:thebal06jb}), which
contribute to this increment, are depicted by the double arrows labeled I to 
IV in Fig.~\ref{fig:6}, with the arrows I and II representing the term 
$\mathbb{F}_{1}(x; [F])$, and III and IV the term $\mathbb{F}_{2}(x; [F])$.

If we set $F(x', x''; x) \equiv
\check{J}_{-}(x',x'';x)$ in expressions (\ref{eq:thebal01}) and
(\ref{eq:thebal06jb}), we find from Eq.~(\ref{eq:thebal01dz}) for the relaxing
thermoballistic spin-polarized current
\begin{equation}
\frac{d}{dx} \check{\mathfrak{J}}_{-}(x) = - \frac{2 v_{e}}{l_{s}}
\check{\mathfrak{n}}_{-}(x) + \check{\mathfrak{D}}_{-}(x) .
\label{eq:thebal06}
\end{equation}
In the first term on the right-hand side of this equation, we have introduced 
the relaxing thermoballistic spin-polarized density 
$\check{\mathfrak{n}}_{-}(x)$ by using the balance equation
(\ref{eq:baleq01zs}) to replace the derivatives 
$\partial \check{J}_{-}(x', x''; x) / \partial x$ which we encounter when 
differentiating the integrals in expression (\ref{eq:thebal01}).

The thermoballistic currents and densities, Eq.~(\ref{eq:thebal01}), are
expressed, via the corresponding ballistic currents and densities,
Eqs.~(\ref{eq:ball004jvx})--(\ref{eq:netbal02xy}), in terms of two dynamical
functions, {\em viz.}, the mean spin function $\tilde{A}(x)$ and the spin
accumulation function $A(x)$, or, equivalently, the average chemical potential
$\tilde{\mu}(x)$ and the splitting of the spin-resolved chemical potentials,
$\mu_{-}(x)$ [see Eqs.~(\ref{eq:eqden03zk}), (\ref{eq:spfra01fv}), and
(\ref{eq:ball004jhm})].  To implement the thermoballistic concept, we must
establish algorithms for calculating $\tilde{A}(x)$ and $A(x)$ in terms of the
intrinsic parameters of the semiconducting system, like momentum and spin
relaxation lengths, as well as of the external parameters, like applied voltage
and spin polarization in the external leads.

Before establishing these algorithms within the thermoballistic approach, i.e.,
for arbitrary momentum relaxation length $l$, we directly evaluate the
thermoballistic currents and densities in the drift-diffusion regime (small
$l$) and for the ballistic case ($l \rightarrow \infty$).  The results will
turn out to be equal to the respective standard physical expressions summarized
in Sec.~\ref{sec:dibatran}, which demonstrates that the thermoballistic
description indeed bridges the gap between the drift-diffusion and ballistic
descriptions of carrier transport.

\subsubsection{Drift-diffusion regime}

\label{sec:driftlim}

In the {\em drift-diffusion regime}, when $l/S \ll 1$ and $l/l_{s} \ll 1$,
nonzero contributions to the thermoballistic currents and densities defined by
Eq.~(\ref{eq:thebal01}) arise only from very short ballistic intervals 
$[x', x'']$  enclosing the point $x$, so that $x - x'$ and $x'' - x$
are infinitesimals and only the double integral over $x'$ and $x''$
contributes.

Then, from Eq.~(\ref{eq:ball004juy}), the persistent part of the total
thermoballistic density, $\overstar{\mathfrak{n}}_{+}(x)$, is seen to reduce to
the total equilibrium density $n_{+}(x)$ [see Eq.~(\ref{eq:ball004jux})],
\begin{equation}
\overstar{\mathfrak{n}}_{+}(x) = n_{+}(x) ,
\label{eq:ball004jul}
\end{equation}
with $n_{+}(x)$ expressed in terms of the functions $\tilde{A}(x)$ and $A(x)$, 
\begin{equation}
n_{+}(x) = \frac{N_{c}}{2} B_{+}(x) [\tilde{A}(x) + \smfrac{1}{2} P(x) A(x)] .
\label{eq:ball004jum}
\end{equation}
Further, since ${\cal D}^{m}_{+}(x',x'';x) = 0$ in the drift-diffusion regime, 
the relaxing part of the joint total ballistic density,
$\check{\mathfrak{n}}_{+}(x', x''; x)$ [see Eq.~(\ref{eq:netbal02xy})],
vanishes. Consequently, we have
\begin{equation}
\check{\mathfrak{n}}_{+}(x) = 0
\label{eq:ball004jus}
\end{equation}
for the relaxing part of the total thermoballistic density,
$\check{\mathfrak{n}}_{+}(x)$, and hence
\begin{equation}
\mathfrak{n}_{+}(x) \equiv \overstar{\mathfrak{n}}_{+}(x) +
\check{\mathfrak{n}}_{+}(x) = n_{+}(x)
\label{eq:ball004jup}
\end{equation}
for the total thermoballistic density $\mathfrak{n}_{+}(x)$ in the
drift-diffusion limit.

The persistent part of the thermoballistic spin-polarized density,
$\overstar{\mathfrak{n}}_{-}(x)$, is obtained from Eq.~(\ref{eq:ball004juy}) as
\begin{equation}
\overstar{\mathfrak{n}}_{-}(x) = \frac{N_{c}}{2} B_{-}(x) [ \tilde{A}(x)
+ \smfrac{1}{2} P(x) A(x)] ,
\label{eq:ball004juz}
\end{equation}
and the relaxing part of the thermoballistic spin-polarized density,
$\check{\mathfrak{n}}_{-}(x)$, from Eq.~(\ref{eq:netbal02xy}) as
\begin{eqnarray}
\check{\mathfrak{n}}_{-}(x) &=& \frac{N_{c}}{4}  [B_{+}(x) - P(x)
B_{-}(x)] A(x) \nonumber \\ &=&  \frac{N_{c}}{4} B_{+}(x) Q^{2}(x)
A(x) ,
\label{eq:ball004jua}
\end{eqnarray}
with $Q(x)$ given by Eq.~(\ref{eq:spfra17c}). Then, re-expressing the
spin-polarized equilibrium density  $n_{-}(x)$ [see Eq.~(\ref{eq:ball004jux})]
in the form
\begin{equation}
n_{-}(x) = \frac{N_{c}}{2} [B_{-}(x) \tilde{A}(x) + \smfrac{1}{2}
B_{+}(x) A(x)] ,
\label{eq:ball004jvz}
\end{equation}
we have
\begin{equation}
\mathfrak{n}_{-}(x) \equiv \overstar{\mathfrak{n}}_{-}(x) +
\check{\mathfrak{n}}_{-}(x) = n_{-}(x)
\label{eq:ball004jut}
\end{equation}
for the thermoballistic spin-polarized density $\mathfrak{n}_{-}(x)$ in the
drift-diffusion limit.

The net total ballistic current $J_{+}(x', x'')$ is found from
Eq.~(\ref{eq:ball004jvx}),  by expanding to first order in $x' - x$ and $x'' -
x$, in the form
\begin{equation}
J_{+}(x',x'') =  \frac{v_{e} N_{c}}{2} B_{+}(x) \hat{A}(x) (x' - x'') ,
\label{eq:ballim30}
\end{equation}
where
\begin{equation}
\hat{A}(x) = \frac{d \tilde{A}(x)}{dx} + \smfrac{1}{2} P(x) \frac{d
A(x)}{dx} .
\label{eq:ballim30aha}
\end{equation}
Now, evaluating the double integral over $x'$ and $x''$ in
Eq.~(\ref{eq:thebal01}) for $F(x', x''; x) = x' - x''$ and taking the limit
$\delta \rightarrow 0$ at fixed $l > 0$, we are left with a factor $- 2l$, so 
that the drift-diffusion limit of the total thermoballistic current
$\mathfrak{J}_{+}(x)$ is given by
\begin{eqnarray}
\mathfrak{J}_{+}(x)  &=& - v_{e} N_{c} l B_{+}(x) \hat{A}(x) \nonumber \\
 &=& - \frac{\nu N_{c}}{2 \beta e} B_{+}(x) \hat{A}(x) \equiv J . 
\label{eq:ballim20}
\end{eqnarray}
Here, we have used relation (\ref{eq:ballim20a}) to introduce the electron
mobility $\nu$, and we have identified the constant, total thermoballistic 
current in 
the drift-diffusion regime with the (conserved) total physical current $J$. 
Then, using Eqs.~(\ref{eq:ball004jum}) and (\ref{eq:ball004jvz}), we can 
express $J$ in terms of the equilibrium densities $n_{\pm}(x)$ [see
Eqs.~(\ref{eq:ball004jup}) and (\ref{eq:ball004jut})] in the form
\begin{eqnarray}
J &=& - \frac{\nu}{e} \left[ n_{+}(x) \frac{d E_{c}(x)}{dx} + \frac{1}{\beta}
\frac{d n_{+}(x)}{dx} + \smfrac{1}{2} n_{-}(x) \frac{d \Delta (x)}{dx} \right]
. \nonumber \\
\label{eq:ballim20cc}
\end{eqnarray}
For zero spin splitting, $\Delta (x) = 0$, this expression becomes
equivalent to expression (\ref{eq:drude008}) for the total current in the
standard drift-diffusion model.  However, by contrast with the latter model, we
have obtained Eq.~(\ref{eq:ballim20cc}) without invoking the Einstein relation
(\ref{eq:drude011}).  This feature can be traced back to the probabilistic
description underlying the thermoballistic approach (see
Sec.~\ref{sec:probappr}), which allows the diffusion current to be directly
expressed in terms of the collision time (and hence of the momentum relaxation
length $l$) [see Eq.~(\ref{eq:prob006})].

To obtain an explicit expression for the mean spin function $\tilde{A}(x)$, and
hence for the average chemical potential $\tilde{\mu}(x)$, we observe that in
the drift-diffusion regime the total thermoballistic density
$\mathfrak{n}_{+}(x)$ is equal to the total equilibrium density
$n_{+}(x)$ [see Eq.~(\ref{eq:ball004jup})], which, in turn, is related to
$\tilde{A}(x)$ via Eq.~(\ref{eq:ball004jum}). Using Eq.~(\ref{eq:ballim30aha}) 
to solve Eq.~(\ref{eq:ballim20}) for $d \tilde{A}(x)/dx$, integrating over 
the interval $[x_{1}, x]$, and using 
Eqs.~(\ref{eq:eqden60})--(\ref{eq:spfra17c}) to simplify the integrals, we can
express $\tilde{\mu}(x)$ in the form
\begin{eqnarray}
e^{\beta \tilde{\mu}(x)} = e^{\beta \tilde{\mu}(x_{1})} &-& \frac{\beta e
J}{\nu N_{c}} \int_{x_{1}}^{x} dx' e^{\beta E_{c}(x')} Q(x') \nonumber \\ &-&
\smfrac{1}{2} [P(x) A(x) - P(x_{1}) A(x_{1})] \nonumber \\ &-& \frac{\beta}{4}
\int_{x_{1}}^{x} dx' \frac{d \Delta (x')}{dx'} Q^{2}(x') A(x') ,
\nonumber \\
\label{eq:ballim20kf}
\end{eqnarray}
which generalizes expression (\ref{eq:diff003}).  Integration over the interval
$[x, x_{2}]$ leads to another expression for $\tilde{\mu}(x)$, which is 
different in form, but numerically equal to expression (\ref{eq:ballim20kf}).
The drift-diffusion form of the spin accumulation function $A(x)$ in expression
(\ref{eq:ballim20kf}) is determined by a differential equation [see
Eq.~(\ref{eq:ballim26xx}) below].

We now set $x = x_{2}$ in Eq.~(\ref{eq:ballim20kf}) and identify the boundary
values of $\tilde{\mu}(x)$ at the interface positions $x_{1,2}$ with
the values, $\mu_{1,2}$, of the equilibrium chemical potential at the
contact side of the contact-semiconductor interfaces,
\begin{equation}
\tilde{\mu}(x_{1,2}) = \mu_{1,2}
\label{eq:ballim05}
\end{equation}
[see Eq.~(\ref{eq:diff005})], so that from Eq.~(\ref{eq:eqden03zk})
\begin{equation}
\tilde{A}(x_{1,2})  = e^{\beta \mu_{1,2}} \equiv \eta_{1,2} .
\label{eq:ballim05ups}
\end{equation}
Similarly, we identify the boundary values of $A(x)$ with external values 
$A_{1,2}$,
\begin{equation}
A(x_{1,2}) = A_{1,2} .
\label{eq:ballim05s}
\end{equation}
We then obtain the drift-diffusion form of the current-voltage characteristic,
which, for simplicity, is written down here for the case of constant spin
splitting, when $d \Delta (x)/dx = 0$, $P(x) = P$, and $Q(x) = Q$,
\begin{eqnarray}
J &=& \frac{\nu N_{c}}{\beta e Q \tilde{S}} e^{-\beta [E_{c}^{m}(x_{1}, x_{2})
- \mu_{1}]} \nonumber \\ &\ & \times [ 1 - e^{- \beta e V} + \smfrac{1}{2}
e^{- \beta \mu_{1}} P (A_{1} - A_{2})] 
 .
\label{eq:diff006hsa}
\end{eqnarray}
Here, we have used Eqs.~(\ref{eq:diff004}) and (\ref{eq:diff008}), 
respectively, to introduce the voltage bias $V$ and the effective sample length
$\tilde{S}$.

For zero spin splitting, when $P(x) = 0$ and $Q(x) = 1$, expressions
(\ref{eq:ballim20kf}) and (\ref{eq:diff006hsa}) become equivalent to the
expressions (\ref{eq:diff003}) and (\ref{eq:diff006}), respectively, in the
standard drift-diffusion model.

The drift-diffusion limit of the persistent thermoballistic spin-polarized
current $\overstar{\mathfrak{J}}_{-}(x)$ immediately follows from
Eqs.~(\ref{eq:netbal06}) and (\ref{eq:ballim20}) as
\begin{equation}
\overstar{\mathfrak{J}}_{-}(x)  = - \frac{\nu N_{c}}{2 \beta e} B_{-}(x)
\hat{A}(x) = J P(x) ,
\label{eq:ballim2012}
\end{equation}
so that $\overstar{\mathfrak{J}}_{-}(x) = 0$ for zero spin splitting.

Proceeding in analogy to the derivation of $\mathfrak{J}_{+}(x)$, we find the
drift-diffusion limit of the relaxing thermoballistic spin-polarized current
$\check{\mathfrak{J}}_{-}(x)$ from Eqs.~(\ref{eq:netbal02}) and
(\ref{eq:thebal01}) in the form
\begin{equation}
\check{\mathfrak{J}}_{-}(x) = - \frac{\nu N_{c}}{4 \beta e} B_{+}(x)
Q^{2}(x) \frac{d A(x)}{dx} .
\label{eq:ballim21}
\end{equation}
Using Eq.~(\ref{eq:ball004jua}), we can express $\check{\mathfrak{J}}_{-}(x)$
in terms of the analogous thermoballistic density
$\check{\mathfrak{n}}_{-}(x)$,
\begin{equation}
\check{\mathfrak{J}}_{-}(x) = - \frac{\nu}{e} \left[\hat{B}_{+}(x)
\check{\mathfrak{n}}_{-}(x) + \frac{1}{\beta} \frac{d
\check{\mathfrak{n}}_{-}(x)}{dx} \right] ,
\label{eq:ballim2010}
\end{equation}
where
\begin{equation}
\hat{B}_{+}(x) = - \frac{1}{\beta} \frac{d\ln \bm{(} B_{+}(x) Q^{2}(x)
\bm{)}}{dx} ,
\label{eq:ballim24xy}
\end{equation}
so that $\hat{B}_{+}(x) = dE_{c}(x)/dx$ for zero spin splitting.

Inserting expressions (\ref{eq:ballim21}) and (\ref{eq:ball004jua}) for
$\check{\mathfrak{J}}_{-}(x)$ and $\check{\mathfrak{n}}_{-}(x)$, respectively,
in the general balance equation (\ref{eq:ballim24new}), we obtain a homogeneous
second-order differential equation for the spin accumulation function $A(x)$,
\begin{equation}
\frac{d^{2} A(x)}{dx^{2}} - \beta \hat{B}_{+}(x) \frac{dA(x)}{dx} -
\frac{1}{L^{2}_{s}} A(x) = 0 ,
\label{eq:ballim26xx}
\end{equation}
where
\begin{equation}
L_{s} = \sqrt{l l_{s}}
\label{eq:ballim26}
\end{equation}
is the spin diffusion length. Equation (\ref{eq:ballim26xx}) can be converted
into an analogous equation for the density $\check{\mathfrak{n}}_{-}(x)$,
\begin{eqnarray}
\frac{d^{2} \check{\mathfrak{n}}_{-}(x)}{dx^{2}} &+& \beta \frac{d}{dx} [
\hat{B}_{+}(x) \check{\mathfrak{n}}_{-}(x) ] - \frac{1}{L^{2}_{s}}
\check{\mathfrak{n}}_{-}(x) = 0 , \nonumber \\
\label{eq:ballim25}
\end{eqnarray}
where Eq.~(\ref{eq:ball004jua}) has been used. For zero spin splitting, this 
equation generalizes, by including arbitrarily
shaped potential energy profiles, the drift-diffusion equation commonly used to
describe electric-field effects in spin-polarized transport in semiconductors
[see Eq.~(2.8) of Ref.~\onlinecite{yuf02a};\ in that equation, the spin
diffusion length appears as $\sqrt{D \tau_{s}}$, with $D$ an effective
diffusion coefficient and $\tau_{s}$ the spin relaxation time].  Note that, in
contrast to Eq.~(\ref{eq:ballim26xx}) for $A(x)$, Eq.~(\ref{eq:ballim25}) for $
\check{\mathfrak{n}}_{-}(x)$ contains terms proportional to the first and
second derivatives of $E_{c}(x)$ [see Eq.~(\ref{eq:ballim24xy})].

\subsubsection{Ballistic limit}

\label{sec:ballim}

In the strictly {\em ballistic limit}, when $l \rightarrow \infty$, there are
no points of local thermodynamic equilibrium in the interval $[x_{1}, x_{2}]$,
and the average chemical potential $\tilde{\mu} (x)$ is not defined
inside this interval.  Expression (\ref{eq:thebal01}) reduces to
\begin{equation}
\mathfrak{F}(x) = F(x_{1},x_{2};x) ,
\label{eq:ballim01}
\end{equation}
i.e., the thermoballistic currents and densities are given by the corresponding
expressions for the net ballistic currents and joint ballistic densities (see
Sec.~\ref{sec:netball}), evaluated at $x' = x_{1}$ and $x'' = x_{2}$.  In
expressions (\ref{eq:ball004jvx}), (\ref{eq:netbal02}), (\ref{eq:ball004juy}),
and (\ref{eq:netbal02xy}), the boundary values of the mean spin function,
$\tilde{A}(x_{1,2})$, and those of the spin accumulation function,
$A(x_{1,2})$, are to be identified with the corresponding values in the 
contacts [see Eqs.~(\ref{eq:ballim05ups}) and (\ref{eq:ballim05s}), 
respectively].

From Eq.~(\ref{eq:ball004jvx}), we then obtain the total thermoballistic
current $\mathfrak{J}_{+}(x)$ in the form
\begin{eqnarray}
\mathfrak{J}_{+}(x) &\equiv& \mathfrak{J}_{+} = J_{+}(x_{1},x_{2}) \nonumber \\
&=& \frac{v_{e} N_{c}}{2} B^{m}_{+}(x_{1},x_{2}) [\eta^{-}_{12} +
\smfrac{1}{2} P^{m}(x_{1},x_{2}) A^{-}_{12}] , \nonumber \\
\label{eq:ballim02}
\end{eqnarray}
where
\begin{equation}
\eta^{\pm}_{12} = \eta_{1} \pm \eta_{2} ,
\label{eq:ballim04}
\end{equation}
with $\eta_{1,2}$ defined by Eq.~(\ref{eq:ballim05ups}), and
\begin{equation}
A^{\pm}_{12} = A_{1} \pm A_{2} .
\label{eq:ballim04st}
\end{equation}
For zero spin splitting, expression (\ref{eq:ballim02}) becomes equivalent to
expression (\ref{eq:ball008}) for the total current in the ballistic transport
model.  The ballistic limit of the persistent thermoballistic spin-polarized
current $\overstar{\mathfrak{J}}_{-}(x)$ is obtained from $\mathfrak{J}_{+}(x)
$ by replacing in expression (\ref{eq:ballim02}) the quantity
$B^{m}_{+}(x_{1},x_{2})$ with $B^{m}_{-}(x_{1},x_{2})$ [see
Eqs.~(\ref{eq:emtra09}) and (\ref{eq:netbal06})].

For the total thermoballistic density $\mathfrak{n}_{+}(x)$, we have, defining
\begin{eqnarray}
{\cal A}^{\pm}_{12}(x_{1}, x_{2}; x) &\equiv&  A_{1} e^{-{\cal
C}^{m}(x_{1},x_{2};x_{1},x)/l_{s}} \nonumber \\ &\ & \pm A_{2} e^{-{\cal
C}^{m}(x_{1},x_{2};x,x_{2})/l_{s}} 
\label{eq:ballim04uv}
\end{eqnarray}
and using Eqs.~(\ref{eq:netbal03ax})--(\ref{eq:netbal02xy}),
\begin{eqnarray}
\mathfrak{n}_{+}(x) &=& n_{+}(x_{1},x_{2};x) = \overstar{n}_{+}(x_{1},x_{2};x)
+ \check{n}_{+}(x_{1},x_{2};x) \nonumber \\ &=& \frac{N_{c}}{4} \{
D^{m}_{+}(x_{1},x_{2};x) [\eta^{+}_{12} + \smfrac{1}{2}
P^{m}(x_{1},x_{2}) A_{12}^{+}] \nonumber \\ &\ & + \smfrac{1}{2} {\cal
D}^{m}_{+}(x_{1},x_{2};x) {\cal A}^{+}_{12}(x_{1}, x_{2}; x)\} .
\label{eq:ballim03}
\end{eqnarray}
For zero spin splitting, this becomes equivalent to expression
(\ref{eq:ball007za}).

Further, using Eq.~(\ref{eq:netbal02}) and (\ref{eq:netbal02xy}), we express
the relaxing thermoballistic spin-polarized current
$\check{\mathfrak{J}}_{-}(x)$ and the corresponding density
$\check{\mathfrak{n}}_{-}(x)$ as
\begin{eqnarray}
\check{\mathfrak{J}}_{-}(x) &=& \check{J}_{-}(x_{1}, x_{2}; x) \nonumber \\ &=&
\frac{v_{e} N_{c}}{4} B^{m}(x_{1},x_{2}) {\cal A}^{-}_{12}(x_{1}, x_{2}; x)
\label{eq:ballim21a}
\end{eqnarray}
and
\begin{eqnarray}
\check{\mathfrak{n}}_{-}(x) &=& \check{n}_{-}(x_{1}, x_{2}; x) \nonumber \\ &=&
\frac{N_{c}}{8} {\cal D}^{m}_{-}(x_{1},x_{2};x){\cal A}^{+}_{12}(x_{1},
x_{2}; x) , 
\label{eq:ballim21b}
\end{eqnarray}
respectively.  In view of Eq.~(\ref{eq:baleq01zs}), the current
$\check{\mathfrak{J}}_{-}(x)$ and density $\check{\mathfrak{n}}_{-}(x)$ in the
ballistic limit are trivially connected by the general balance equation
(\ref{eq:ballim24new}).

\subsubsection{Thermoballistic energy dissipation}

The stochastic equilibration of the electrons that occurs during their motion 
across the sample is associated with the dissipation of energy, i.e., the net
transfer of energy out of the ensemble of conduction band electrons into a 
reservoir (``heat bath'') of electrons in thermodynamic equilibrium.

To describe this transfer within the thermoballistic concept, we introduce 
the (conserved) ballistic {\em energy} currents $\mathbb{E}^{l,r}(x',x'')$ 
generated by thermal electron emission at the end-points $x'$ and $x''$, 
respectively, of the ballistic interval $[x', x'']$. Neglecting spin degrees 
of freedom, we have for the current $\mathbb{E}^{l}(x',x'')$,  
by an obvious modification of expression (\ref{eq:ball004}) for the electron 
current $J^{l}(x_{1},x_{2})$, writing $\epsilon = m^{\ast} v_{x}^{2}/2$,
\begin{eqnarray}
\mathbb{E}^{l} (x',x'') &=& \frac{4 \pi m^{\ast}}{\beta h^{3}} 
\int_{0}^{\infty} d\epsilon \epsilon e^{- \beta [\epsilon + E_{c}(x') 
- \tilde{\mu}(x')]} \nonumber \\ &\ & \times \Theta \bm{(} \epsilon - 
E_{b}^{l}(x', x'') \bm{)} , 
\label{eq:thec01}
\end{eqnarray} 
where, in analogy to Eq.~(\ref{eq:diff007}),
\begin{equation}
E_{b}^{l}(x', x'') = E_{c}^{m}(x', x'') - E_{c}(x') 
\label{eq:thec02}
\end{equation} 
is the maximum barrier height of the potential energy profile $E_{c}(x)$  
relative to its value at $x'$. Evaluating the integral in expression 
(\ref{eq:thec01}), we obtain, in extension of Eq.~(\ref{eq:ball004f}),    
\begin{eqnarray}
\mathbb{E}^{l} (x',x'') &=& v_{e} N_{c} \left[ \frac{1}{\beta} + 
E_{b}^{l}(x', x'')\right]  e^{- \beta [ E_{c}^{m}(x', x'') - \tilde{\mu}(x')]} 
. \nonumber \\
\label{eq:thec03}
\end{eqnarray} 
For the ballistic energy current $\mathbb{E}^{r} (x',x'')$, we have   
\begin{equation}
\mathbb{E}^{r} (x',x'') = - \mathbb{E}^{l} (x'',x') ,
\label{eq:thec03gyl}
\end{equation} 
in line with Eq.~(\ref{eq:ball004rfd}) for the corresponding ballistic 
electron current, and for the net ballistic energy current in the interval 
$[x', x'']$,
\begin{equation}
\mathbb{E}(x', x'') = \mathbb{E}^{l}(x', x'') + \mathbb{E}^{r}(x', x'') , 
\label{eq:thec00gjb}
\end{equation} 
in line with Eq.~(\ref{eq:ball004zgx}) for the net ballistic electron current.

Now, we again introduce a ``reference coordinate'' $x$ inside the sample, 
whose meaning, however, differs from that of the coordinate $x$ in the 
definition (\ref{eq:thebal01}) of the thermoballistic currents and densities.
While in that definition, $x$ is a coordinate {\em inside ballistic 
intervals}, we here consider $x$ a coordinate characterizing the position of a
{\em point of local thermodynamic equilibrium}. More precisely, we consider a
collection of such points, with density $dx/l$, in an interval $dx$ centered 
about $x$, in which ballistic energy currents are absorbed from, and emitted 
towards, either side. The equilibrium point $x$, where incoming electron 
currents are completely equilibrated and the outgoing currents are solely 
determined by the parameters of the reservoir (i.e., by the chemical 
potential), dynamically separates the two sample partitions to the left and 
right of $x$.
   
We denote by $\mathfrak{W}_{in}(x)$ the (kinetic) energy transferred per unit 
volume and unit time {\em into} the reservoir by absorption of electrons 
at the point $x$, and correspondingly by $\mathfrak{W}_{out}(x)$ the change in
energy of the reservoir due to thermal electron emission {\em out of} it. The 
{\em net} energy $\mathfrak{W}(x)$, i.e., the energy dissipated locally at the
 point $x$, is then given by 
\begin{equation}
\mathfrak{W}(x) = \mathfrak{W}_{in}(x) +\mathfrak{W}_{out}(x) .
\label{eq:thec00ijv}
\end{equation}
The energies $\mathfrak{W}_{in}(x)$ and $\mathfrak{W}_{out}(x)$ are each
composed of two parts (see also Fig.~\ref{fig:6}).  One part of 
$\mathfrak{W}_{in}(x)$ is given 
by the weighted sum of the energy currents $\mathbb{E}^{l}(x', x'')$ over all
ballistic intervals $[x', x'']$ lying to the left of $x$ and having their right
end at $x'' = x$, where the currents  are absorbed. The other part is expressed
analogously in terms of the currents $\mathbb{E}^{r}(x, x'')$ in ballistic 
intervals to the right of the absorption point $x$. For the energy 
$\mathfrak{W}_{in}(x) dx$ dissipated in an interval $dx$ centered around $x$, 
we then have         
\begin{equation}
\mathfrak{W}_{in}(x) dx  = \frac{dx}{l} \{ 
\mathbb{F}_{1}(x; [\mathbb{E}^{l}]) + \mathbb{F}_{2}(x; [\mathbb{E}^{r}]) \} .
\label{eq:thec00kcg}
\end{equation} 
Here, the functionals $\mathbb{F}_{1}(x; [\mathbb{F}])$ and 
$\mathbb{F}_{2}(x; [\mathbb{F}])$ are given, for an arbitrary ballistic 
current $\mathbb{F}(x', x''; x)$, by Eqs.~(\ref{eq:thec00ohx}) and  
(\ref{eq:thec00gma}), respectively.
The energy current $\mathfrak{W}_{out}(x) dx$ is obtained from 
Eq.~(\ref{eq:thec00kcg}) by interchanging the role of $\mathbb{E}^{l}(x', x'')$
and $\mathbb{E}^{r}(x', x'')$,
\begin{equation}
\mathfrak{W}_{out}(x) dx  = \frac{dx}{l} \{ \mathbb{F}_{1}(x; [\mathbb{E}^{r}])
+ \mathbb{F}_{2}(x; [\mathbb{E}^{l}]) \} ,
\label{eq:thec00tuz}
\end{equation} 
which comprises all ballistic energy currents emitted at the point $x$ towards 
either side. For the dissipated energy $\mathfrak{W}(x)$, we then have 
\begin{equation}
\mathfrak{W}(x)  = \frac{1}{l} \{ 
\mathbb{F}_{1}(x; [\mathbb{E}]) + \mathbb{F}_{2}(x; [\mathbb{E}]) \} ,
\label{eq:thec00phy}
\end{equation} 
where the net ballistic energy current $\mathbb{E}(x', x'')$ is given by 
Eq.~(\ref{eq:thec00gjb}).

We note that in obtaining expression (\ref{eq:thec00phy}) for 
$\mathfrak{W}(x)$ we have {\em not} drawn on 
the thermoballistic energy current $\mathfrak{E}(x)$ that results from 
identifying in Eq.~(\ref{eq:thebal01}) the function $F(x', x''; x)$ with net 
ballistic energy current $\mathbb{E}(x', x'')$. Naively, one might expect 
that the dissipated energy $\mathfrak{W}(x)$ can be represented by the 
derivative $(\partial / \partial x^{+} +\partial / 
\partial x^{-})\mathfrak{E}(x) $ given by Eq.~(\ref{eq:thebal06jb}). 
However, comparing expressions (\ref{eq:thec00phy}) and  (\ref{eq:thebal06jb}),
one observes that the contributions of the ballistic energy currents 
from the two sides of $x$ add up in the former, and are subtracted from one 
another in the latter. The result (\ref{eq:thec00phy})
can be obtained from the thermoballistic energy current $\mathfrak{E}(x)$ in
the form
\begin{equation}
\mathfrak{W}(x) = - \left( \frac{\partial}{\partial x^{+}} - 
\frac{\partial}{\partial x^{-}} \right) \mathfrak{E}(x) ,
\label{eq:thec00cfk}
\end{equation}
at variance with expression (\ref{eq:thebal06jb}). 

We do not write down here the general expression for $\mathfrak{W}(x)$ in terms
of the average chemical potential $\tilde{\mu} (x)$ obtained by using 
expressions (\ref{eq:thec03})--(\ref{eq:thec00gjb}) 
in Eq.~(\ref{eq:thec00phy}), and confine ourselves to considering the ballistic
limit and the drift-diffusion regime. Owing to the overall factor $1/l$ in the 
right-hand side of Eq.~(\ref{eq:thec00phy}), $\mathfrak{W}(x)$ vanishes in the
{\em ballistic limit} $l \rightarrow \infty$, which reflects the complete 
absence of equilibration in this limit. In the {\em drift-diffusion regime}, 
on the other hand, when $l \ll S$, only the integral terms contribute. 
We evaluate these terms for the special case of a homogeneous sample subjected 
to a constant external electric field of magnitude ${\cal E}$ directed 
antiparallel to the $x$-axis, so that the potential energy profile has the form
\begin{equation}
E_{c}(x) = E_{c}(x_{1}) - \frac{\epsilon}{\beta} (x - x_{1}) ,
\label{eq:difeq01}
\end{equation}
where 
\begin{equation}
\epsilon = \beta e {\cal E} .
\label{eq:difeq01erd}
\end{equation}
For this profile, the average chemical
potential $\tilde{\mu}(x)$ is found from Eqs.~(\ref{eq:diff003}) and 
(\ref{eq:diff006}), using the relation $\epsilon S = \beta e V$, to run 
parallel to the profile,
\begin{equation}
\tilde{\mu}(x) = [\mu_{1} - E_{c}(x_{1})] + E_{c}(x) . 
\label{eq:the03b}
\end{equation}
Therefore, the total equilibrium electron density is constant, $n(x) = 
n(x_{1})$. The net ballistic energy current $\mathbb{E}(x', x'')$ can now 
be expressed as
\begin{eqnarray}
\mathbb{E}(x', x'') &=& \frac{v_{e}}{\beta} n(x_{1}) 
\{ 1 - e^{- \epsilon (x'' - x')]} \left[ 1 + \epsilon (x'' - x') \right] 
\}  \nonumber \\ &\approx& \frac{v_{e}}{2 \beta} n(x_{1}) 
\epsilon^{2} (x''- x')^{2} ,
\label{eq:the03c}
\end{eqnarray}
where the second, approximate equation holds in the zero-bias limit $\epsilon S
\ll 1$.
Inserting the approximate representation of $\mathbb{E}(x', x'')$ in the 
integral terms of expression (\ref{eq:thec00phy}), we find for the locally 
dissipated energy  $\mathfrak{W}(x)$ in the drift-diffusion regime
\begin{equation}
\mathfrak{W}(x) \equiv \mathfrak{W} =  \frac{2 v_{e}}{\beta} l  n(x_{1}) 
\epsilon^{2} = \sigma \left( \frac{V}{S} \right)^{2}  , 
\label{eq:thec04}
\end{equation}
where we have introduced the conductivity $\sigma$ via Eqs.~(\ref{eq:drude006})
and (\ref{eq:ballim20a}). For the total energy $W = \mathfrak{W} A S$  
dissipated per unit time (``heat production'') in a sample with cross-sectional
area $A$ and resistance $R = S/\sigma A$, we then recover the Ohmic expression 
$W = V^{2}/R$.

\subsection{Synopsis of the thermoballistic concept}

\label{sec:thersyno}

We conclude this section with a synopsis of the concept underlying the 
thermoballistic description of charge carrier transport in semiconductors, 
in which we briefly comment on its basic ingredients and elucidate the 
physical content of its formal structure. This will be followed by an 
assessment of its merits as well as its weaknesses.

\subsubsection{Ingredients and physical content}

\label{sec:revcom}

The basic ingredients of the (semiclassical) thermoballistic concept are the 
ballistic carrier currents and densities which are constructed within the 
following framework. (i) Thermal emission of carriers occurs at points of local
thermodynamic equilibrium  randomly distributed over the sample. (ii) The 
equilibrium points link ``ballistic transport intervals'' across which the 
emitted carriers move ballistically under the influence of potential energy 
profiles arising from internal and external electrostatic potentials. During
their ballistic motion, the carriers undergo spin relaxation controlled by a 
ballistic spin relaxation length $l_{s}$. (iii) At the end-points of the 
ballistic intervals, instantaneous ``point-like'' thermalization 
(``absorption'') of the carriers takes place. Here, we invoke the picture of 
``reflectionless contacts''\cite{lan87, dat95,imr99}, according to which the 
ballistic carriers that enter a contact (representing a ``bath'' with an 
effectively infinite number of transverse modes) are completely absorbed there,
after having been emitted from a similar contact at the opposite end of the 
sample. In the thermoballistic concept, the emission and absorption of
carriers are treated in this way at {\em all} points of local thermodynamic 
equilibrium inside the sample. At these points, the absorption of carriers into
the bath is complete (``reflectionless''), but at the same instant carriers are
emitted out of the bath into the ballistic intervals on either side.The 
opposing collision-free currents emitted at either end into the ballistic 
interval combine to form  a {\em net ballistic current} inside the interval,
with an associated {\em joint ballistic density}.

The random distribution of points of local thermodynamic equilibrium is 
mirrored in a random partitioning of the length of the sample into ballistic
intervals, where each partition defines a ``ballistic configuration'', a 
central notion of the concept. The ballistic carrier currents and densities in 
these configurations are 
assembled to form the corresponding {\em thermoballistic currents and 
densities}. These are constructed, at a reference position $x$ located 
arbitrarily inside the  sample, by performing weighted summations, with 
weights controlled by a momentum relaxation length $l$, over the net 
ballistic currents and joint ballistic densities in all ballistic intervals 
containing the point $x$. The thermoballistic currents and densities constitute
the key element of the thermoballistic concept.

The physical content of the thermoballistic transport mechanism can be 
exhibited by analyzing the underlying formalism with regard to the 
intertwined effects of thermal electron emission and ballistic motion.
Let us consider expressions (\ref{eq:ball004jvx})--(\ref{eq:netbal02xy}) for the 
net ballistic currents and joint ballistic densities, which are 
essentially composed of two factors each. On the one hand, they contain the 
nonlocal barrier factors $B^{m}_{\pm}(x', x'')$ and  $B^{m}(x', x'')$ in the 
expressions for the total and persistent spin-polarized currents and the 
relaxing spin-polarized current, respectively, and the factors 
$D^{m}_{\pm}(x', x'';x)$ and  $D^{m}(x', x'';x)$ in the analogous expressions
for the densities. These factors describe the collision-free motion of the 
electrons across the ballistic interval $[x', x'']$, which is essentially 
determined by the potential energy profiles $E_{\uparrow, \downarrow} 
(x)$ inside the interval. They represent the {\em ballistic 
attribute} of thermoballistic transport. The factors in brackets, on the other
hand, contain terms depending on the 
average chemical potentials $\tilde{\mu}(x')$,  $\tilde{\mu}(x'')$ and the spin
accumulation functions $A(x')$, $A(x'')$, which are directly related to the
spin-resolved chemical potentials $\mu_{\uparrow, \downarrow} (x')$,
$\mu_{\uparrow, \downarrow} (x'')$ at the end-points $x', x''$ of the ballistic
interval. These factors, which describe the thermal emission 
(``thermal activation'') of the ballistic currents  at the points of local 
thermodynamic equilibrium at $x'$ and $x''$, respectively, represent the 
{\em thermal attribute} of  thermoballistic transport. The term in brackets is
the ``activation term''. [The joint appearance of ballistic and thermal 
attributes shows that the term ``thermoballistic'' indeed provides an 
appropriate characterization of our approach.]

The contributions of the ballistic currents and densities  
(\ref{eq:ball004jvx})--(\ref{eq:netbal02xy}), summarily denoted by 
$F(x', x''; x)$, to the corresponding thermoballistic currents and densities 
$\mathfrak{F}(x)$ are to be read from Eq.~(\ref{eq:thebal01}):\ they are given 
by the current (or density) $F(x', x''; x)$ in the interval $[x', x'']$, 
multiplied by the probability $e^{- (x'' - x')/l}$ that the electrons traverse 
this interval ballistically. 

The momentum relaxation length $l$ controls the magnitude of the ballistic 
contribution to the entire transport process. At the same time, it determines 
the average number of collisions, $S/l$, in a sample of length $S$. In the 
{\em ballistic limit}, when $l \rightarrow \infty$ and there is no point of 
local thermodynamic equilibrium inside the sample, the transport is purely
ballistic between the end-points $x_{1}$ and $x_{2}$, and only the first term
on the right-hand side of Eq.~(\ref{eq:thebal01}) contributes. In the opposite 
limit, when $l \rightarrow 0$, the points of local thermodynamic equilibrium at
which the electrons are equilibrated, lie infinitesimally close to one another.
Then only the double integral in Eq.~(\ref{eq:thebal01}) survives, and we 
arrive at
\begin{equation}
\mathfrak{J}_{+}(x) = -  v_{e} N_{c} l B_{+}(x) 
\frac{d}{dx} e^{\beta \tilde{\mu}(x)}
\label{eq:syn02}
\end{equation}
for the total thermoballistic current $\mathfrak{J}_{+}(x) = J$ [see 
Eq.~(\ref{eq:ballim20})]. This current is essentially given 
in terms of equilibrium quantities, a property that characterizes the 
{\em drift-diffusion limit}. Expression (\ref{eq:syn02}) has the form of the 
current in the standard drift-diffusion approach [see Eqs.~(\ref{eq:drude006f})
and (\ref{eq:diff001})], 
\begin{equation}
\mathfrak{J}_{+}(x) = - \frac{\nu}{e} n_{+}(x) \frac{d \tilde{\mu}(x)}{dx} ,
\label{eq:syn03}
\end{equation}
where $\nu$ is the electron mobility given by Eq.~(\ref{eq:ballim20a}), and
$n_{+}(x)$ the equilibrium electron density given by Eq.~(\ref{eq:ball004jum}).
In that approach, the momentum relaxation length $l$ is nonzero (so that $\nu$ 
remains nonzero), but small compared with the length scales over which the 
other parameters vary appreciably. The activation term reduces to a derivative,
and the relation between the total thermoballistic current and the average 
chemical potential becomes a local one.

\subsubsection{Merits and weaknesses}

\label{sec:accsho}

The principal merit of the  thermoballistic concept is that it allows to 
establish a consistent and transparent formalism for bridging, within the 
semiclassical approximation, the gap between the standard drift-diffusion and 
ballistic descriptions of charge carrier transport in semiconductors. While 
incorporating basic features of these descriptions, the thermoballistic concept
consistently unifies and generalizes them by introducing random partitionings
of the sample length into ballistic configurations. 

The concept is transparent in a twofold way. First, as shown above, a lucid 
interpretation of its physical content can be given in terms of ballistic and
thermal attributes. Second, owing to its semiclassical character,\cite{semcl} 
the concept allows the effects of the different parameters describing a 
semiconducting system to be clearly distinguished. In the implementation of 
this concept, explicit equations  for various transport quantities can be 
derived, and simple solutions can be obtained in important special cases.
The merit of transparency of the thermoballistic concept carries with it some 
simplifications and weak points which, however, in many cases can be remedied, 
albeit at the cost of increased complexity:\ they are not detrimental to the 
concept as a whole.

While the formulation of the full thermoballistic concept given here 
describes semiclassical transport in nondegenerate semiconducting systems, we
have demonstrated within the prototype model how effects of electron tunneling 
and degeneracy can be taken into account. Quantum interference effects in the 
electron motion are not treated explicitly, but they may be assumed to be
implicitly incorporated via an extended interpretation  of the mean free path
(or momentum relaxation length) $l$, which from the outset has been taken as a
phenomenological parameter. By treating it formally as the average distance 
that the carriers travel without collision between points of {\em complete} 
thermodynamic equilibrium, as in the relaxation time approximation,\cite{ash76} it 
simulates the effects of incomplete equilibration due to elastic or 
inelastic impurity scattering, of dimensionality, and, in the extreme, of 
quantal phase correlations. 

Indeed, in this work the momentum relaxation length $l$ is the determining 
parameter in
which a great diversity of detail is subsumed. It is introduced as a constant,
so it must include in an average way the effect of spatial variations in 
the internal and external parameters characterizing the semiconducting system;\
in particular, this constant is chosen to be independent of the potential 
energy profile, to which, however, it should be related in a self-consistent
way.  Moreover, the choice of the momentum relaxation length $l$,
rather than the relaxation time $\tau$, is also merely one of convenience for
the stationary treatment in this work. We may work with position-dependent 
momentum relaxation lengths, but this would increase the complexity of the 
formalism and obscure the general line of argument.

The spin relaxation mechanism in terms of the ballistic spin relaxation length
$l_{s}$ is again a phenomenological one, having certain similarities with the
D'yakonov-Perel' mechanism.\cite{dya71} We assume spin relaxation to occur only
during the ballistic electron motion;\ however, simultaneous spin and 
momentum relaxation could be taken into account by introducing additional terms
in the spin balace equation.   

In principle, the thermoballistic concept allows a fully three-dimensional 
treatment of bipolar carrier transport to be implemented. However, in the 
present paper, in order to keep the formalism manageable, we work within a 
narrowed framework. First, we confine ourselves to unipolar transport, dealing 
specifically with electron transport in a spin-split conduction 
band. Second, we consider three-dimensional ``plane-parallel'' samples whose 
parameters (in particular, the average density of the scattering centers 
associated with impurities) do not vary in the directions perpendicular to the 
transport direction (the $x$-direction). Nevertheless, electron transport in 
this kind of  sample depends on the number of dimensions, $n$, via 
``no-scattering probabilities'' $p_{n}(x)$ [see Sec.~II of 
Ref.~\onlinecite{lip05}]. Here, in order to be able to write down physically transparent 
formulae, we use one-dimensional no-scattering probabilities of the form 
$p_{1}(x) = e^{- x/l}$ (see Sec.~\ref{sec:probappr}).

\section{Thermoballistic approach:\ Implementation}

\label{sec:therimpl}

Having presented, in the preceding section, the concept underlying the
thermoballistic approach, we now turn to the implementation of this concept. We
begin by establishing the physical conditions from which the algorithms for 
calculating the dynamical functions, {\em viz.}, the average chemical potential
$\tilde{\mu}(x)$ [via the mean spin function $\tilde{A}(x)$] and the spin 
accumulation function $A(x)$, are developed. Thereafter, these algorithms will 
be described in detail.

\subsection{Physical conditions determining the dynamical functions}

\label{sec:condi} 

We call {\em physical conditions} (i) a relation introduced to connect the 
total thermoballistic current $\mathfrak{J}_{+}(x)$ with the cognate total 
physical current $J$, which allows us to calculate the function
$\tilde{A}(x)$, and (ii) an assumption concerning the detailed spin relaxation
mechanism, which leads to the determination of the function $A(x)$.

\subsubsection{Determination of the mean spin function}

\label{sec:detmsf}

The current $\mathfrak{J}_{+}(x) \equiv \mathfrak{J}_{+}(x_{1}, x_{2}; x; l)$, 
owing to its construction in terms of ballistic currents averaged over 
random ballistic configurations (with weights controlled by the 
momentum relaxation length $l$), is to be interpreted as an ``ensemble 
average'' of the electron current at the point $x$.  This average is spatially
varying, and, therefore, it is the {\em spatial average} of the ensemble 
average over the length of the sample which is to be identified with the 
(constant) total 
physical current $J_{+}$ inside the sample.  By current conservation at 
$x_{1}$ and $x_{2}$, this current is equal to the total current $J$ in the 
left and right leads, so that we have
\begin{equation}
\frac{1}{x_{2} - x_{1}} \int_{x_{1}}^{x_{2}} dx \mathfrak{J}_{+}(x_{1},
x_{2}; x; l) = J_{+} \equiv J .
\label{eq:thebal04abc}
\end{equation}
In this condition, the current $\mathfrak{J}_{+}(x_{1}, x_{2}; x; l)$ is 
defined at a ballistic point $x$ located inside an ensemble of ballistic 
intervals $[x', x'']$, where $x'$ and $x''$, unlike $x$, are points of local 
thermodynamic equilibrium [see Eqs.~(\ref{eq:thebal00}) and 
(\ref{eq:thebal01})]. The integration over $x$ starts at the fixed equilibrium
point $x_{1}$ and ends at the fixed equilibrium point $x_{2}$. Now, just as we 
have introduced the  {\em ballistic} reference point $x$, we consider here an
{\em equilibrium} point of reference, i.e., a point of local thermodynamic 
equilibrium anywhere inside the sample, which, again, is labeled  by 
the coordinate $x$ $(x_{1} < x < x_{2})$. Such a point acts
in the same way as the fixed equilibrium points $x_{1}$ and $x_{2}$.
The ballistic current entering $x$, say, from the left, is completely 
absorbed, whereupon a thermal current is instantaneously emitted 
to either side of $x$. The same happens to the current entering the 
equilibrium point $ x$ from the right. However, in contrast to the ``true'', 
externally controlled equilibrium
points $x_{1,2}$, which are located at the contact side of the 
contact-semiconductor interfaces, no Sharvin-type interface resistance (see 
Sec.~\ref{sec:sharvin} below) appears at $x$.
   
Within this scheme, we introduce the thermoballistic current 
$\mathfrak{J}_{+}^{(1)}(x_{1},x; \xi; l)$, with $x_{1} < \xi < x \leq x_{2}$, 
where $x$ is an equilibrium point,
and $\xi$ a ballistic point. In conformance with Eq.~(\ref{eq:thebal04abc}),
the spatial average of this current over the range $[x_{1}, x]$ is again, by 
current conservation at $x_{1}$, equal to the physical current in the left
lead, so that 
\begin{equation}
\frac{1}{x - x_{1}} \int_{x_{1}}^{ x } d \xi \mathfrak{J}_{+}^{(1)}(x_{1}, 
x; \xi; l) =  J  .
\label{eq:thebal04ohg}
\end{equation}
Analogously, we have for the thermoballistic current $\mathfrak{J}_{+}^{(2)} 
(x,x_{2}; \xi; l)$ in the range $[x, x_{2}]$, using similar arguments as
above, 
\begin{equation}
\frac{1}{x_{2} - x} \int_{x}^{x_{2}} d\xi \mathfrak{J}_{+}^{(2)}(x,x_{2}; 
\xi; l) = J 
\label{eq:thebal04zul}
\end{equation}
($x_{1} \leq x < \xi \leq x_{2}$). Equations (\ref{eq:thebal04ohg}) and 
(\ref{eq:thebal04zul}), when expressed in
terms of the function $\tilde{A}(x)$ via Eqs.~(\ref{eq:ball004jvx}) and 
(\ref{eq:thebal01}), lead to two different Volterra-type integral equations
with solutions $\tilde{A}_{1}(x)$  and $\tilde{A}_{2}(x)$, respectively. 
Trivially, by satisfying Eqs.~(\ref{eq:thebal04ohg}) and 
(\ref{eq:thebal04zul}), these solutions also satisfy condition 
(\ref{eq:thebal04abc}). They define, for given spin accumulation function 
$A(x)$, two different total equilibrium densities $n_{+}^{(1)}(x)$  and 
$n_{+}^{(2)}(x)$, respectively, via Eq.~(\ref{eq:ball004jum}). The total 
equilibrium density at $x$ is obtained as the mean value of these,
\begin{equation}
n_{+}(x) = \smfrac{1}{2} [ n_{+}^{(1)}(x) +  n_{+}^{(2)}(x) ] ,
\label{eq:thebal05lg}
\end{equation}
so that, using Eq.~(\ref{eq:ball004jum}) again, we have for the unique
(thermoballistic) mean spin function
\begin{equation}
\tilde{A}(x) \equiv e^{\beta \tilde{\mu}(x)} = \smfrac{1}{2} 
[ \tilde{A}_{1}(x) +  \tilde{A}_{2}(x) ] .
\label{eq:thebal05pk}
\end{equation}
The physical conditions (\ref{eq:thebal04ohg}) and (\ref{eq:thebal04zul}),
together with Eq.~(\ref{eq:thebal05lg}), determine the procedure for 
calculating the average chemical potential $\tilde{\mu}(x)$. Using the mean
value (\ref{eq:thebal05lg}) as the point of
departure for constructing $\tilde{\mu}(x)$ reflects the fact that this
function expresses an intrinsic property of the semiconducting sample, with
no preference for one or the other of the sample ends at $x_{1}$ and $x_{2}$.
With $\tilde{A}(x)$ given by Eq.~(\ref{eq:thebal05pk}), it follows from 
Eqs.~(\ref{eq:ball004jvx}) and (\ref{eq:thebal01}) that the unique 
thermoballistic current $\mathfrak{J}_{+}(x_{1}, x_{2}; x; l)$ is given by
\begin{eqnarray}
\mathfrak{J}_{+}(x_{1}, x_{2}; x; l) &=&  \smfrac{1}{2} 
[\mathfrak{J}_{+}^{(1)}(x_{1}, x_{2}; x; l) + 
\mathfrak{J}_{+}^{(2)}(x_{1}, x_{2}; x; l) ], \nonumber \\
\label{eq:thebal04lfs}
\end{eqnarray}
a symmetric combination as in Eq.~(\ref{eq:thebal05pk}).

We note that in Ref.~\onlinecite{lip05} we have applied a different procedure
for constructing the average chemical potential $\tilde{\mu}(x)$. There, the 
combination of the functions $\tilde{A}_{1}(x)$ and $\tilde{A}_{2}(x)$ was not
chosen to be symmetric as in Eq.~(\ref{eq:thebal05pk}), but was determined
by the requirement that the values of the unique total thermoballistic current 
at the two sample ends be equal,
\begin{equation} 
\mathfrak{J}_{+}(x_{1}, x_{2}; x_{1}^{+};l) 
= \mathfrak{J}_{+}(x_{1}, x_{2}; x_{2}^{-};l) .
\label{eq:thebal04gcx}
\end{equation}
This condition derives from postulating that the position dependence of the 
total thermoballistic current is compensated by that of a 
``background current'',\cite{lip03,lip05} such that these currents add up to 
the (conserved) total physical current. The background current is assumed to
be fed by sources and sinks whose effect averages out to zero when the current 
is integrated over the length of the sample. In the present work, this 
hypothesis has been abandoned for being unphysical.

\subsubsection{Determination of the spin accumulation function}

\label{sec:detsaf}

In the thermoballistic transport mechanism, we assume that spin relaxation
takes place only {\em inside} the ballistic intervals (see
Sec.~\ref{sec:balance}), and that an infinitesimal shift of the end-points of
the ballistic intervals does not affect the current
$\check{\mathfrak{J}}_{-}(x)$.  Accordingly, we set the term arising from
differentiating the limits of integration in expression (\ref{eq:thebal01}) for
$\check{\mathfrak{J}}_{-}(x)$ equal to zero,
\begin{equation}
\check{\mathfrak{D}}_{-}(x) = 0 .
\label{eq:thebal06rt}
\end{equation}
From Eq.~(\ref{eq:thebal06}), the relaxing thermoballistic spin-polar\-ized
current and density are then seen to be connected by the balance equation
\begin{equation}
\frac{d}{dx} \check{\mathfrak{J}}_{-}(x) + \frac{2 v_{e}}{l_{s}}
\check{\mathfrak{n}}_{-}(x) = 0 ,
\label{eq:ballim24new}
\end{equation}
which is of the same form as the balance equation (\ref{eq:baleq01zs})
connecting the relaxing spin-polarized current and density in the individual
ballistic intervals.

The {\em  spin accumulation function} $A(x)$, and hence the relaxing
thermoballistic spin-polarized current $\check{\mathfrak{J}}_{-}(x)$, are
determined by condition (\ref{eq:thebal06rt}).
When written in terms of $A(x)$ by using expression (\ref{eq:netbal02}) in
Eq.~(\ref{eq:thebal06jb}), this condition turns into a linear, inhomogeneous
Fredholm-type integral equation of the second kind\cite{mor53} for $A(x)$.  The
explicit form of this equation and its conversion, in a specific case of
particular importance, into a differential equation will be the subject of
Sec.~\ref{sec:accufunc}.

\subsection{Average chemical potential}

\label{sec:resisfunc}

The average chemical potential $\tilde{\mu} (x)$ is determined, via expression 
(\ref{eq:thebal05pk}) for the mean spin function $\tilde{A}(x)$, by the 
solutions $\tilde{A}_{1,2}(x)$ of the integral equations (\ref{eq:thebal04ohg})
and (\ref{eq:thebal04zul}), respectively, which are conveniently solved in 
terms of ``resistance functions''.

\subsubsection{The resistance functions}

\label{sec:resf}

We begin by considering Eq.~(\ref{eq:thebal04ohg}). To solve it for the 
function $\tilde{A}_{1}(x)$, we define functions $\tilde{\mathfrak{A}}_{1}(x)$
and $\mathfrak{A}(x)$ via
\begin{equation}
\tilde{\mathfrak{A}}_{1}(x) =  \frac{v_{e} N_{c}}{J} 
\frac{B_{+}^{m}(x_{1}, x_{2})}{2}  \tilde{A}_{1}(x)
\label{eq:inteq04scp}
\end{equation}
and
\begin{equation}
\mathfrak{A}(x) =  \frac{v_{e} N_{c}}{J} \frac{B_{+}^{m}(x_{1}, x_{2})}{2} A(x)
,
\label{eq:inteq04lil}
\end{equation}
where the spin accumulation function $A(x)$ is assumed to be given.
Furthermore, we introduce the ``resistance function'' 
$\tilde{\mathfrak{R}}_{1}(x)$ [the choice of this name will be substantiated
in Sec.~\ref{sec:currvolt} below] as 
\begin{equation}
\tilde{\mathfrak{R}}_{1}(x) = \tilde{\mathfrak{A}}_{1}(x_{1}) -
\tilde{\mathfrak{A}}_{1}(x) ,
\label{eq:inteq02}
\end{equation}
along with the function
\begin{equation}
\mathfrak{R}_{1}(x) =  \smfrac{1}{2} [\mathfrak{A}(x_{1}) -
\mathfrak{A}(x)] 
\label{eq:inteq02ax}
\end{equation}
[for given $A(x)$, $\mathfrak{R}_{1}(x)$ is a given function as well].

Now, expressing the net total ballistic current $J_{+}(x', x'')$  given by
Eq.~(\ref{eq:ball004jvx}) in terms of the functions 
$\tilde{\mathfrak{R}}_{1}(x)$ and $\mathfrak{R}_{1}(x)$, we obtain from 
Eq.~(\ref{eq:thebal01}) the total thermoballistic  current  
$\mathfrak{J}_{+}^{(1)}(x_{1}, x; \xi; l)$ to be inserted in 
Eq.~(\ref{eq:thebal04ohg}) in the form 
\begin{equation}
\mathfrak{J}_{+}^{(1)}(x_{1},x;\xi;l) = \tilde{\mathfrak{J}}_{+}^{(1)}(x_{1}, 
x; \xi; l) + \mathfrak{J}_{-}^{(1)}(x_{1}, x; \xi; l) .
\label{eq:inteq04lhn}
\end{equation}
Here,
\begin{eqnarray}
\tilde{\mathfrak{J}}_{+}^{(1)}(x_{1}, x; \xi; l)  &=& J 
\left\{ \rule{0mm}{6mm} w_{+}(x_{1}, x; l) \tilde{\mathfrak{R}}_{1}(x) \right. 
\nonumber \\ 
&\ & \hspace{-2.2cm} + \int_{x_{1}}^{\xi^{-}} \frac{dx'}{l} w_{+}(x', x; l) 
[\tilde{\mathfrak{R}}_{1}(x) -\tilde{\mathfrak{R}}_{1}(x')]
\nonumber \\ &\ & \hspace{-2.2cm}  + \int_{\xi^{+}}^{x} \frac{dx''}{l} 
w_{+}(x_{1}, x''; l) [\tilde{\mathfrak{R}}_{1}(x'') - 
\tilde{\mathfrak{R}}_{1}(x_{1})]  \nonumber \\ &\ & \hspace{-2.2cm} + \left. 
\int_{x_{1}}^{\xi^{-}} \frac{dx'}{l} \int_{\xi^{+}}^{x} \frac{dx''}{l} 
w_{+}(x', x''; l) [\tilde{\mathfrak{R}}_{1}(x'') - 
\tilde{\mathfrak{R}}_{1}(x')] \right\} , \nonumber \\
\label{eq:thebal01tzg}
\end{eqnarray}
while the term $\mathfrak{J}_{-}^{(1)}(x_{1}, x; \xi; l)$ is obtained from 
expression (\ref{eq:thebal01tzg}) by replacing $\tilde{\mathfrak{R}}_{1}(x)$ 
with $\mathfrak{R}_{1}(x)$ throughout, and $w_{+}(x', x''; l)$ with 
$w_{-}(x', x''; l)$, where 
\begin{eqnarray}
w_{\pm}(x',x'';l) &=&  e^{- |x' - x''|/l} 
\frac{B^{m}_{\pm}(x',x'')}{B^{m}_{+}(x_{1},x_{2})} \nonumber \\ &\equiv& 
w_{\pm}(x'',x';l) . 
\label{eq:thebal02}
\end{eqnarray}
To evaluate condition (\ref{eq:thebal04ohg}), we use the  relation 
\begin{eqnarray}
&\ & \hspace{-1.5cm} \int_{x_{1}}^{x} d\xi \int_{x_{1}}^{\xi} dx' 
\int_{\xi}^{x} dx'' F(x', x'') \nonumber \\  &\ & = \int_{x_{1}}^{x} dx' 
\int_{x'}^{x} dx'' (x'' - x')  F(x', x'') 
\label{eq:thebal02gfc}
\end{eqnarray}
to carry out the integration over $\xi$, so that the condition can be 
expressed in the explicit form
\begin{eqnarray}
\mathfrak{K}_{+}(x_{1}, x;x;l) \tilde{\mathfrak{R}}_{1}(x) &+& \int_{x_{1}}^{x}
\frac{d x'}{l} \mathfrak{K}_{+}(x_{1}, x; x';l) 
\tilde{\mathfrak{R}}_{1}(x') \nonumber \\ &=& - \Omega_{-}^{(1)} 
(x_{1},x; l) .
\label{eq:inteq04ftx}
\end{eqnarray}
Here, the inhomogeneity $\Omega_{-}^{(1)} (x_{1}, x; l)$ is given by  
\begin{eqnarray}
\Omega_{-}^{(1)} (x_{1}, x; l) &=& \frac{x - x_{1}}{l} +
\mathfrak{K}_{-}(x_{1}, x;x;l) \mathfrak{R}_{1}(x) \nonumber \\ &\
& + \int_{x_{1}}^{x} \frac{d x'}{l} \mathfrak{K}_{-}(x_{1},x;x';l)
\mathfrak{R}_{1}(x') , \nonumber \\
\label{eq:inteq03}
\end{eqnarray}
and the integral kernels $\mathfrak{K}_{\pm}(x_{1}, x;x';l)$ are defined as
\begin{eqnarray}
\mathfrak{K}_{\pm}(x_{1}, x; x';l) &=& u_{\pm}(x', x;l) - 
u_{\pm}(x_{1}, x';l) \nonumber \\ &\ & + \int_{x_{1}}^{x} \frac{d x''}{l} 
u_{\pm}(x', x'' ;l) ,
\label{eq:inteq06opa}
\end{eqnarray}
with
\begin{eqnarray}
u_{\pm}(x',x'';l)  &=& \frac{x'' - x'}{l} w_{\pm}(x',x'';l) \nonumber \\ 
&\equiv& - u_{\pm}(x'',x';l) .
\label{eq:inteq07}
\end{eqnarray}
Equation (\ref{eq:inteq04ftx}) is a linear, inhomogeneous, Volterra-type 
integral equation of the second kind\cite{mor53} in the range $x_{1} < x 
\leq x_{2}$.

The kernels $\mathfrak{K}_{+}(x_{1}, x; x';l)$ depend
on the potential energy profiles $E_{\uparrow,
\downarrow}(x)$ solely via the quantities $B^{m}_{\pm}(x',x'')$, and are
independent of the total physical current $J$.  Nonetheless, for nonzero spin
splitting, a $J$-dependence of the resistance function
$\tilde{\mathfrak{R}}_{1}(x)$ can arise, via the function
$\mathfrak{R}_{1}(x)$, from a nonlinear $J$-dependence of the spin accumulation
function $A(x)$ [see Secs.~\ref{sec:accufunc} and \ref{sec:currden} below].  In
the zero-bias limit, when $A(x)$ is proportional to $J$, the function
$\mathfrak{R}_{1}(x)$, and hence $\tilde{\mathfrak{R}}_{1}(x)$, become
independent of $J$.  For zero spin splitting, when $B^{m}_{-}(x',x'') = 0$, we
have 
\begin{equation}
\Omega_{-}^{(1)} (x_{1}, x; l)= \frac{x - x_{1}}{l} , 
\label{eq:inteq07zgq}
\end{equation}
and $\tilde{\mathfrak{R}}_{1}(x)$ does not depend on $A(x)$.

The resistance function $\tilde{\mathfrak{R}}_{1}(x)$ is discontinuous at 
$x=x_{1}$:\ we have
\begin{equation}
\tilde{\mathfrak{R}}_{1}(x_{1}) = 0
\label{eq:inteq08}
\end{equation}
from Eq.~(\ref{eq:inteq02}), whereas we obtain
\begin{equation}
\tilde{\mathfrak{R}}_{1}(x_{1}^{+}) = 
\frac{B_{+}^{m}(x_{1},x_{2})}{B_{+}(x_{1})} - P(x_{1})
\mathfrak{R}_{1}(x_{1}^{+})
\label{eq:inteq09}
\end{equation}
by expanding Eq.~(\ref{eq:inteq04ftx}) to first order in $x - x_{1}$ (we assume
the potential energy profiles $E_{\uparrow , \downarrow}(x) $ to be continuous
in the interval $[x_{1}, x_{2}]$).

For arbitrary functions $E_{c}(x)$ and $\Delta(x)$, the calculation of the
resistance function $\tilde{\mathfrak{R}}_{1}(x)$ for a chosen parameter set 
consists of {\em three, consecutive} steps:\ (i) calculation of the spin 
accumulation function $A(x)$ as solution of the integral equation 
(\ref{eq:spiacc02}) [see Sec.~\ref{sec:accufunc} below], using the boundary 
conditions (\ref{eq:ballim05s}) with given values 
$A_{1,2}$ at the contact-semiconductor interfaces;\ (ii) with $A(x)$ as input 
in expression (\ref{eq:inteq02ax}), calculation of the inhomogeneity 
$\Omega_{-}^{(1)}(x_{1}, x; l)$ from Eq.~(\ref{eq:inteq03}); (iii) using 
$\Omega_{-}^{(1)} (x_{1}, x; l)$ in the Volterra equation 
(\ref{eq:inteq04ftx}), calculation of $\tilde{\mathfrak{R}}_{1}(x)$ by, in 
general, numerical methods.

Turning now to the function $\tilde{\mathfrak{A}}_{2}(x)$, we proceed as for
$\tilde{\mathfrak{A}}_{1}(x)$. In analogy to Eq.~(\ref{eq:inteq04ftx}) for the
resistance function $\tilde{\mathfrak{R}}_{1}(x)$, we introduce a resistance 
function $\tilde{\mathfrak{R}}_{2}(x) $ as
\begin{equation}
\tilde{\mathfrak{R}}_{2}(x) = \tilde{\mathfrak{A}}_{2}(x) -
\tilde{\mathfrak{A}}_{2}(x_{2}) ,
\label{eq:spiche02}
\end{equation}
for which we obtain from Eq.~(\ref{eq:thebal04zul}) a Volterra-type 
integral equation in the range $x_{1} \leq 
x < x_{2}$,
\begin{eqnarray}
\mathfrak{K}_{+}(x, x_{2};x;l) \tilde{\mathfrak{R}}_{2}(x) &+& \int_{x}^{x_{2}}
\frac{d x'}{l} \mathfrak{K}_{+}(x, x_{2} ; x';l)
\tilde{\mathfrak{R}}_{2}(x') \nonumber \\ &=& \Omega_{-}^{(2)} (x, x_{2}; l) ,
\label{eq:inteq04iz}
\end{eqnarray}
with the inhomogeneity $\Omega_{-}^{(2)} (x, x_{2}; l)$ expressed in terms of
the function
\begin{equation}
\mathfrak{R}_{2}(x) = \smfrac{1}{2} [\mathfrak{A}(x) - \mathfrak{A}(x_{2})] 
\label{eq:inteq02im}
\end{equation}
as
\begin{eqnarray}
\Omega_{-}^{(2)} (x, x_{2}; l) &=& \frac{x_{2} - x}{l} +
\mathfrak{K}_{-}(x, x_{2};x;l) \mathfrak{R}_{2}(x) \nonumber \\ &\
& + \int_{x}^{x_{2}} \frac{d x'}{l} \mathfrak{K}_{-}(x,x_{2};x';l)
\mathfrak{R}_{2}(x') . \nonumber \\
\label{eq:inteq03rdl}
\end{eqnarray}
The kernels $\mathfrak{K}_{\pm}(x,x_{2};x';l)$ are obtained by replacing in  
expression (\ref{eq:inteq06opa}) for $\mathfrak{K}_{\pm}(x_{1},x;x';l)$ the 
pair of arguments $x_{1},x$ with $x,x_{2}$.

The resistance function $\tilde{\mathfrak{R}}_{2}(x)$ is discontinuous
at $x =x _{2}$,
\begin{equation}
\tilde{\mathfrak{R}}_{2}(x_{2}) = 0 ,
\label{eq:spiche05}
\end{equation}
\begin{equation}
\tilde{\mathfrak{R}}_{2}(x_{2}^{-})  = 
\frac{B_{+}^{m}(x_{1},x_{2})}{B_{+}(x_{2})} - P(x_{2})
\mathfrak{R}_{2}(x_{2}^{-}) ,
\label{eq:spiche06}
\end{equation}
in analogy to the discontinuity of the function $\tilde{\mathfrak{R}}_{1}(x)$ 
at $x = x_{1}$ [see Eqs.~(\ref{eq:inteq08}) and (\ref{eq:inteq09})].

As to the calculation of $\tilde{\mathfrak{R}}_{2}(x)$ from
Eq.~(\ref{eq:inteq04iz}), the foregoing discussion regarding the calculation of
$\tilde{\mathfrak{R}}_{1}(x)$ from Eq.~(\ref{eq:inteq04ftx}) applies {\em
mutatis mutandis}.  Alternatively, one may calculate
$\tilde{\mathfrak{R}}_{2}(x)$ by using the relation
\begin{equation}
\tilde{\mathfrak{R}}_{2}(x) = \tilde{\mathfrak{R}}_{1}^{*}(x_{1} + x_{2} - x) ,
\label{eq:spiche04}
\end{equation}
where  $\tilde{\mathfrak{R}}_{1}^{*}(x)$ is the solution of
Eq.~(\ref{eq:inteq04ftx}) corresponding to spatially reversed potential energy
profiles, 
\begin{equation}
E^{*}_{\uparrow , \downarrow}(x) = E_{\uparrow , \downarrow}(x_{1} + x_{2} - 
x) ,
\label{eq:spiche04uhm}
\end{equation}
using as input a spin accumulation function
$A(x)$ calculated from Eq.~(\ref{eq:spiacc02}) below with the reversed profiles
and with the boundary values $A_{1}$ and $A_{2}$ interchanged.

The calculation of the resistance functions $\tilde{\mathfrak{R}}_{1}(x)$ and
$\tilde{\mathfrak{R}}_{2}(x)$ simplifies considerably if the spin splitting is
constant over the sample, $\Delta (x) \equiv \Delta$ [while $E_{c}(x)$ still
may be arbitrary].  With
\begin{equation}
E_{\uparrow , \downarrow}(x) = E_{c}(x) \pm \smfrac{1}{2} \Delta ,
\label{eq:66hjc}
\end{equation}
we have
\begin{equation}
E_{\uparrow , \downarrow}^{m}(x', x'') = E_{c}^{m}(x', x'') \pm \smfrac{1}{2}
\Delta ,
\label{eq:66ixy}
\end{equation}
and hence for the functions $B^{m}_{\pm}(x',x'')$ from Eqs.~(\ref{eq:ball004u})
and (\ref{eq:emtra09}), using Eqs.~(\ref{eq:eqden06})--(\ref{eq:spfra17c}),
\begin{equation}
B^{m}_{+}(x',x'')  = \frac{1}{Q} B^{m}_{0}(x',x'') 
\label{eq:66x}
\end{equation}
and
\begin{equation}
B^{m}_{-}(x',x'') = P B^{m}_{+}(x',x'') = \frac{P}{Q} B^{m}_{0}(x',x'') .
\label{eq:66y}
\end{equation}
Here, we have defined
\begin{equation}
B^{m}_{0}(x',x'')  = 2 e^{-\beta E_{c}^{m}(x', x'')} ,
\label{eq:66asa}
\end{equation}
\begin{equation}
Q \equiv (1 - P^{2})^{1/2} = \frac{1}{\cosh(\beta \Delta /2)} ,
\label{eq:66blg}
\end{equation}
and
\begin{equation}
P = - \tanh(\beta \Delta /2)
\label{eq:66zxs}
\end{equation}
is the static spin polarization.

Considering, for instance, the calculation of $\tilde{\mathfrak{R}}_{1}(x)$, we
then find from Eqs.~(\ref{eq:inteq06opa}) for the kernels
$\mathfrak{K}_{\pm}(x_{1}, x;x';l)$, using expressions (\ref{eq:66x}) and
(\ref{eq:66y}) in Eq.~(\ref{eq:thebal02}),
\begin{equation}
\mathfrak{K}_{+}(x_{1}, x;x';l) = \frac{1}{Q} \mathfrak{K}_{0}(x_{1},
x;x';l)
\label{eq:inteq06wuf}
\end{equation}
and
\begin{eqnarray}
\mathfrak{K}_{-}(x_{1}, x;x';l) &=&  P \mathfrak{K}_{+}(x_{1}, x;x';l)
\nonumber \\ &=& \frac{P}{Q} \mathfrak{K}_{0}(x_{1}, x;x';l) ,
\label{eq:inteq06waf}
\end{eqnarray}
where the reduced kernel $\mathfrak{K}_{0}(x_{1}, x;x';l)$ corresponds to 
$\Delta = 0$.           

Now, inserting expressions (\ref{eq:inteq06wuf}) and (\ref{eq:inteq06waf}) in
Eqs.~(\ref{eq:inteq04ftx}) and (\ref{eq:inteq03}), respectively, we can 
rewrite the integral equation (\ref{eq:inteq04ftx}) in the form
\begin{eqnarray}
&\ & \hspace{-0.3cm}\mathfrak{K}_{0}(x_{1}, x;x;l)
\tilde{\mathfrak{R}}_{10}(x)
+ \int_{x_{1}}^{x} \frac{d x'}{l} \mathfrak{K}_{0}(x_{1}, x; x' ;l)
\tilde{\mathfrak{R}}_{10}(x') \nonumber \\ &\ & \hspace{4.0cm} = - \frac{x -
x_{1}}{l}
\label{eq:inteq04aha}
\end{eqnarray}
$(x_{1} < x \leq x_{2})$, where
\begin{equation}
\tilde{\mathfrak{R}}_{10}(x) = \frac{1}{Q} [ \tilde{\mathfrak{R}}_{1}(x) + P
\mathfrak{R}_{1}(x)] ,
\label{eq:inteq09wu}
\end{equation}
and the function $\mathfrak{R}_{1}(x)$ is independent of $\Delta$ (see
Sec.~\ref{sec:intequat} below).  As a solution of Eq.~(\ref{eq:inteq04aha}),
the function $\tilde{\mathfrak{R}}_{10}(x)$ is a universal function that
describes an intrinsic property of the semiconducting sample;\ it is determined
by the potential energy profile $E_{c}(x)$ and the momentum relaxation length
$l$, and does not depend on the spin splitting $\Delta$ and the spin
accumulation function $A(x)$.

The resistance function $\tilde{\mathfrak{R}}_{1}(x)$ for constant spin
splitting can now be expressed as
\begin{equation}
\tilde{\mathfrak{R}}_{1}(x) = Q \tilde{\mathfrak{R}}_{10}(x) - P
\mathfrak{R}_{1}(x) .
\label{eq:inteq09wq}
\end{equation}
The calculation here separates into {\em two, independent} steps:\ (i) solution
of Eq.~(\ref{eq:inteq04aha}) for $\tilde{\mathfrak{R}}_{10}(x)$;\ (ii)
calculation of $A(x)$ as solution of the integral equation (\ref{eq:spiacc02})
and determination of $\mathfrak{R}_{1}(x)$ via Eqs.~(\ref{eq:inteq04lil}) and
(\ref{eq:inteq02ax}).  In particular, for zero spin splitting, we have 
$\tilde{\mathfrak{R}}_{1}(x) = \tilde{\mathfrak{R}}_{10}(x).$

\subsubsection{Thermoballistic average chemical potential}

\label{sec:equichem}

Introducing the mean value $\tilde{\mathfrak{A}}(x)$ of the functions 
$\tilde{\mathfrak{A}}_{1,2}(x)$, which both satisfy condition 
(\ref{eq:thebal04abc}),
\begin{equation}
\tilde{\mathfrak{A}}(x) = \smfrac{1}{2} [\tilde{\mathfrak{A}}_{1}(x) +
\tilde{\mathfrak{A}}_{2}(x)] 
\label{eq:inteq09plb}
\end{equation}
[see Eq.~(\ref{eq:thebal05pk})], we observe that $\tilde{\mathfrak{A}}(x)$ 
satisfies condition (\ref{eq:thebal04abc}) as well.  Using 
Eqs.~(\ref{eq:inteq02}) and (\ref{eq:spiche02}), we can express 
$\tilde{\mathfrak{A}}(x)$ in the form
\begin{eqnarray}
\tilde{\mathfrak{A}}(x) &=& \smfrac{1}{2} [\tilde{\mathfrak{A}}_{1}(x_{1})
+ \tilde{\mathfrak{A}}_{2}(x_{2})] - \smfrac{1}{2} [\tilde{\mathfrak{R}}_{1}(x) -
\tilde{\mathfrak{R}}_{2}(x)] . \nonumber \\
\label{eq:spiche08gs}
\end{eqnarray}
Now, writing down this expression for $x = x_{1}$ and $x = x_{2}$,
respectively, and using Eqs.~(\ref{eq:inteq04scp}), (\ref{eq:inteq08}), and
(\ref{eq:spiche05}) together with the boundary conditions 
(\ref{eq:ballim05ups}), we add the resulting two expressions to obtain for the
quantity $\eta_{12}^{+}$ defined by Eq.~(\ref{eq:ballim04}),
\begin{eqnarray}
\eta_{12}^{+} &=&  \frac{J}{v_{e} N_{c}} \frac{2}{B_{+}^{m}(x_{1}, x_{2})} 
\nonumber \\ &\ & \times \{ \tilde{\mathfrak{A}}_{1}(x_{1}) +
\tilde{\mathfrak{A}}_{2}(x_{2}) - \smfrac{1}{2} 
[\tilde{\mathfrak{R}}_{1}(x_{2}) - \tilde{\mathfrak{R}}_{2}(x_{1})] \} .
\nonumber \\
\label{eq:spiche16}
\end{eqnarray}
On the other hand, subtracting the two expressions, we have
\begin{equation}
\eta_{12}^{-} = \frac{J}{v_{e} N_{c}} \frac{2}{B_{+}^{m}(x_{1}, x_{2})} 
\tilde{\mathfrak{R}} ,
\label{eq:spiche14}
\end{equation}
with the parameter $\tilde{\mathfrak{R}}$ defined as
\begin{equation}
\tilde{\mathfrak{R}} = \smfrac{1}{2} [\tilde{\mathfrak{R}}_{1}(x_{2}) +
\tilde{\mathfrak{R}}_{2}(x_{1}) ] .
\label{eq:spiche15}
\end{equation}
Then, using Eqs.~(\ref{eq:spiche16}) and (\ref{eq:spiche14}), we can eliminate
from expression (\ref{eq:spiche08gs}) the dependence on the boundary values $
\tilde{\mathfrak{A}}_{1}(x_{1})$ and $ \tilde{\mathfrak{A}}_{2}(x_{2})$ of the
functions $\tilde{\mathfrak{A}}_{1,2}(x)$.  Introducing the function
\begin{eqnarray}
\tilde{\mathfrak{R}}_{-}(x) &=&  \tilde{\mathfrak{R}}_{1}(x) -
\smfrac{1}{2} \tilde{\mathfrak{R}}_{1}(x_{2}) - [ \tilde{\mathfrak{R}}_{2}(x) -
\smfrac{1}{2} \tilde{\mathfrak{R}}_{2}(x_{1}) ] , \nonumber \\
\label{eq:spiche18}
\end{eqnarray}
so that, in view of Eqs.~(\ref{eq:inteq08}) and (\ref{eq:spiche05}),
\begin{equation}
\tilde{\mathfrak{R}}_{-}(x_{1,2}) = \mp \tilde{\mathfrak{R}} ,
\label{eq:spiche15ug}
\end{equation}
we find for the mean spin function $\tilde{A}(x)$, using 
Eq.~(\ref{eq:inteq04scp}),
\begin{equation}
\tilde{A}(x) = \frac{ \eta^{+}_{12}}{2} -
\frac{\tilde{\mathfrak{R}}_{-}(x)}{2 \tilde{\mathfrak{R}}} \eta^{-}_{12}
\label{eq:spiche17}
\end{equation}
for $x_{1}< x < x_{2}$, and 
\begin{equation}
\tilde{A}(x_{1}) =  \eta_{1} , \; \tilde{A}(x_{2}) =  \eta_{2} .
\label{eq:spiche17kdn}
\end{equation}
Equivalently, we write
\begin{equation}
e^{\beta \tilde{\mu}(x)} = \smfrac{1}{2} (e^{\beta \mu_{1}} + e^{\beta
\mu_{2}}) - \frac{\tilde{\mathfrak{R}}_{-}(x)}{2 \tilde{\mathfrak{R}}} 
(e^{\beta \mu_{1}} - e^{\beta \mu_{2}}) 
\label{eq:spiche17zg}
\end{equation}
for $x_{1}< x < x_{2}$, and 
\begin{equation}
\tilde{\mu}(x_{1}) = \mu_{1} , \; \tilde{\mu}(x_{2}) = \mu_{2} .
\label{eq:spiche17pub}
\end{equation}
Expression (\ref{eq:spiche17zg}) represents the final, general expression for 
the (local) thermoballistic average chemical potential $\tilde{\mu}(x)$ inside 
the sample [henceforth, we drop the attribute ``thermoballistic'' when 
referring to $ \tilde{\mu}(x)$]. It is seen to be uniquely determined by the 
values of the {\em external} parameters at the contact sides of the 
contact-semiconductor interfaces, {\em viz.,} 
the values $\mu_{1,2}$ of the equilibrium chemical potential and 
the values $A_{1,2}$ of the spin accumulation function [which enter via the 
resistance functions $\tilde{\mathfrak{R}}_{1}(x)$ and 
$\tilde{\mathfrak{R}}_{2}(x)$]. Comparing expression (\ref{eq:spiche17zg}) to
expression (\ref{eq:prochem06}) for the chemical potential $\mu(x)$ in the
prototype thermoballistic model,  we observe a formally identical structure,
but differences in the explicit forms of the functions 
${\cal R}_{-}(x)$ and $\mathfrak{R}_{-}(x)$  and of the reduced resistances 
${\cal R}$ and $\mathfrak{R}$.   

An explicit expression for $\tilde{\mu}(x)$ can be obtained if $l/S \gg 1$
(``ballistic regime''), in which case the probability for an electron to 
suffer a collision when traversing the sample of length $S$, $1 - e^{-S/l}
\approx S/l$, is vanishingly small.  Then, keeping terms of order $S/l$, we
have from Eq.~(\ref{eq:inteq04ftx})
\begin{equation}
\tilde{\mathfrak{R}}_{1}(x) = \frac{B^{m}_{+}(x_{1}, x_{2})}{B^{m}_{+}(x_{1},
x)} - P^{m}(x_{1}, x) \mathfrak{R}_{1}(x) ,
\label{eq:inteq53z}
\end{equation}
and similarly from Eq.~(\ref{eq:inteq04iz})
\begin{equation}
\tilde{\mathfrak{R}}_{2}(x) = 
\frac{B^{m}_{+}(x_{1}, x_{2})}{B^{m}_{+}(x, x_{2})} - P^{m}(x, x_{2})
\mathfrak{R}_{2}(x) .
\label{eq:inteq53jh}
\end{equation}
Since from Eqs.~(\ref{eq:ballim05s}), (\ref{eq:inteq04scp}),
(\ref{eq:inteq02ax}), and (\ref{eq:inteq02im})
\begin{equation}
\mathfrak{R}_{1}(x_{2}) = \mathfrak{R}_{2}(x_{1}) = \frac{v_{e} N_{c}}{2 J}
\frac{B^{m}_{+}(x_{1}, x_{2})}{2} A_{12}^{-} ,
\label{eq:inteq53cb}
\end{equation}
we now find from Eq.~(\ref{eq:spiche15})
\begin{equation}
\tilde{\mathfrak{R}} = 1 - \frac{v_{e} N_{c}}{2J} 
\frac{B^{m}_{+}(x_{1}, x_{2})}{2} P^{m}(x_{1}, x_{2}) A_{12}^{-} .
\label{eq:inteq53pl}
\end{equation}
For the function $\tilde{\mathfrak{R}}_{-}(x)$, we have from
Eq.~(\ref{eq:spiche18}), using Eq.~(\ref{eq:inteq53cb}),
\begin{equation}
\tilde{\mathfrak{R}}_{-}(x) =  \tilde{\mathfrak{R}}_{1}(x) -
\tilde{\mathfrak{R}}_{2}(x) .
\label{eq:spiche18fv}
\end{equation}
With expressions (\ref{eq:inteq53pl}) and (\ref{eq:spiche18fv}) inserted in
Eq.~(\ref{eq:spiche17zg}), we obtain the average chemical potential
$\tilde{\mu}(x)$ in explicit form. In the particular
case of zero spin splitting, when $B^{m}_{+}(x', x'') = 2 e^{-\beta
E_{c}^{m}(x', x'')}$ and $P^{m}(x', x'') = 0$, we have
\begin{equation}
\frac{\tilde{\mathfrak{R}}_{-}(x)}{\tilde{\mathfrak{R}}} = 
e^{-\beta E_{c}^{m}(x_{1}, x_{2})}[e^{\beta E_{c}^{m}(x_{1}, x)} -
e^{\beta E_{c}^{m}(x, x_{2})}] .
\label{eq:spiche17ld}
\end{equation}
This result, here obtained in the ballistic regime, $l/S \gg 1$, contrasts with
the (strict) ballistic limit, $l \rightarrow \infty$, considered in 
Sec.~\ref{sec:ballim},
when there is absolutely no point of local thermodynamic equilibrium inside
the sample.  In this extreme case, no meaning can be attached to an average
chemical potential, and there is nothing a calculation of such a quantity can
be based upon.

In Ref.~\onlinecite{lip05}, we have presented and discussed numerical results
for $\tilde{\mu}(x)$, calculated with the potential energy profile 
(\ref{eq:difeq01}) for values of the ratio $l/S$ ranging between $10^{-2}$ to
$10^{2}$. If calculated with the procedure adopted in the present work (see
the remarks at the end of Sec.~\ref{sec:detmsf}), the results for  
$\tilde{\mu}(x)$ would differ quantitatively from the former ones, but would 
agree with those in all qualitative respects.

\subsubsection{Sharvin interface resistance}

\label{sec:sharvin}

The discontinuities in the resistance functions $\tilde{\mathfrak{R}}_{1}(x)$ 
and $\tilde{\mathfrak{R}}_{2}(x)$ at the points of local thermodynamic 
equilibrium $x = x_{1}$ and $x = x_{2}$, respectively, give rise to 
discontinuities in the function $\tilde{\mathfrak{R}}_{-}(x)$ both at $x = 
x_{1}$ and $x = x_{2}$.  From Eq.~(\ref{eq:spiche18}), we find, using 
Eqs.~(\ref{eq:inteq08}) and (\ref{eq:inteq09}),
\begin{eqnarray}
\tilde{\mathfrak{R}}_{-}(x_{1}^{+}) - \tilde{\mathfrak{R}}_{-}(x_{1}) &=& 
\tilde{\mathfrak{R}}_{1}(x_{1}^{+}) \nonumber \\ &=& 
\frac{B_{+}^{m}(x_{1}, x_{2})}{B_{+}(x_{1})} -
P(x_{1}) \mathfrak{R}_{1}(x_{1}^{+}) , \nonumber \\
\label{eq:sharv01}
\end{eqnarray}
and similarly, using Eqs.~(\ref{eq:spiche05}) and (\ref{eq:spiche06}),
\begin{eqnarray}
\tilde{\mathfrak{R}}_{-}(x_{2}) - \tilde{\mathfrak{R}}_{-}(x_{2}^{-}) &=& 
\tilde{\mathfrak{R}}_{2}(x_{2}^{-}) \nonumber \\ &=& 
\frac{B_{+}^{m}(x_{1}, x_{2})}{B_{+}(x_{2})} -
P(x_{2}) \mathfrak{R}_{2}(x_{2}^{-}) . \nonumber \\
\label{eq:sharv02}
\end{eqnarray}
Therefore, according to Eq.~(\ref{eq:spiche17zg}), the average chemical
potential $\tilde{\mu}(x)$ exhibits discontinuities at these points
as well (``Sharvin effect'';\ see Ref.~\onlinecite{sha65}),
\begin{eqnarray}
e^{\beta [\tilde{\mu}(x_{1}^{+}) - \mu_{1}]} - 1 &=& - \frac{\eta_{12}^{-} e^{-
\beta \mu_{1}} }{2} \frac{\tilde{\mathfrak{R}}_{1}
(x_{1}^{+})}{\tilde{\mathfrak{R}}} \nonumber \\ &=& - \smfrac{1}{2} \beta e^{2}
 J \tilde{\rho}_{1} , 
\label{eq:sharv03}
\end{eqnarray}
\begin{eqnarray}
e^{\beta [\tilde{\mu}(x_{2}^{-}) - \mu_{2}]} - 1 &=& \frac{\eta_{12}^{-} e^{-
\beta \mu_{2}} }{2} \frac{\tilde{\mathfrak{R}}_{2}
(x_{2}^{-})}{\tilde{\mathfrak{R}}} \nonumber \\ &=&  \smfrac{1}{2} \beta e^{2}
J \tilde{\rho}_{2} , 
\label{eq:sharv04}
\end{eqnarray}
where we have used Eqs.~(\ref{eq:spiche14}) and (\ref{eq:spiche15}) to
introduce the total physical current $J$. The quantities
$\tilde{\rho}_{1,2}$ are the {\em Shar\-vin interface
resistances},\cite{sha65,lip05}
\begin{equation}
\tilde{\rho}_{1} = \frac{2}{\beta e^{2} v_{e} N_{c} B_{+}^{m}(x_{1}, x_{2}) 
e^{\beta \mu_{1}}} \tilde{\mathfrak{R}}_{1}(x_{1}^{+}) ,
\label{eq:sharv05}
\end{equation}
\begin{equation}
\tilde{\rho}_{2} = \frac{2}{\beta e^{2} v_{e} N_{c } B_{+}^{m}(x_{1}, x_{2}) 
e^{\beta \mu_{2}}}\tilde{\mathfrak{R}}_{2}(x_{2}^{-}) .
\label{eq:sharv05nh}
\end{equation}
For zero spin splitting, when $P(x_{1,2}) = 0$, so that
$\tilde{\mathfrak{R}}_{1}(x_{1}^{+}) = B_{+}^{m}(x_{1}, x_{2}) /B_{+}(x_{1})$ 
and $\tilde{\mathfrak{R}}_{2}(x_{2}^{-}) =  B_{+}^{m}(x_{1}, x_{2}) / 
B_{+}(x_{2})$, we can express
$\tilde{\rho}_{1,2}$ in the form
\begin{equation}
\tilde{\rho}_{1,2} = \frac{1}{\beta e^{2} v_{e} n_{+}^{(0)}(x_{1,2})} ,
\label{eq:sharv05kl}
\end{equation}
where
\begin{eqnarray}
n_{+}^{(0)}(x_{1,2}) &=&  \frac{N_{c}}{2} B_{+}(x_{1,2}) e^{\beta \mu_{1,2}}
\nonumber \\ &=& N_{c} e^{- \beta [E_{c}(x_{1,2}) - \mu_{1,2}]}
\label{eq:sharv05tf}
\end{eqnarray}
are the total equilibrium electron densities at the semiconductor side of the
contact-semi\-cond\-uctor interfaces [see Eqs.~(\ref{eq:ball004jux}) and
(\ref{eq:ball004jum})]. Equations (\ref{eq:sharv03}) and 
(\ref{eq:sharv04}), respectively, are analogous to 
Eqs.~(\ref{eq:prochem06amf}) and (\ref{eq:prochem06pky}) for the 
discontinuities of the chemical potential $\mu (x)$ in the prototype
thermoballistic model.

\subsubsection{Current-voltage characteristic and magnetoresistance}

\label{sec:currvolt}

The {\em current-voltage characteristic} of the thermoballistic transport model
is obtained from Eq.~(\ref{eq:spiche14}) in the form
\begin{equation}
J = \frac{v_{e} N_{c}}{2} B_{+}^{m}(x_{1}, x_{2}) e^{\beta \mu_{1}} 
\frac{1}{\tilde{\mathfrak{R}}} ( 1 - e^{- \beta e V} ) ,
\label{eq:sharv06}
\end{equation}
where
\begin{equation}
V = \frac{\mu_{1} - \mu_{2}}{e}
\label{eq:ballim09}
\end{equation}
is the voltage bias between the metal contacts (we assume the voltage drop
across the contacts to be negligibly small).  In the zero-bias limit, when
$\beta e V \ll 1$, we have
\begin{eqnarray}
R &\equiv& \left.  \frac{V}{eJ} \right|_{J \rightarrow 0} = \frac{2}{\beta 
e^{2} v_{e} N_{c} B_{+}^{m}(x_{1}, x_{2}) e^{\beta \mu_{1}}} 
\tilde{\mathfrak{R}} \nonumber \\ &=& \tilde{\rho}_{1} 
\frac{\tilde{\mathfrak{R}}}{{\tilde{\mathfrak{R}}_{1}(x_{1}^{+})}} 
\label{eq:sharv07tf}
\end{eqnarray}
for the resistance times cross-sectional area of the sample. Comparing the 
current-voltage characteristic (\ref{eq:sharv06}) to that of the (spinless) 
prototype thermoballistic model, Eq.~(\ref{eq:gendr004a}), one sees 
that the (dimensionless) parameter $\tilde{\mathfrak{R}}$ directly corresponds
to the reduced resistance ${\cal R}$ of the prototype model, 
Eq.~(\ref{eq:gendr003}). Therefore, $\tilde{\mathfrak{R}}$ is here also called 
the {\em reduced resistance} of the sample, and the functions 
$\tilde{\mathfrak{R}}_{1}(x)$ and $\tilde{\mathfrak{R}}_{2}(x)$ from
which it is derived, the {\em resistance functions}.

The reduced resistance $\tilde{\mathfrak{R}}$ is a central element of the
thermoballistic description of electron transport in semiconductors.  It
comprises the effect of the detailed shape of the potential energy profiles as
well as that of the momentum relaxation length.  Moreover, via its
dependence on the spin accumulation function $A(x)$, it takes into account the
effect of spin relaxation.  As exemplified by expression
(\ref{eq:inteq53pl}) for $\tilde{\mathfrak{R}}$, which holds in the ballistic
regime, the reduced resistance depends on the total current $J$ unless $A(x)$
is proportional to $J$.  The explicit form of $\tilde{\mathfrak{R}}$ in the
drift-diffusion regime can be obtained by rewriting Eq.~(\ref{eq:ballim20kf})
for $x = x_{2}$ in the form of a current-voltage characteristic and comparing 
it to the characteristic (\ref{eq:sharv06}).

Defining now the {\em relative magnetoresistance} $R_{m}$ of the semiconducting
sample\cite{zim65,rot94} as
\begin{equation}
R_{m} \equiv \frac{R - R_{0}}{R_{0}} ,
\label{eq:sharv07yg}
\end{equation}
where $R_{0}$ is the resistance at zero external magnetic field, i.e., at zero
spin splitting, we have from Eq.~(\ref{eq:sharv07tf})
\begin{equation}
R_{m} = \frac{\tilde{\mathfrak{R}} -
\tilde{\mathfrak{R}}_{0}}{\tilde{\mathfrak{R}}_{0}} .
\label{eq:sharv07xbs}
\end{equation}
The reduced resistance at zero spin splitting, $\tilde{\mathfrak{R}}_{0}$, is
obtained by solving Eq.~(\ref{eq:inteq04aha}) for the function
$\tilde{\mathfrak{R}}_{10}(x)$ and the analogous equation for
$\tilde{\mathfrak{R}}_{20}(x)$ and using Eq.~(\ref{eq:spiche15}),
\begin{equation}
\tilde{\mathfrak{R}}_{0} = \smfrac{1}{2} [\tilde{\mathfrak{R}}_{10}(x_{2})
+ \tilde{\mathfrak{R}}_{20}(x_{1}) ] .
\label{eq:spiche15axa}
\end{equation}
For {\em constant} splitting, we find from Eq.~(\ref{eq:spiche15}), using
Eq.~(\ref{eq:inteq09wq}) and the analogous equation for
$\tilde{\mathfrak{R}}_{2}(x)$ as well as Eq.~(\ref{eq:inteq53cb}),
\begin{equation}
\tilde{\mathfrak{R}} = Q \tilde{\mathfrak{R}}_{0} - \frac{v_{e} N_{c}}{4 J}
B_{+}^{m}(x_{1}, x_{2}) P A^{-}_{12} ,
\label{eq:spiche15lll}
\end{equation}
and hence
\begin{equation}
R_{m} = Q  - 1  - \frac{v_{e} N_{c}}{4 J \tilde{\mathfrak{R}}_{0}}
 B_{+}^{m}(x_{1}, x_{2}) P A^{-}_{12} 
\label{eq:sharv07lz}
\end{equation}
for the magnetoresistance.

\subsection{Spin accumulation function}

\label{sec:accufunc}

The net relaxing ballistic spin-polarized current $\check{J}_{-}(x',x'';x)$
[see Eq.~(\ref{eq:netbal02})] and the joint relaxing ballistic spin-polarized
density $\check{n}_{-}(x',x'';x)$ [see Eq.~(\ref{eq:netbal02xy})], and thus
also the corresponding thermoballistic current $\check{\mathfrak{J}}_{-}(x)$
and density $\check{\mathfrak{n}}_{-}(x)$ evaluated from
Eq.~(\ref{eq:thebal01}), are dynamically determined solely by the spin
accumulation function $A(x)$. The calculation of this function is, in
general, prerequisite to a complete determination of the average chemical
potential $\tilde{\mu} (x)$ [see Sec.~\ref{sec:resisfunc}].

\subsubsection{Integral equation}

\label{sec:intequat}

Following Sec.~\ref{sec:thermcurr}, we invoke condition (\ref{eq:thebal06rt}),
which is equivalent to the spin balance equation (\ref{eq:ballim24new}), and
insert expression (\ref{eq:netbal02}) for the quantity $F(x', x''; x)$ in 
Eq.~(\ref{eq:thebal06jb}) to obtain for the spin accumulation function $A(x)$ 
an integral equation of the form
\begin{eqnarray}
\check{\cal W}(x_{1},x;l,l_{s}) A_{1} &+& \check{\cal W}(x,x_{2};l,l_{s}) A_{2}
\nonumber \\ &-& \check{W}(x;x_{1},x_{2};l) A(x) \nonumber \\ &+&
\int_{x_{1}}^{x_{2}} \frac{d x'}{l} \check{\cal W}(x',x;l,l_{s}) A(x') =
0 , \nonumber \\
\label{eq:spiacc02}
\end{eqnarray}
where
\begin{equation}
\check{\cal W}(x', x'';l,l_{s}) = \check{w}(x', x'';l) e^{- {\cal C}^{m}
(x',x'')/l_{s}}
\label{eq:spiacc03}
\end{equation}
and
\begin{eqnarray}
\check{W}(x;x_{1},x_{2};l) &=& \check{w}(x_{1},x;l) + \check{w}(x,x_{2};l)
\nonumber \\  &\ & + \int_{x_{1}}^{x_{2}} \frac{d x'}{l} \check{w}(x',x;l) .
\label{eq:spiacc04}
\end{eqnarray}
Here, the values $A(x_{1,2})$ of the spin accumulation function at the
interface positions $x_{1,2}$ have been set equal to the external
values, $A_{1,2}$, at the contact side of the contact-semiconductor 
interfaces [see Eq.~(\ref{eq:ballim05s})]. The function 
${\cal C}^{m} (x', x'')$ is defined as
\begin{equation}
{\cal C}^{m} (x', x'') \equiv {\cal C}^{m}(x',x'';x',x'')
\label{eq:spiacc06}
\end{equation}
[see Eq.~(\ref{eq:baleq06})], and
\begin{eqnarray}
\check{w}(x',x'';l) &=&  e^{- |x' - x''|/l} \frac{B^{m}(x',x'')}{B^{m}_{+} 
(x_{1},x_{2})}  \nonumber \\ &\equiv& \check{w}(x'',x';l) , 
\label{eq:thebal03}
\end{eqnarray}
with $B^{m}(x',x'')$ given by Eq.~(\ref{eq:spirel15}).

With the first two terms in Eq.~(\ref{eq:spiacc02}) acting as an inhomogeneity,
this equation is a linear, inhomogeneous, Fredholm-type integral equation of
the second kind\cite{mor53} for the spin accumulation function $A(x)$.  The
corresponding homogeneous equation is solved by $A(x) \equiv 0$ only, so that
the solution $A(x)$ of Eq.~(\ref{eq:spiacc02}) for $x_{1} < x < x_{2}$ is
linear and homogeneous in $A_{1}$ and $A_{2}$.  Just like the average chemical
potential $\tilde{\mu}(x)$, the spin accumulation function $A(x)$ exhibits the
Sharvin effect, i.e., it is not, in general, continuous at the interfaces with
the contacts, $A(x_{1}^{+}) - A_{1} \neq 0, A_{2} - A(x_{2}^{-}) \neq 0$.  This
will be demonstrated in Sec.~\ref{sec:diffequa} below by way of a specific
example.  The solution of Eq.~(\ref{eq:spiacc02}) is found, in general,
numerically by applying matrix methods after discretization.

For constant spin splitting, $\Delta(x) \equiv \Delta$, when the functions
$E^{m}_{\uparrow, \downarrow} (x', x'')$ are given by Eq.~(\ref{eq:66ixy}), the
function $B^{m}(x',x'')$ defined by Eq.~(\ref{eq:spirel15}) reduces to
\begin{equation}
B^{m}(x',x'') = Q B^{m}_{0}(x',x'') ,
\label{eq:thebal03ihi}
\end{equation}
with $B^{m}_{0}(x',x'')$ given by Eq.~(\ref{eq:66asa}), and $Q$ by
Eq.~(\ref{eq:66blg}). Further, using Eqs.~(\ref{eq:66hjc}) and (\ref{eq:66ixy})
in expression (\ref{eq:ball004n}), we see that $\Delta$ cancels out in the
functions $C^{m}_{\uparrow, \downarrow} (x', x''; x)$,
\begin{equation}
C^{m}_{\uparrow, \downarrow} (x', x''; x) \equiv C^{m}_{0} (x', x''; x) .
\label{eq:thebal03olw}
\end{equation}
Inserting this in Eq.~(\ref{eq:ball004p}), we then find, via
Eqs.~(\ref{eq:balden15}), (\ref{eq:baleq03}), and (\ref{eq:baleq06}), the
function ${\cal C}^{m} (x', x'')$ defined by Eq.~(\ref{eq:spiacc06}) to be
independent of $\Delta$,
\begin{equation}
{\cal C}^{m}(x',x'') \equiv {\cal C}^{m}_{0}(x',x'') .
\label{eq:thebal03qfw}
\end{equation}
With this result, in conjunction with Eqs.~(\ref{eq:thebal03}) and
(\ref{eq:thebal03ihi}), used in Eqs.~(\ref{eq:spiacc03}) and
(\ref{eq:spiacc04}), the spin splitting $\Delta$ is seen to drop out from the
integral equation (\ref{eq:spiacc02}), i.e., its solution $A(x)$ does not
depend on $\Delta$.

\subsubsection{Differential equation}

\label{sec:diffequa}

In {\em homogeneous} semiconductors without space charge, the conduction band 
edge potential is constant.  In this case, the integral equation for $A(x)$ can
be converted into a differential equation.  Assuming the electrons to be driven
by an external electric field of magnitude ${\cal E}$ directed antiparallel to 
the $x$-axis, the
potential energy profile (\ref{eq:eqden01}) for zero spin splitting, $\Delta
(x) = 0$, is given by Eq.~(\ref{eq:difeq01}). [According to the above 
discussion, the inclusion of a nonzero, constant spin splitting would not alter
the results.]  For the profile (\ref{eq:difeq01}), the function 
$C^{m}(x',x'';x)$ defined by Eq.~(\ref{eq:baleq03}) takes the form
\begin{eqnarray}
C^{m}(x',x'';x) &\equiv& C^{m} \bm{(}\epsilon (x-x') \bm{)} \nonumber \\ &=&
e^{\epsilon (x - x')} {\rm erfc}\bm{(}[\epsilon (x - x')]^{1/2}\bm{)} 
\nonumber \\
\label{eq:difeq02}
\end{eqnarray}
$(\epsilon = \beta e {\cal E})$, as can be shown by using 
Eqs.~(\ref{eq:ball004p}), (\ref{eq:ball004n}), and (\ref{eq:balden15}).

Using expression (\ref{eq:difeq02}), via Eq.~(\ref{eq:baleq06}), in
Eq.~(\ref{eq:spiacc06}), the integral equation (\ref{eq:spiacc02}) can now be 
reduced to
\begin{eqnarray}
f_{1}(x-x_{1}) A_{1} &+& f_{2}(x_{2}-x) A_{2} - f(x-x_{1}) A(x)
\nonumber \\ &+& \int_{x_{1}}^{x} \frac{dx'}{l} f_{1}(x-x') A(x')
\nonumber \\ &+& \int_{x}^{x_{2}} \frac{dx'}{l} f_{2}(x'- x) A(x') = 0 ,
\nonumber \\
\label{eq:difeq04}
\end{eqnarray}
where
\begin{equation}
f_{1}(x) =  e^{-[\epsilon + 1/l + \mathfrak{c} (\epsilon x)/l_{s}] x} ,
\label{eq:difeq05}
\end{equation}
\begin{equation}
f_{2}(x) =  e^{-[1/l + \mathfrak{c} (\epsilon x)/l_{s}] x} ,
\label{eq:difeq06}
\end{equation}
and
\begin{equation}
f(x) = \frac{1}{1 + \epsilon l} \{ 2 + \epsilon l [ 1 + e^{-(\epsilon + 1/l)x}
] \} .
\label{eq:difeq07}
\end{equation}
The function $\mathfrak{c}(\zeta)$ defined by
\begin{equation}
\mathfrak{c}(\zeta) = \frac{1}{\zeta} \int_{0}^{\zeta} d \zeta' C^{m}(\zeta')
\label{eq:difeq08}
\end{equation}
obeys the relations $0 < \mathfrak{c}(\zeta) \leq 1$,
$\mathfrak{c}(\zeta) \rightarrow1 $ for $\zeta \rightarrow 0$, and
$\mathfrak{c}(\zeta) \sim 2 (\pi \zeta)^{-1/2}$ for $\zeta \rightarrow \infty$.

We can convert the inhomogeneous integral equation (\ref{eq:difeq04}) into a 
homogeneous integrodifferential equation for $A(x)$ by supplementing 
Eq.~(\ref{eq:difeq04}) with the equations obtained by forming its first and 
second derivative with respect to $x$,  and subsequently eliminating from this 
set of three equations the boundary values $A_{1}$ and $A_{2}$. [In principle, 
a similar procedure could also be applied to the general equation 
(\ref{eq:spiacc02}), but this does not seem to lead to any advantage.]
If we replace in the integral
equation (\ref{eq:difeq04}) the function $\mathfrak{c}(\zeta)$ with a 
position-independent average value $\bar{\mathfrak{c}}$, the functions 
$f_{1}(x)$ and $f_{2}(x)$ become pure exponentials, and the corresponding
integrodifferential equation reduces to a homogeneous second-order 
differential equation of the form
\begin{equation}
b_{0}(x) \frac{d^{2} A(x)}{dx^{2}} + b_{1}(x) \frac{d A(x)}{dx} + b_{2}(x) A(x)
= 0 .
\label{eq:difeq09}
\end{equation}
Here,
\begin{equation}
b_{0}(x) = 2 + \epsilon l [ 1 + b(x)] ,
\label{eq:difeq10}
\end{equation}
\begin{equation}
b_{1}(x) =  \epsilon (2 + \epsilon l) [ 1 - b(x)] ,
\label{eq:difeq11}
\end{equation}
and
\begin{equation}
b_{2}(x) = \frac{\tilde{l} - l}{l \tilde{l}^{2}} \{ 2 + \epsilon (l + \tilde{l}
+ \epsilon l \tilde{l}) [1 + b(x)] \} ,
\label{eq:difeq12}
\end{equation}
where
\begin{equation}
b(x) = e^{-(\epsilon + 1/l)(x-x_{1})}
\label{eq:difeq13}
\end{equation}
and
\begin{equation}
\frac{1}{\tilde{l}} = \frac{1}{l} + \frac{\bar{\mathfrak{c}}}{l_{s}} ,
\label{eq:difeq14a}
\end{equation}
with the ballistic spin relaxation length $l_{s}$ given by
Eq.~(\ref{eq:spirel01}). [Note that  expression 
(\ref{eq:difeq12}) for the function $b_{2}(x)$ differs from the corresponding 
expression (3.48) of Ref.~\onlinecite{lip05}, which was derived by introducing
the average value $\bar{\mathfrak{c}}$ in the integrodifferential equation, 
rather than in the original integral equation.] Since, owing to the presence of
the factor $e^{-x/l}$ in the functions
$f_{1}(x)$ and $f_{2}(x)$, only the values of $\mathfrak{c}(\epsilon x)$ in
the range $0 \leq x \alt l$ contribute appreciably, we may choose
$\bar{\mathfrak{c}}$ as the average of $\mathfrak{c}(\epsilon x)$ over an
$x$-interval of length equal to the momentum relaxation length $l$,
\begin{equation}
\bar{\mathfrak{c}}  = \frac{1}{l} \int_{0}^{l} dx \mathfrak{c}(\epsilon
x) \equiv  \frac{1}{\epsilon l} \int_{0}^{\epsilon l} d \zeta  \ln ( \epsilon
l/\zeta) C^{m}(\zeta) .
\label{eq:difeq15}
\end{equation}
In the right-hand integral in this equation, the range of small $\zeta$,
when $C^{m}(\zeta) \approx 1$, is emphasized because of the weight factor
$ \ln ( \epsilon l/\zeta)$.  

For large $\epsilon l$ (i.e., in the ballistic regime
and/or for strong fields), the variation of $\mathfrak{c}(\epsilon x)$ with $x$
becomes essential.  However, for the purpose of demonstrating the principal
effects of the transport mechanism, the approximation in terms of a constant
average value $\bar{\mathfrak{c}}$ of the function $\mathfrak{c} (\zeta)$, in
conjunction with the choice (\ref{eq:difeq15}) for $\bar{\mathfrak{c}}$,
appears to be sufficiently accurate.

In the drift-diffusion regime, when $l/S \ll 1$, $l/l_{s} \ll 1$, and $\epsilon
l \ll 1$, we have $\bar{\mathfrak{c}} = 1$ and hence $\tilde{l} = \bar{l}$, 
where
\begin{equation}
\frac{1}{\bar{l}} = \frac{1}{l} + \frac{1}{l_{s}} .
\label{eq:difeq14c}
\end{equation}
Equation (\ref{eq:difeq09}) then reduces to
\begin{equation} \frac{d^{2} A(x)}{dx^{2}} + \epsilon \frac{d A(x)}{dx} -
\frac{1}{L_{s}^{2}} A(x) = 0 ,
\label{eq:difeq16}
\end{equation}
where $L_{s}$ is the spin diffusion length given by Eq.~(\ref{eq:ballim26}).
Setting $\hat{B}_{+}(x) = - \epsilon /\beta$ in Eq.~(\ref{eq:ballim26xx}), we
find Eq.~(\ref{eq:difeq16}) to agree with the former equation, which was
obtained by directly evaluating the thermoballistic current
$\check{\mathfrak{J}}_{-}(x)$ and density $\check{\mathfrak{n}}_{-}(x)$ in the
drift-diffusion limit, and which leads to the standard form (\ref{eq:ballim25})
of the drift-diffusion equation for spin-dependent transport.

In the zero-bias limit $\epsilon \rightarrow 0$, the integral equation
(\ref{eq:difeq04}) reduces to
\begin{eqnarray}
e^{-(x-x_{1})/\bar{l}} A_{1} &+& e^{-(x_{2}-x)/\bar{l}} A_{2} - 2 A(x)
\nonumber \\ &+& \int_{x_{1}}^{x_{2}} \frac{dx'}{l} e^{-|x- x'|/\bar{l}} A(x')
= 0 , \nonumber \\
\label{eq:difeq17}
\end{eqnarray}
from which we obtain the differential equation
\begin{equation}
\frac{d^{2}A(x)}{dx^{2}} - \frac{1}{L^{2}} A(x) = 0 .
\label{eq:difeq18}
\end{equation}
Here,
\begin{equation}
L = \sqrt{\bar{l} l_{s}} = \frac{L_{s}}{\sqrt{1 + l/l_{s}}}
\label{eq:difeq19}
\end{equation}
is the generalization of the spin diffusion length $L_{s}$ given by
Eq.~(\ref{eq:ballim26}).  The length $L$ becomes equal to the latter length,
$L = L_{s}$, in the drift-diffusion regime, when $l/l_{s} \ll 1$, and to the 
ballistic spin relaxation length, $L = l_{s}$, in the ballistic case, when 
$l/l_{s} \rightarrow \infty$ and hence $\bar{l} = l_{s}$.

For $x_{1}< x < x_{2}$, the general solution of Eq.~(\ref{eq:difeq18}) reads
\begin{equation}
A(x) = C_{1} e^{-(x-x_{1})/L} + C_{2} e^{-(x_{2}-x)/L} .
\label{eq:difeq20}
\end{equation}
Inserting this expression for $A(x)$ in Eq.~(\ref{eq:difeq17}) and setting  
$x = x_{1}$ and $x = x_{2}$, we obtain, using
Eq.~(\ref{eq:ballim05s}), two linear equations for the coefficients, with the 
solution
\begin{equation}
C_{1} = \frac{1}{D} [ (1+ \gamma) e^{S/L} A_{1} - (1- \gamma) A_{2} ] ,
\label{eq:difeq21}
\end{equation}
\begin{equation}
C_{2} = - \frac{1}{D} [ (1- \gamma) A_{1}- (1+ \gamma) e^{S/L} A_{2} ] .
\label{eq:difeq22}
\end{equation}
Here,
\begin{equation}
D = 2 [( 1+ \gamma^{2}) \sinh(S/L) + 2 \gamma \cosh(S/L)] ,
\label{eq:difeq23}
\end{equation}
with
\begin{equation}
\gamma = \frac{L}{l_{s}} = \frac{\bar{l}}{L} = \sqrt{\frac{l}{l+l_{s}}} \leq
1 .
\label{eq:difeq24}
\end{equation}
It then follows from Eqs.\ (\ref{eq:difeq20})--(\ref{eq:difeq24}) that $A(x)$
is discontinuous at $x = x_{1}$ and $x = x_{2}$,
\begin{equation}
\Delta A_{1} \equiv A(x_{1}^{+}) - A_{1} = - \smfrac{1}{2} (g A_{1} - h A_{2}) ,
\label{eq:difeq25}
\end{equation}
\begin{equation}
\Delta A_{2} \equiv A_{2} - A(x_{2}^{-}) = - \smfrac{1}{2} (h A_{1} - g A_{2}) ,
\label{eq:difeq26}
\end{equation}
where
\begin{equation}
g = h [ \cosh(S/L) + \gamma \sinh(S/L) ] \leq 1 ,
\label{eq:difeq27}
\end{equation}
with
\begin{equation}
h = \frac{4 \gamma}{D} \leq \frac{2}{1 + \gamma} e^{-S/L} .
\label{eq:difeq28}
\end{equation}
In the drift-diffusion regime, when $L = L_{s}$ and $\gamma \ll 1$, we have
\begin{equation}
A(x) = A_{1} e^{- (x-x_{1})/L_{s}} +  A_{2} e^{-(x_{2}-x)/L_{s}} ,
\label{eq:difeq29}
\end{equation}
and in the ballistic case, when $L = l_{s}$ and $\gamma = 1$,
\begin{equation}
A(x) = \smfrac{1}{2} [ A_{1} e^{-(x-x_{1})/l_{s}} + A_{2} e^{-(x_{2}-x)/l_{s}}] .
\label{eq:difeq30}
\end{equation}
At $x = x_{1}$, for example, the discontinuity of $A(x)$ is $\Delta A_{1}
= A_{2} \exp(-S/L_{s})$ in the drift-diffusion regime, and $\Delta A_{1} =
\smfrac{1}{2}[ - A_{1} + A_{2} \exp(-S/l_{s})]$ in the ballistic case.

\subsection{Current and density spin polarizations}

\label{sec:currden}

The position dependence of the current and density spin polarizations as well
as the magnetoresistance are the physical quantities of principal interest in
the study of spin-polarized electron transport in paramagnetic semiconducting
systems.

\subsubsection{Spin polarizations in the semiconductor}

In the thermoballistic approach, we define the {\em persistent} current
spin polarization in the semiconducting sample, $\overstar{P}_{J}(x)$, in terms
of the persistent thermoballistic spin-polarized current
$\overstar{\mathfrak{J}}_{-}(x)$ and the total physical current $J$ as
\begin{equation}
\overstar{P}_{J}(x) =
\frac{\overstar{\mathfrak{J}}_{-}(x)}{J}
\label{eq:65}
\end{equation}
$(x_{1} \leq x \leq x_{2})$.  Analogously, the {\em relaxing} current spin
polarization $\check{P}_{J}(x)$ is defined in terms of the relaxing
thermoballistic spin-polarized current $\check{\mathfrak{J}}_{-}(x)$ and the
total current $J$ as
\begin{equation}
\check{P}_{J}(x) = \frac{\check{\mathfrak{J}}_{-}(x)}{J} .
\label{eq:65ax}
\end{equation}
For the {\em total} current spin polarization $P_{J}(x)$, we then have
\begin{equation}
P_{J}(x) \equiv \overstar{P}_{J}(x) + \check{P}_{J}(x) =
\frac{\overstar{\mathfrak{J}}_{-}(x) + \check{\mathfrak{J}}_{-}(x)}{J} .
\label{eq:91xx}
\end{equation}
[Note that in Refs.~\onlinecite{lip05} and \onlinecite{lip09}, we have defined 
the current spin polarization in terms of the total {\em thermoballistic} 
current $\mathfrak{J}_{+}(x)$, rather than in terms of the total {\em physical}
current $J$.]

The {\em persistent} density spin polarization $\overstar{P}_{n}(x)$ and the 
{\em relaxing} density spin polarization $\check{P}_{n}(x)$ are introduced by 
replacing in expressions (\ref{eq:65}) and (\ref{eq:65ax}) the
thermoballistic currents $\overstar{\mathfrak{J}}_{-}(x)$ and
$\check{\mathfrak{J}}_{-}(x)$ with the respective densities
$\overstar{\mathfrak{n}}_{-}(x)$ and $\check{\mathfrak{n}}_{-}(x)$, and 
the current $J$ in the denominator of those expressions with the 
thermoballistic joint total density $\mathfrak{n}_{+}(x)$. Hence, we obtain  
\begin{equation}
P_{n}(x) \equiv \overstar{P}_{n}(x) + \check{P}_{n}(x) =
\frac{\overstar{\mathfrak{n}}_{-}(x) + 
\check{\mathfrak{n}}_{-}(x)}{\mathfrak{n}_{+}(x)} 
\label{eq:91olsb}
\end{equation}
for the {\em total} density spin polarization $P_{n}(x)$. 

In expression (\ref{eq:91xx}) for the total current spin polarization inside
the semiconductor, the current $\overstar{\mathfrak{J}}_{-}(x)$ depends on the
reduced spin accumulation function $\breve{A}(x)$, and hence linearly on the
boundary values $\breve{A}_{1,2}$, via the dependence of the total ballistic
current $J_{+}(x',x'')$ on $\breve{A}(x)$,
\begin{equation}
\overstar{\mathfrak{J}}_{-}(x) \equiv \overstar{\mathfrak{J}}_{-}(x;
\breve{A}_{1}, \breve{A}_{2})
\label{eq:91zack}
\end{equation}
[see Eqs.~(\ref{eq:ball004jvx}) and (\ref{eq:netbal06});\ here, we have assumed
$J_{+}(x',x'')$ to be expressed in terms of $\breve{A}(x)$, rather than $A(x)$,
using Eq.~(\ref{eq:spfra01fv})].  The relaxing thermoballistic spin-polarized
current $\check{\mathfrak{J}}_{-}(x)$ is determined by the ballistic current
$\check{J}_{-}(x',x'';x)$ given by Eq.~(\ref{eq:netbal02}), which is a
linear-homogeneous functional of $\breve{A}(x)$.  Hence,
$\check{\mathfrak{J}}_{-}(x)$ is linear and {\em homogeneous} in the boundary 
values $\breve{A}_{1,2}$, so that we can write
\begin{eqnarray}
\check{\mathfrak{J}}_{-}(x) &\equiv&  \check{\mathfrak{J}}_{-}(x; 
\breve{A}_{1}, \breve{A}_{2} ) \nonumber \\ &=& \frac{v_{e} N_{c}}{2} B_{+}(x) 
[ {\cal F}_{1}(x) e^{\beta \mu_{1}} \breve{A}_{1} + {\cal F}_{2}(x) 
e^{\beta \mu_{2}} \breve{A}_{2} ] , \nonumber \\
\label{eq:91x}
\end{eqnarray}
with (dimensionless) ``formfactors'' ${\cal F}_{1}(x)$ and ${\cal F}_{2}(x)$.
Then, we can express the values of the total current spin polarization at the 
semiconductor side of the interfaces at $x_{1,2}$ as
\begin{eqnarray}
P_{J}^{(sc)}(x_{1,2}) &=&
\overstar{P}_{J}(x_{1,2}; \breve{A}_{1}, \breve{A}_{2}) + \frac{v_{e} N_{c}
B_{+}(x_{1,2})} {2 J} \nonumber \\ &\ &  \times [ {\cal
F}_{1}(x_{1,2}) e^{\beta \mu_{1}} \breve{A}_{1} + {\cal F}_{2}(x_{1,2})
e^{\beta \mu_{2}} \breve{A}_{2} ] . \nonumber \\
\label{eq:66c}
\end{eqnarray}
The boundary values $\breve{A}_{1,2}$ of the reduced spin accumulation
function, which are undetermined as yet, are to be expressed in
terms of the current $J$ and the parameters of the system by matching the
current spin polarization in the semiconductor to the corresponding
polarization in the contacts (see Sec.~\ref{sec:match} below).

\subsubsection{Current spin polarization in the contacts}

\label{sec:pollead}

In order to keep the formulation sufficiently general, we assume the
semiconducting sample to be connected to (semi-infinite) {\em ferromagnetic}
metal contacts, which are treated as fully degenerate Fermi systems.

In the left contact located in the range $x < x_{1}$, the spin-resolved
chemical potentials $\mu_{\uparrow , \downarrow}(x)$  have the
form\cite{yuf02a}
\begin{equation}
\mu_{\uparrow , \downarrow}(x) = \frac{e^{2} J}{\sigma_{+}^{(l)}} (x_{1} -
x) \pm \frac{c_{l}} {\sigma_{\uparrow , \downarrow}^{(l)}}
e^{-(x_{1}-x)/L_{s}^{(l)}} .
\label{eq:50}
\end{equation}
Here, $L_{s}^{(l)}$ is the spin diffusion length, and $\sigma_{\uparrow ,
\downarrow}^{(l)}$ are the ($x$-independent) spin-up and spin-down bulk 
conductivities, respectively. Setting $x = x_{1}^{-}$ in Eqs.~(\ref{eq:50}), 
we obtain
\begin{equation}
c_{l} = \frac{\sigma_{\uparrow}^{(l)}
\sigma_{\downarrow}^{(l)}}{\sigma_{+}^{(l)}} \mu_{-}(x_{1}^{-}) .
\label{eq:pollead01}
\end{equation}
With the spin-resolved currents $J_{\uparrow , \downarrow}(x)$ given by
\begin{equation}
J_{\uparrow , \downarrow}(x) =  - \frac{\sigma_{\uparrow ,
\downarrow}^{(l)}}{e^{2}} \frac{d \mu_{\uparrow , \downarrow}(x)}{dx} ,
\label{eq:zeropol01}
\end{equation}
we then find for the current spin polarization in the left contact, using
$J_{+}(x) \equiv J$,
\begin{eqnarray}
P_{J}(x) &\equiv& \frac{J_{-}(x)}{J_{+}(x)} \nonumber \\ &=& P_{l} 
- \frac{G_{l}}{2 e^{2}J} \mu_{-}(x_{1}^{-}) e^{-(x_{1}-x)/L_{s}^{(l)}} . 
\nonumber \\
\label{eq:57lala}
\end{eqnarray}
Here,
\begin{equation}
P_{l} \equiv \frac{\sigma_{-}^{(l)}}{\sigma^{(l)}_{+}}
\label{eq:57tyg}
\end{equation}
is the bulk (current or density) spin polarization, and the parameter
\begin{equation}
G_{l} = \frac{4 \sigma_{\uparrow}^{(l)}
\sigma_{\downarrow}^{(l)}}{\sigma_{+}^{(l)} L_{s}^{(l)}} =
\frac{\sigma_{+}^{(l)}}{L_{s}^{(l)}} \left( 1 - P_{l}^{2} \right) , 
\label{eq:57xyz}
\end{equation}
which has the dimension of interface conductance, characterizes the 
spin-dependent transport in the ferromagnet.

For the current spin polarization in the right contact located in the range 
$x > x_{2}$,  we have
\begin{equation}
P_{J}(x) = P_{r}  + \frac{G_{r}}{2 e^{2} J} \mu_{-}(x_{2}^{+})
e^{-(x-x_{2})/L_{s}^{(r)}} , 
\label{eq:57zzz}
\end{equation}
which follows by replacing in Eq.~(\ref{eq:50}) the coordinate $x$ with 
$x_{1} + x_{2} - x$, and the labels "l" attached to the parameters in
Eqs.~(\ref{eq:50})--(\ref{eq:57xyz}) with "r".

When spin-selective interface resistances are absent, the chemical-potential
splitting $\mu_{-}(x)$ is continuous at the interfaces, 
\begin{equation}
\mu_{-}(x_{1}^{-}) = \mu_{-}^{(ct)}(x_{1}) ,
\label{eq:57zxl}
\end{equation}
\begin{equation}
\mu_{-}^{(ct)}(x_{2}) = \mu_{-}(x_{2}^{+}) ,
\label{eq:57ugv}
\end{equation}
where $\mu_{-}^{(ct)}(x_{1,2})$ are its values at the contact side of the 
interfaces. The latter values are to be identified with the corresponding 
values $\mu_{-}^{(sc)}(x_{1,2})$ at the semiconductor side of the interface, 
\begin{eqnarray}
\mu_{-}^{(ct)}(x_{1,2}) &=& \mu_{-}^{(sc)}(x_{1,2}) \nonumber \\
&=& \frac{1}{\beta} \ln \left( \frac{1 + \breve{A}_{1,2}/2}{1
-\breve{A}_{1,2}/2} \right) , 
\label{eq:68}
\end{eqnarray}
where the right-hand equations have been obtained by solving
Eq.~(\ref{eq:spfra01hl}) for $\mu_{-}(x')$, setting $x' = x_{1,2}$, and
defining 
\begin{equation}
\breve{A}_{1,2} = \breve{A}(x_{1,2})
\label{eq:57gfd}
\end{equation}
for the boundary values of the reduced spin accumulation function 
$\breve{A}(x')$.  With Eqs.~(\ref{eq:57zxl})--(\ref{eq:68}) used in 
Eqs.~(\ref{eq:57lala}) and (\ref{eq:57zzz}), respectively, we now have for the
current spin polarizations at the contact side of the interfaces
\begin{equation}
P_{J}^{(ct)}(x_{1}) = P_{l} - \frac{G_{l}}{2 \beta e^{2} J} \ln \left( 
\frac{1 + \breve{A}_{1}/2}{1 -\breve{A}_{1}/2} \right) ,
\label{eq:57angi}
\end{equation}
\begin{equation}
P_{J}^{(ct)}(x_{2}) = P_{r} + \frac{G_{r}}{2  \beta e^{2} J} \ln \left( \frac{1 +
\breve{A}_{2}/2}{1 -\breve{A}_{2}/2} \right) ,
\label{eq:57guido}
\end{equation}
which are to be identified with the corresponding polarizations 
$P_{J}^{(sc)}(x_{1,2})$ at the semiconductor side of the interfaces [see 
Eq.~(\ref{eq:66zara}) below].

Spin-selective interface resistances $\rho_{\uparrow , \downarrow}^{(1,2)}$
give rise to discontinuities of the spin-resolved chemical potentials
on the contact sides of the interfaces.\cite{ras00,smi01,fer01,ras02,kra03} 
At $x = x_{1}$, for example, the discontinuity has the form
\begin{equation}
\mu_{\uparrow , \downarrow}(x_{1}^{-}) - \mu_{\uparrow , 
\downarrow}^{(ct)}(x_{1}) = e^{2} J_{\uparrow , \downarrow}(x_{1}) 
\rho_{\uparrow , \downarrow}^{(1)} .
\label{eq:ratz}
\end{equation}
The interface resistances $\rho_{\uparrow , \downarrow}^{(1)}$  are located 
between $x = x_{1}^{-}$ and $x_{1}$ in the ferromagnetic contact, adjacent to 
the Sharvin interface resistance $\tilde{\rho}_{1}$ 
[see Eq.~(\ref{eq:sharv05})] between $x = x_{1}$ and $x_{1}^{+}$ in the
semiconductor.  The quantity $\mu_{-}(x_{1}^{-})$ to be substituted in
Eq.~(\ref{eq:57lala}) is obtained from Eqs.~(\ref{eq:50}),
(\ref{eq:zeropol01}), and (\ref{eq:57xyz}) as
\begin{eqnarray}
\mu_{-}(x_{1}^{-}) &=& \frac{1}{1 + G_{l} \rho_{+}^{(1)}/4 } \nonumber \\ &\ &
\times \left\{ \mu_{-}^{(ct)}(x_{1}) + \frac{e^{2} J}{2} \left[ P_{l} 
\rho_{+}^{(1)} + \rho_{-}^{(1)} \right] \right\} . \nonumber \\
\label{eq:karol}
\end{eqnarray}
The same procedure applies {\em mutatis mutandis} to the interface at $x =
x_{2}$. The connection of $\mu_{-}^{(ct)}(x_{1,2})$ with the interface values 
of the reduced spin accumulation function, $\breve{A}_{1,2}$, is again
given by Eqs.~(\ref{eq:68}).

\subsubsection{Matching the polarizations at the interfaces}

\label{sec:match}

To evaluate the current and density spin polarizations all across the 
contact-semiconductor system, we have to relate the boundary values 
of the reduced spin accumulation function, $\breve{A}_{1,2}$, to the total 
physical current $J$ and the internal parameters characterizing the system. 
To this end, we exploit the continuity of the total current spin polarization 
$P_{J}(x)$ across the contact-semiconductor interfaces at $x_{1,2}$,
\begin{equation}
P_{J}^{(ct)}(x_{1,2}) = P_{J}^{(sc)}(x_{1,2}) .
\label{eq:66zara}
\end{equation}
Using expressions (\ref{eq:57angi}) and (\ref{eq:57guido}) for 
$P_{J}^{(ct)}(x_{1,2})$ and expressions (\ref{eq:66c}) for 
$P_{J}^{(sc)}(x_{1,2})$, we obtain a pair of coupled nonlinear equations for 
$\breve{A}_{1,2}$, which we will not write down here in their general form.

In the zero-bias limit, when $|\beta \mu_{-}(x)| \ll 1$, we have
$|\breve{A}_{1,2}| \ll 1$ [see Eq.~(\ref{eq:spfra01hl})], so that 
$P_{J}^{(ct)}(x_{1})$ and $P_{J}^{(ct)}(x_{2})$ become linear in 
$\breve{A}_{1}$ and $\breve{A}_{2}$, respectively.  Further, to first
order, the term involving the reduced spin accumulation function $\breve{A}(x)$
in the persistent ballistic spin-polarized current $\overstar{J}_{-}(x',x'')$
[see Eq.~(\ref{eq:netbal06}), with $J_{+}(x', x'')$ expressed in terms of 
$\breve{A}(x)$ via Eqs.(\ref{eq:spfra01hl}) and (\ref{eq:ball004jhm})] can be
neglected, so that this current becomes independent of $\breve{A}_{1,2}$. The 
same then holds for the corresponding thermoballistic current, 
$\overstar{\mathfrak{J}}_{-}(x)$, as well as for the persistent current spin 
polarization $\overstar{P}_{J}(x)$ derived therefrom. Hence, the polarizations 
$P_{J}^{(sc)}(x_{1,2})$ are linear in $\breve{A}_{1,2}$, and the coupled 
equations for $\breve{A}_{1,2}$ become linear.  Since $\mu_{1} = \mu_{2}$ in 
the zero-bias limit, they can be expressed, using Eq.~(\ref{eq:ball004jux}), 
in the form
\begin{eqnarray}
\left[ \frac{G_{l}}{ 2 \beta e^{2}} + v_{e} n_{+}(x_{1}) {\cal F}_{1}(x_{1}) 
\right] \breve{A}_{1} +  v_{e} n_{+}(x_{1}) {\cal F}_{2}(x_{1}) \breve{A}_{2} 
\nonumber \\ &\ & \hspace{-4.0cm} = P_{l} J - 
\overstar{\mathfrak{J}}_{-}(x_{1}) ,
\label{eq:66e}
\end{eqnarray}
\begin{eqnarray}
v_{e} n_{+}(x_{2}) {\cal F}_{1}(x_{2}) \breve{A}_{1} 
- \left[\frac{G_{r}}{2 \beta e^{2}} - v_{e} n_{+}(x_{2}) {\cal F}_{2}(x_{2})
\right] \breve{A}_{2}  \nonumber \\ &\ & \hspace{-4.0cm} = P_{r} J - 
\overstar{\mathfrak{J}}_{-}(x_{2}) .
\label{eq:66ezg}
\end{eqnarray}
The current $\overstar{\mathfrak{J}}_{-}(x)$ is proportional to $J$, so that
the quantities $\breve{A}_{1,2}$ are, as solutions of Eqs.~(\ref{eq:66e}) and
(\ref{eq:66ezg}), proportional to $J$ as well.  Hence, the relaxing 
thermoballistic spin-polarized current 
$\check{\mathfrak{J}}_{-}(x)$ is also proportional to $J$, so that the total 
current spin polarization $P_{J}(x)$ calculated from Eq.~(\ref{eq:91xx}) is, 
in the zero-bias limit, independent of the total physical current $J$.

Summing up, we obtain the total current spin polarization along the entire
heterostructure, $P_{J}(x)$, as follows.  In the metal contacts, it is 
given by expressions (\ref{eq:57lala}) and (\ref{eq:57zzz}), respectively, 
with $\mu_{-}(x_{1}^{-})$ and $\mu_{-}(x_{2}^{+})$ expressed in terms of 
$\breve{A}_{1,2}$ via Eqs.~(\ref{eq:57zxl})--(\ref{eq:68}).  In the 
semiconductor, it is given by expression (\ref{eq:91xx}) in terms of the 
persistent, $\overstar{\mathfrak{J}}_{-}(x)$, and relaxing, 
$\check{\mathfrak{J}}_{-}(x)$, thermoballistic spin-polarized currents, where
the formfactors ${\cal F}_{1,2}(x)$ entering the latter current are determined
by the solution of the integral equation (\ref{eq:spiacc02}) for 
the spin accumulation function $A(x)$. The total density spin polarization 
$P_{n}(x)$ is obtained in an analogous way.

\section{Examples}

This section deals with the application of the thermoballistic approach to 
various specific examples.  The emphasis is on spin-polarized transport in 
heterostructures including nonmagnetic and magnetic semiconductors.  

\subsection{Homogeneous semiconductor at zero bias}

\label{sec:homsem}

In considering the case of a homogeneous semiconductor at zero bias, we
illustrate the thermoballistic formalism for a particularly simple example and,
moreover, collect results needed for treating heterostructures involving
homogeneous layers of nonmagnetic and magnetic semiconductors (see
Secs.~\ref{sec:fmnmshet} and \ref{sec:dmsnmshe} below).

For a homogeneous semiconducting layer at zero bias, the potential energy
profiles $E_{\uparrow , \downarrow}(x)$ are given by
\begin{equation}
E_{\uparrow , \downarrow}(x) \equiv  E_{c} \pm \smfrac{1}{2} \Delta
\label{eq:66iiit}
\end{equation}
$(x_{1} \leq x \leq x_{2})$.  Specializing the development presented in
Sec.~\ref{sec:resf} for constant spin splitting to the case $E_{c}(x)
\equiv E_{c}$, we find from Eq.~(\ref{eq:inteq06opa}) the reduced $(\Delta = 
0)$ kernel $\mathfrak{K}_{0}(x_{1}, x; x';l)$ in the explicit form
\begin{equation}
\mathfrak{K}_{0}(x_{1}, x; x'; l) = Q [e^{-(x' - x_{1})/l} -
e^{-(x - x')/l}] .
\label{eq:inteq06woft}
\end{equation}
From the corresponding solution $\tilde{\mathfrak{R}}_{10}(x)$ of
Eq.~(\ref{eq:inteq04aha}) and the solution of the analogous equation for the
function $\tilde{\mathfrak{R}}_{20}(x)$, we then obtain for $x > x_{1}$ and $x
< x_{2}$, respectively, using Eq.~(\ref{eq:inteq09wq}),
\begin{equation}
\tilde{\mathfrak{R}}_{1,2}(x) =  1 \pm \frac{x - x_{1,2}}{2l} 
- P  \mathfrak{R}_{1,2}(x)
\label{eq:inteq09zdt}
\end{equation}
for the resistance functions $\tilde{\mathfrak{R}}_{1,2}(x)$ of the homogeneous
semiconductor;\ at $x = x_{1}$ and $x = x_{2}$, on the other hand, we have
$\tilde{\mathfrak{R}}_{1}(x_{1}) = \tilde{\mathfrak{R}}_{2}(x_{2}) = 0$;\ see
Eqs.~(\ref{eq:inteq08}) and (\ref{eq:spiche05}).

For the reduced resistance $\tilde{\mathfrak{R}}$, we then find from
Eq.~(\ref{eq:spiche15}), using Eq.~(\ref{eq:inteq53cb}),
\begin{equation}
\tilde{\mathfrak{R}} = 1 + \frac{S}{2l} - \frac{v_{e}N_{c}}{2 J} \frac{P}{Q}
e^{- \beta E_{c}} A_{12}^{-} ,
\label{eq:66zick}
\end{equation}
and for the function $\tilde{\mathfrak{R}}_{-}(x)$ from
Eq.~(\ref{eq:spiche18}), using Eqs.~(\ref{eq:ballim04st}),
(\ref{eq:inteq04lil}), (\ref{eq:inteq02ax}), and (\ref{eq:inteq02im}),
\begin{eqnarray}
\tilde{\mathfrak{R}}_{-}(x) &=&  \frac{x - (x_{1} + x_{2})/2}{l} 
\nonumber \\ &\ & + \frac{v_{e}N_{c}}{J} \frac{P}{Q} 
e^{- \beta E_{c}} [A(x) - \smfrac{1}{2} A_{12}^{+}] 
\label{eq:66zora}
\end{eqnarray}
for $x_{1} < x < x_{2}$, whereas from Eq.~(\ref{eq:spiche15ug}),
$\tilde{\mathfrak{R}}_{-}(x_{1,2}) = \mp \tilde{\mathfrak{R}}$.
The spin accumulation function $A(x)$ to be inserted in this expression is
given by Eqs.~(\ref{eq:ballim05s}) and (\ref{eq:difeq20}).

Inserting expressions (\ref{eq:66zick}) and (\ref{eq:66zora}) in
Eq.~(\ref{eq:spiche17zg}) and using Eq.~(\ref{eq:spiche14}), we find from
Eq.~(\ref{eq:ball004jvx}) for the net total ballistic current in a homogeneous
semiconductor at zero bias
\begin{equation}
J_{+}(x_{1}, x_{2}) = J  \left( 1 + \frac{S}{2l} \right) ,
\label{eq:66jafz}
\end{equation}
\begin{equation}
J_{+}(x', x_{2}) = J \left( \frac{1}{2} + \frac{x_{2} - x'}{2l} \right)
\label{eq:66jkwq}
\end{equation}
for $x' > x_{1}$,
\begin{equation}
J_{+}(x_{1}, x'') = J \left( \frac{1}{2} + \frac{x'' - x_{1}}{2l} \right)
\label{eq:66jpcb}
\end{equation}
for $x'' < x_{2}$, and
\begin{equation}
J_{+}(x', x'') = J \frac{x'' - x'}{2l}
\label{eq:66juhu}
\end{equation}
for $x_{1} < x' < x'' < x_{2}$, in which the dependence on $A(x)$ present in
$\tilde{\mathfrak{R}}$ and $\tilde{\mathfrak{R}}_{-}(x)$ has cancelled out
completely. At this point, it is appropriate to note that the dependence of the
ballistic current $J_{+}(x', x'')$ on the end-point coordinates $x', x''$ 
exhibited here contradicts the basic assumption of the 
prototype thermoballistic model as expressed by Eq.~(\ref{eq:gendr001qal}).   
  
Now, using expressions (\ref{eq:66jafz})--(\ref{eq:66juhu})
in Eq.~(\ref{eq:thebal01}), we find
\begin{equation}
\mathfrak{J}_{+}(x) \equiv\mathfrak{J}_{+} = J ,
\label{eq:66z}
\end{equation}
i.e., the total thermoballistic current is equal to the (constant) total
physical current. Further, using Eq.~(\ref{eq:netbal06}), we have
\begin{equation}
\overstar{\mathfrak{J}}_{-}(x) \equiv \overstar{\mathfrak{J}}_{-} = P J
\label{eq:66zyxv}
\end{equation}
for the persistent thermoballistic spin-polarized current, and hence from
Eq.~(\ref{eq:65}) for the persistent current spin polarization
\begin{equation}
\overstar{P}_{J}(x) \equiv \overstar{P}_{J} =
\frac{\overstar{\mathfrak{J}}_{-}}{J} = P ,
\label{eq:66zb}
\end{equation}
thereby retrieving the static spin polarization.

Turning  now to the determination of the relaxing thermoballistic 
spin-polarized current $\check{\mathfrak{J}}_{-}(x)$ in a homogeneous 
semiconductor at zero bias, we have from Eq.~(\ref{eq:netbal02}) for the 
corresponding ballistic current
\begin{eqnarray}
\check{J}_{-}(x',x'';x)  &=& \frac{v_{e} N_{c}}{2} Q e^{- \beta E_{c}}
\nonumber \\ &\ & \hspace{-0.2cm} \times [ A(x') e^{-(x - x')/l_{s}} - A(x'')
e^{-(x'' -x)/l_{s}}] \nonumber \\
\label{eq:netbal02ihk}
\end{eqnarray}
$(x_{1} \leq x' < x'' \leq x_{2})$. Inserting this expression in
Eq.~(\ref{eq:thebal01}) and subsequently using the derivative of 
Eq.~(\ref{eq:difeq17}) with respect to $x$, we obtain  
$\check{\mathfrak{J}}_{-}(x)$ in the form
\begin{equation}
\check{\mathfrak{J}}_{-}(x) = - v_{e} N_{c} Q e^{- \beta E_{c}} \bar{l}
\frac{dA(x)}{dx}
\label{eq:netbal02fln}
\end{equation}
[see Eq.~(4.11) of Ref.~\onlinecite{lip05};\ the factor 2 appearing in the
right-hand side of the latter equation again reflects the fact that the
normalization of $A(x)$ used there differs from that used in the present
article].  From Eqs.~(\ref{eq:91xx}), (\ref{eq:66zyxv}), and 
(\ref{eq:netbal02fln}), we now have for the total current spin polarization
\begin{equation}
P_{J}(x) = P - \frac{v_{e} N_{c}}{J} Q e^{- \beta E_{c}} \bar{l} 
\frac{dA(x)}{dx} .
\label{eq:66ze}
\end{equation}
[Note that owing to its normalization to the total thermoballistic current, the
expression for $P_{J}(x)$ given by Eq.~(136) of Ref.~\onlinecite{lip09} differs
from expression (\ref{eq:66ze});\ see the remark following Eq.~(\ref{eq:91xx}) 
above.]

As to the density spin polarization in a homogeneous semiconductor, we have
from Eq.~(\ref{eq:ball004juy}) for the persistent part, $\overstar{n}_{+}(x',
x''; x) \equiv \overstar{n}_{+}(x', x'')$, of the joint total  ballistic
density  
\begin{equation}
\overstar{n}_{+}(x_{1}, x_{2}) =  \frac{N_{c}}{2} \frac{e^{- \beta E_{c}}}{Q} 
\left( \eta_{12}^{+} + \frac{P}{2} A_{12} \right) ,
\label{eq:66kohj}
\end{equation}
\begin{equation}
\overstar{n}_{+}(x', x_{2}) = \overstar{n}_{+}(x_{1}, x_{2})
 - \frac{J}{2 v_{e}} \left( \frac{1}{2} + \frac{x' - x_{1}}{2l} \right)
\label{eq:66fpkv}
\end{equation}
for $x' > x_{1}$,
\begin{equation}
\overstar{n}_{+}(x_{1}, x'') = \overstar{n}_{+}(x_{1}, x_{2}) + 
\frac{J}{2 v_{e}}  \left( \frac{1}{2} + \frac{x_{2} - x''}{2l} \right)
\label{eq:66dfsx}
\end{equation}
for $x'' < x_{2}$, and
\begin{eqnarray}
\overstar{n}_{+}(x', x'') &=& \overstar{n}_{+}(x_{1}, x_{2}) - 
\frac{J}{2 v_{e}} \frac{x' - x_{1} - ( x_{2} - x'')}{2l} \nonumber \\ 
\label{eq:66fsnb}
\end{eqnarray}
for $x_{1} < x' < x'' < x_{2}$. For the relaxing part of the joint total 
ballistic density, $\check{n}_{+}(x', x''; x) \equiv \check{n}_{+}(x', x'')$,
we find from Eq.~(\ref{eq:netbal02xy})
\begin{equation}
\check{n}_{+}(x', x'') = 0 ,
\label{eq:66salk}
\end{equation}
so that
\begin{equation}
n_{+}(x',x'') = \overstar{n}_{+}(x', x'') .
\label{eq:66trfc}
\end{equation}
Observing Eq.~(\ref{eq:66trfc}), we now use expressions 
(\ref{eq:66kohj})--(\ref{eq:66fsnb}) in Eq.~(\ref{eq:thebal01}) to obtain for
the thermoballistic joint total density
\begin{eqnarray}
\mathfrak{n}_{+}(x) &=& \overstar{\mathfrak{n}}_{+}(x) \nonumber \\ &=&
 \overstar{n}_{+}(x_{1}, x_{2}) - \frac{J}{2 v_{e}} \left( \frac{x}{l} - 
\frac{x_{1} + x_{2}}{2l} \right) , \nonumber \\
\label{eq:66gvya}
\end{eqnarray}
with $\overstar{n}_{+}(x_{1}, x_{2})$ given explicitly by 
Eq.~(\ref{eq:66kohj}). Since 
\begin{equation}
\overstar{n}_{-}(x', x'') = P \overstar{n}_{+}(x', x'')
\label{eq:66tfld}
\end{equation}
from Eq.~(\ref{eq:ball004juy}), we have for the persistent part of the joint 
ballistic spin-polarized density
\begin{equation}
\overstar{\mathfrak{n}}_{-}(x) = P \overstar{\mathfrak{n}}_{+}(x)
= P \mathfrak{n}_{+}(x) ,
\label{eq:66plfn}
\end{equation}
and hence for the persistent density spin polarization
\begin{equation}
\overstar{P}_{n}(x) = \frac{\overstar{\mathfrak{n}}_{-}(x)}{\mathfrak{n}_{+}
(x)} = P ,
\label{eq:66hcqw}
\end{equation}
in agreement with the corresponding current spin polarization.

By inserting expression (\ref{eq:netbal02fln}) for the corresponding current 
$\check{\mathfrak{J}}_{-}(x)$ in the balance equation (\ref{eq:ballim24new}) 
and using Eq.~(\ref{eq:difeq18}) for the spin accumulation function $A(x)$,
the relaxing thermoballistic spin-polarized density $\check{\mathfrak{n}}_{-}
(x)$ for a homogeneous semiconductor is readily obtained in the form
\begin{equation}
\check{\mathfrak{n}}_{-}(x) = \frac{N_{c}}{2} Q e^{- \beta E_{c}} A(x) , 
\label{eq:66zbsy}
\end{equation}
so that we have for the total density spin polarization
\begin{equation}
P_{n}(x) = P +  \frac{N_{c}}{2 \mathfrak{n}_{+}(x)} Q e^{- \beta E_{c}} A(x) ,
\label{eq:66gxky}
\end{equation}
with $\mathfrak{n}_{+}(x)$ given by Eq.~(\ref{eq:66gvya}). Now, omitting in 
expression (\ref{eq:66gvya}) for $\mathfrak{n}_{+}(x)$ the term proportional 
to $J$, so that $\mathfrak{n}_{+}(x) = \overstar{n}_{+}(x_{1}, x_{2})$, we can 
use expression (\ref{eq:66gxky}) for $P_{n}(x)$ in Eq.~(\ref{eq:66ze}) to 
express the current spin polarization $P_{J}(x)$ in terms of $P_{n}(x)$ in the 
form
\begin{equation}
P_{J}(x) = P - \frac{2 v_{e} \overstar{n}_{+}(x_{1}, x_{2})}{J} \bar{l} 
\frac{d P_{n}(x)}{dx} .
\label{eq:66fgns}
\end{equation}
Differentiating this equation and using Eqs.~(\ref{eq:difeq18}) and 
(\ref{eq:66gxky}), we then obtain 
\begin{equation}
P_{n}(x) = P - \frac{J}{2 v_{e} \overstar{n}_{+}(x_{1}, x_{2})} l_{s} 
\frac{d P_{J}(x)}{dx} 
\label{eq:66pgmq}
\end{equation}
for the density spin polarization $P_{n}(x)$ in terms of $P_{J}(x)$.
  
For the magnetoresistance of a homogeneous semiconductor at zero bias, we find
\begin{equation}
R_{m} = Q - 1- \frac{v_{e}N_{c} }{2 J} \frac{P}{Q} e^{- \beta E_{c}} A^{-}_{12}
\frac{2l}{2l + S}
\label{eq:66zbg}
\end{equation}
by inserting the zero-field limit of expression (\ref{eq:66zick}),
\begin{equation}
\tilde{\mathfrak{R}}_{0} =  1 + \frac{S}{2l}  ,
\label{eq:66tfk}
\end{equation}
for $\tilde{\mathfrak{R}}_{0}$ in Eq.~(\ref{eq:sharv07lz}).

\subsection{FM/NMS/FM heterostructures}

\label{sec:fmnmshet}

We now consider heterostructures formed of a {\em homogeneous, nonmagnetic}
semiconducting (NMS) layer and two {\em ferromagnetic} metal (FM) contacts. 
The study of spin-polarized transport in this kind of structure, 
both experimentally and theoretically, marked the beginning of semiconductor 
spintronics\cite{ras00,sch00,fil00,smi01,fer01,ras02,sch02,alb02,alb03,kra03,sch05} 
(for recent surveys of the physics of semiconductor-based spintronic devices,
see Refs.~\onlinecite{tan11,sah11,zut11}).
For zero bias, we evaluate the position dependence of the current and density 
spin polarizations using the results of Secs.~\ref{sec:pollead}, 
\ref{sec:match}, and \ref{sec:homsem}. For nonzero bias, we obtain the spin 
polarizations ``injected'' from the contacts into the semiconductor, i.e., the 
polarizations generated inside the semiconductor in the vicinity of either 
FM/NMS interface regardless of the influence of the opposite interface.

\subsubsection{Zero-bias spin polarizations}

\label{sec:zeropola}

Inside the semiconducting layer, we have from Eq.~(\ref{eq:66ze}) with $P = 0$ 
and $Q = 1$, using expression (\ref{eq:difeq20}) for the spin accumulation 
function $A(x)$, 
\begin{eqnarray}
P_{J}(x) &=&  \frac{v_{e} N_{c}}{J} e^{- \beta E_{c}} \gamma  
[ C_{1} e^{-(x-x_{1})/L} - C_{2} e^{-(x_{2}-x)/L}] , \nonumber \\
\label{eq:40c}
\end{eqnarray}
where $\gamma$ is given by Eq.~(\ref{eq:difeq24}). The coefficients $C_{1,2}$ 
can be expressed via Eqs.~(\ref{eq:difeq21}) and (\ref{eq:difeq22}), and using 
Eq.~(\ref{eq:spfra01hl}), in terms of the values $\breve{A}_{1,2}$ of the 
reduced spin accumulation function $\breve{A}(x)$ on the contact sides of the 
interfaces.

In order to determine the quantities $\breve{A}_{1,2}$, we evaluate expression
(\ref{eq:40c}) on the semiconductor sides of the interfaces, obtaining
\begin{equation}
P_{J}^{(sc)}(x_{1}) \equiv  P_{J}(x_{1}^{+}) = \frac{v_{e} n_{+}^{(0)}}{2 J}  
(g \breve{A}_{1} - h \breve{A}_{2}) ,
\label{eq:61a}
\end{equation}
\begin{equation}
P_{J}^{(sc)}(x_{2}) \equiv P_{J}(x_{2}^{-}) = \frac{v_{e} n_{+}^{(0)}}{2 J} 
(h \breve{A}_{1} - g \breve{A}_{2}) .
\label{eq:61abc}
\end{equation}
Here, we have introduced the total equilibrium density for zero spin splitting 
and zero bias,
\begin{equation}
n_{+}^{(0)} = N_{c} e^{- \beta (E_{c} - \mu_{1})} , 
\label{eq:61ujo}
\end{equation}
which is constant inside the semiconducting layer. The coefficients $g$ and 
$h$ in expressions (\ref{eq:61a}) and (\ref{eq:61abc}) are given by 
Eqs.~(\ref{eq:difeq27}) and (\ref{eq:difeq28}), respectively.  

Using the continuity of $P_{J}(x)$ at the interfaces, Eq.~(\ref{eq:66zara}),
we now equate expressions (\ref{eq:57angi}) and (\ref{eq:61a}), and expressions
(\ref{eq:57guido}) and (\ref{eq:61abc}). [When spin-selective interface
resistances are included, expression (\ref{eq:57angi}) for 
$P_{J}^{(ct)}(x_{1})$ is to be replaced with the more general expression 
obtained from using expression (\ref{eq:karol}) for $\mu_{-}(x_{1}^{-})$ in 
Eq.~(\ref{eq:57lala}), and analogously for $P_{J}^{(ct)}(x_{2})$.]  For zero 
bias, when $|\breve{A}_{1,2}| \ll 1$, this results in a system of coupled 
linear equations for $\breve{A}_{1,2}$ [see Eqs.~(\ref{eq:66e}) and 
(\ref{eq:66ezg})],
\begin{equation}
[g  + \tilde{G}_{l}^{(0)} ] \breve{A}_{1} -  h \breve{A}_{2} =
\frac{2 J}{v_{e} n_{+}^{(0)}} P_{l}  ,
\label{eq:69}
\end{equation}
\begin{equation}
h \breve{A}_{1} - [ g + \tilde{G}_{r}^{(0)} ] \breve{A}_{2} = 
\frac{2  J}{v_{e} n_{+}^{(0)}} P_{r}  ,
\label{eq:69abc}
\end{equation}
where
\begin{equation}
\tilde{G}_{l,r}^{(0)} = \frac{G_{l,r}}{\beta e^{2} v_{e} n_{+}^{(0)}} .
\label{eq:90klm}
\end{equation}
The solutions of Eqs.\ (\ref{eq:69}) and (\ref{eq:69abc}) are found to be
\begin{equation}
\breve{A}_{1} = \frac{2 J}{v_{e} n_{+}^{(0)} \Gamma} \{ [ g +
\tilde{G}_{r}^{(0)} ] P_{l} - h P_{r} \}  ,
\label{eq:72abc}
\end{equation}
\begin{equation}
\breve{A}_{2}  = \frac{2 J}{v_{e} n_{+}^{(0)} \Gamma} \{ h P_{l} -
[ g + \tilde{G}_{l}^{(0)} ] P_{r} \}  ,
\label{eq:72}
\end{equation}
where
\begin{equation}
\Gamma = [ g + \tilde{G}_{l}^{(0)}][ g +  \tilde{G}_{r}^{(0)}] - h^{2}   .
\label{eq:72ijk}
\end{equation}
Expressions (\ref{eq:72abc}) and (\ref{eq:72}) determine the values of the 
reduced spin accumulation function, $\breve{A}_{1}$ and $\breve{A}_{2}$ in 
terms of the current $J$, of the polarizations $P_{l}$ and $P_{r}$ in the 
left and right ferromagnetic contact, respectively, and of material parameters,
such as the conductivities $\sigma_{l,r}$ and the spin diffusion lengths 
$L_{s}^{(l,r)}$ of the contacts [via $\tilde{G}_{l,r}^{(0)}]$ and the momentum
relaxation length $l$ and the spin relaxation length $l_{s}$ in the 
semiconducting sample as well as the sample length $S$ (via $g$ and $h$) and 
the (constant) total equilibrium electron density $n_{+}^{(0)}$.  Since the 
quantities $\breve{A}_{1,2}$ are proportional to the total current $J$, the 
current spin polarization $P_{J}(x)$ is independent of $J$, while the density 
spin polarization $P_{n}(x)$ is proportional to $J$.

The spin polarizations along the entire heterostructure are now obtained 
as follows.  Inside the semiconductor, the current spin polarization 
$P_{J}(x)$ is given by expression (\ref{eq:40c}), with $C_{1,2}$ calculated 
from the solutions $\breve{A}_{1,2}$ of Eqs.~(\ref{eq:72abc}) and 
(\ref{eq:72}). The corresponding density spin polarization $P_{n}(x)$ is then 
readily obtained from Eq.~(\ref{eq:66pgmq}). The expressions for $P_{J}(x)$ in 
the ferromagnetic contacts are provided by Eqs.\ (\ref{eq:57lala}) and 
(\ref{eq:57zzz}), respectively, where the quantities $\mu_{-}(x_{1}^{-})$ and 
$\mu_{-}(x_{2}^{+})$ are calculated from Eq.\ (\ref{eq:karol}) and from its 
analogue for $\mu_{-}(x_{2}^{+})$, respectively. We do not write down the 
density spin polarizations in the ferromagnets, but only mention that they do 
not, in general, match the polarizations $P_{n}(x_{1}^{+})$ and 
$P_{n}(x_{2}^{-})$ on the semiconductor sides of the interfaces.

In Refs.~\onlinecite{lip05} and \onlinecite{lip06}, we have presented results 
of detailed calculations for $P_{J}(x)$ at zero bias for typical parameter 
values of FM/NMS/FM heterostructures, emphasizing the dependence of the 
polarization on the momentum and spin relaxation lengths, $l$ and $l_{s}$, and 
on the spin-selective interface resistances 
$\rho^{(1,2)}_{{\uparrow, \downarrow}}$ (see Sec.~\ref{sec:pollead}).   

\subsubsection{Injected spin polarizations at nonzero bias}

\label{sec:injespin}

We define the ``injected spin polarization'' as the polarization inside 
the semiconductor in the vicinity of the interface, e.g., at $x=x_{1}$,
generated by the bulk polarization $P_{l}$ of the left ferromagnet regardless 
of the influence of the right ferromagnet.  More explicitly, we define the 
injected current and density spin polarizations as the  polarizations 
$P_{J}(x_{1}^{+})$ and $P_{n}(x_{1}^{+})$, respectively, in the limit of 
infinite sample length, $S \rightarrow \infty$. The injected spin polarizations
at $x = x_{1}^{+}$ provide the initial values of the left-generated 
polarizations in the semiconductor, which propagate into the region $x > x_{1}$
while being degraded by the effect of spin relaxation.

We now consider the injected spin polarizations for electron transport in a 
homogenenous semiconductor, driven by an external electric field, i.e., for a 
(spin-degenerate)  potential energy profile of the form (\ref{eq:difeq01}), 
with the parameter $\epsilon$ defined by Eq.~(\ref{eq:difeq01erd}). In 
order to obtain the spin accumulation function $A(x)$ for this case, one has to
solve Eq.~(\ref{eq:difeq09}) [in general, numerically] under the asymptotic 
condition 
\begin{equation}
A(x) \propto e^{-x/\lambda}  \; \; {\rm for} \; \; x \rightarrow \infty .
\label{eq:80gfrs}
\end{equation}
To determine the decay length $\lambda$, we solve Eq.~(\ref{eq:difeq09}) in the
range $x - x_{1} \gg 1/(\epsilon + 1/l)$ in which the $x$-dependence of the 
coefficient functions $b_{0}(x)$, $b_{1}(x)$, and $b_{2}(x)$ arising from the 
function $b(x)$ can be disregarded. This yields
\begin{equation}
\frac{1}{\lambda} =  \frac{\epsilon}{2} + \left[ \frac{\epsilon^{2}}{4} +
\frac{l - \tilde{l}}{l \tilde{l}^{2}} \left( 1+ \epsilon \tilde{l} 
\frac{1 + \epsilon l}{2 + \epsilon l} \right) \right]^{1/2}  .
\label{eq:80}
\end{equation}
Since $l > \tilde{l}$, $\lambda$ is a real number.

The injected current spin polarization is obtained from Eq.~(\ref{eq:91xx})
with $\overstar{\mathfrak{J}}_{-}(x) = 0$. We calculate the relaxing 
thermoballistic spin-polarized current at the interface, 
$\check{\mathfrak{J}}_{-}(x_{1}^{+})$, from the relation 
\begin{equation}
\check{\mathfrak{J}}_{-}(x_{1}^{+}) = \frac{v_{e} N_{c}}{2} \check{W}(x_{1};
x_{1}, x_{2};l) [A_{1} - A(x_{1}^{+})] ,
\label{eq:80ijhb}
\end{equation}
where the function $\check{W}(x; x_{1}, x_{2};l)$ is given by 
Eq.~(\ref{eq:spiacc04}). This relation can be shown to hold, for 
arbitrary sample length $S$ and potential energy profiles 
$E_{\uparrow, \downarrow}(x)$, by (i) evaluating 
$\check{\mathfrak{J}}_{-}(x_{1}^{+})$ from Eq.~(\ref{eq:thebal01}) with 
$F(x', x''; x) \equiv \check{J}_{-} (x', x'';x)$  [see 
Eq.~(\ref{eq:netbal02})] and (ii) eliminating from the resulting expression 
the boundary value $A_{2}$ by using the integral equation (\ref{eq:spiacc02}) 
for $x = x_{1}$. 

For the case considered here, we write down the integral equation 
(\ref{eq:difeq04}) [with $\mathfrak{c}(\epsilon x) = \bar{\mathfrak{c}}$] 
for $x = x_{1}$ and take the limit $x_{2} \rightarrow \infty$ to obtain the 
relation  
\begin{equation}
2 A(x_{1}^{+}) - \bar{A}_{1} = A_{1} ,
\label{eq:80lcvs}
\end{equation} 
where
\begin{equation}
\bar{A}_{1} =  \int_{x_{1}}^{\infty} \frac{dx}{l} e^{-(x-x_{1})/\tilde{l}} A(x)
.
\label{eq:8ozykp}
\end{equation}
Relation (\ref{eq:80lcvs}) fixes the normalization of the spin accumulation 
function $A(x)$ in terms of the boundary value $A_{1}$. Then, using this 
relation in Eq.~(\ref{eq:80ijhb}) with $\check{W}(x_{1}; x_{1}, x_{2}; l) = 2 
e^{- \beta E_{c}(x_{1})}$,  we can express the current 
$\check{\mathfrak{J}}_{-}(x_{1}^{+})$ in the form 
\begin{equation}
\check{\mathfrak{J}}_{-}(x_{1}^{+})  = \frac{v_{e} n_{+}^{(0)}(x_{1})}{2} 
A_{J}^{(1)} \breve{A}_{1}, 
\label{eq:81}
\end{equation}
where the quantity
\begin{equation}
A_{J}^{(1)}  = \frac{A(x_{1}^{+}) - \bar{A}_{1}}{A(x_{1}^{+}) - \bar{A}_{1}/2} 
\label{eq:82}
\end{equation}
is independent of the normalization of $A(x)$, and
\begin{equation}
n_{+}^{(0)}(x_{1}) = N_{c} e^{- \beta[E_{c}(x_{1}) - \mu_{1}]} .
\label{eq:82jpfc}
\end{equation}
In obtaining Eq.~(\ref{eq:81}), we have used Eq.~(\ref{eq:ball004jum}), with 
$P(x) = 0$, as well as Eq.~(\ref{eq:spfra01hl}). 

For the injected current spin polarization, we now have from 
Eq.~(\ref{eq:91xx})
\begin{equation}
P_{J}(x_{1}^{+}) = \frac{\check{\mathfrak{J}}_{-}(x_{1}^{+})}{J}  = \frac{v_{e}
n_{+}^{(0)}(x_{1})}{2J} A_{J}^{(1)} \breve{A}_{1} ,
\label{eq:81ijk}
\end{equation}
which, by continuity, is equal to the polarization $P_{J}(x_{1})$ at the 
contact side of the interface. Then, equating the right-hand sides of 
Eqs.~(\ref{eq:57angi}) and (\ref{eq:81ijk}), we obtain
\begin{equation}
P_{l}  - \frac{G_{l}}{2 \beta e^{2} J}  \ln \left( \frac{1+
\breve{A}_{1}/2}{1 -\breve{A}_{1}/2} \right) = \frac{v_{e} 
n_{+}^{(0)}(x_{1})}{2 J} A_{J}^{(1)} \breve{A}_{1} .
\label{eq:69rst}
\end{equation}
This nonlinear equation for $\breve{A}_{1}$ is to be solved for given values of
the parameters $\epsilon$, $P_{l}$, $G_{l}$, $n_{+}^{(0)}(x_{1})$, $l$, and 
$l_{s}$.

We now turn to the calculation of the injected density spin polarization from 
Eq.~(\ref{eq:91olsb}), where $\overstar{\mathfrak{n}}_{-}(x) = 0$ and 
$\mathfrak{n}_{+}(x) = \overstar{\mathfrak{n}}_{+}(x) $  in the present case. 
Proceeding similarly as for the current $\check{\mathfrak{J}}_{-}(x_{1}^{+})$,
we obtain the relaxing thermoballistic spin-polarized density at the interface,
$\check{\mathfrak{n}}_{-}(x_{1}^{+})$, in the form
\begin{equation}
\check{\mathfrak{n}}_{-}(x_{1}^{+}) = \frac{n_{+}^{(0)}(x_{1})}{4} A_{n}^{(1)} 
\breve{A}_{1} ,
\label{eq:82pqqq}
\end{equation}
where
\begin{equation}
A_{n}^{(1)}  = \frac{A(x_{1}^{+})}{A(x_{1}^{+}) - \bar{A}_{1}/2} .
\label{eq:82pqr}
\end{equation}
For the persistent total thermoballistic density at the interface, 
$\overstar{\mathfrak{n}}_{+}(x_{1}^{+})$, we find by evaluating 
Eq.~(\ref{eq:thebal01}) with $F(x', x''; x) \equiv 
\overstar{n}_{+}(x', x''; x)$ [see Eq.~(\ref{eq:ball004juy})] and taking 
the limit $S \rightarrow \infty$ 
\begin{equation}
\overstar{\mathfrak{n}}_{+}(x_{1}^{+})  = \frac{n_{+}^{(0)}(x_{1})}{2} (1 +  
\overstar{A}_{1}) ,
\label{eq:82pqq}
\end{equation}
where the quantity
\begin{equation}
\overstar{A}_{1} = \int_{x_{1}}^{\infty} \frac{dx}{l} e^{- (x - x_{1})/l} 
e^{\beta [\tilde{\mu}(x) - \mu_{1}]}  
\label{eq:82erkg}
\end{equation}
has been expressed in terms of the average chemical potential $\tilde{\mu}(x)$ 
using Eq.~(\ref{eq:eqden03zk}). The injected density spin polarization now 
follows as
\begin{equation}
P_{n}(x_{1}^{+})  = \frac{\check{\mathfrak{n}}_{-}(x_{1}^{+})}{
\overstar{\mathfrak{n}}_{+}(x_{1}^{+})} = \frac{A_{n}^{(1)} \breve{A}_{1}}{2 
(1 + \overstar{A}_{1})}  ,
\label{eq:81zyx}
\end{equation}
where the boundary value $\breve{A}_{1}$ of the reduced spin accumulation 
function is again to be determined by solving Eq.~(\ref{eq:69rst}).

The dependence of the injected current spin polarization, $P_{J}(x_{1}^{+})$,
on the electric-field parameter $\epsilon$, for $l$-values ranging from the 
drift-diffusion to the ballistic regime and for zero as well as nonzero 
interface resistances $\rho^{(1)}_{\uparrow, \downarrow}$, has been studied 
numerically in Ref.~\onlinecite{lip05}. 

In the drift-diffusion regime, when $l/l_{s} \ll 1$ and $\epsilon l \ll 1$,
the spin accumulation function $A(x)$ is given by the exponentially decreasing
solution of Eq.~(\ref{eq:difeq16}), 
\begin{equation}
A(x) \propto e^{ -(x - x_{1})/L_{s}^{\epsilon}} 
\label{eq:81wqad}
\end{equation}
$(x \geq x_{1})$, with the field-dependent spin diffusion length 
$L_{s}^{\epsilon}$ given by
\begin{equation}
\frac{1}{L_{s}^{\epsilon}} =  \frac{\epsilon}{2} + \left( 
\frac{\epsilon^{2}}{4} + \frac{1}{L_{s}^{2}} \right) ^{1/2} ,
\label{eq:80hui}
\end{equation}
where $L_{s}$ is the spin diffusion length defined by Eq.~(\ref{eq:ballim26}).
From Eqs.~(\ref{eq:82}) and (\ref{eq:82pqr}), respectively, we then have
\begin{equation}
A_{J}^{(1)} = \frac{2 l}{L_{s}^{\epsilon}}
\label{eq:80gjb}
\end{equation}
and 
\begin{equation}
A_{n}^{(1)} = 2 ,
\label{eq:80ild}
\end{equation} 
and hence from Eqs.~(\ref{eq:81ijk}) and (\ref{eq:81zyx}), respectively,
\begin{equation}
P_{J}(x_{1}^{+}) = \frac{v_{e} n_{+}^{(0)}(x_{1})}{J} 
\frac{l}{L_{s}^{\epsilon}} \breve{A}_{1} 
= \frac{1}{2 \epsilon L_{s}^{\epsilon}} \breve{A}_{1}
\label{eq:81jslc}
\end{equation}
and
\begin{equation}
P_{n}(x_{1}^{+})  =  \frac{\breve{A}_{1}}{1 + \overstar{A}_{1}}  = 
\frac{\breve{A}_{1}}{2}
\label{eq:81pfqn}
\end{equation}
for the injected current and density spin polarizations in the drift-diffusion 
regime. In deriving the right-hand equation (\ref{eq:81jslc}) , we have used 
Eq.~(\ref{eq:diff006hsa}), with $P=0$, $Q=1$, and $\beta e V = \epsilon S$, as
well as Eq.~(\ref{eq:82jpfc}), to express the total current $J$ in the form 
\begin{equation}
J  =  2 v_{e} n_{+}^{(0)}(x_{1})  \epsilon l  \equiv \frac{1}{\beta e^{2}} 
\sigma_{+}^{(0)} \epsilon , 
\label{eq:81kalo}
\end{equation}
where 
\begin{equation}
\sigma_{+}^{(0)} = 2 \beta e^{2} v_{e} n_{+}^{(0)}(x_{1}) l  
\label{eq:75zqsp} 
\end{equation}
is the (spin-summed) conductivity of the semiconductor [see 
Eq.~(\ref{eq:drude006f})]. The right-hand equation (\ref{eq:81pfqn}) has been
obtained by using Eq.~(\ref{eq:ballim20kf}) in Eq.~(\ref{eq:82erkg}). The 
boundary value $\breve{A}_{1}$ is determined by the equation
\begin{equation}
P_{l}  - \frac{G_{l}}{2 \sigma_{+}^{(0)} \epsilon}  \ln \left( \frac{1+
\breve{A}_{1}/2}{1 -\breve{A}_{1}/2} \right) =  \frac{1}{2 \epsilon
 L_{s}^{\epsilon}} \breve{A}_{1} ,
\label{eq:69hlbv}
\end{equation}
which follows from Eq.~(\ref{eq:69rst}) by replacing its right-hand side
with the right-hand side of Eq.~(\ref{eq:81jslc}) and inserting expression 
(\ref{eq:81kalo}) for $J$. 
 
The effect of external electric fields on the current spin polarization 
injected at FM/NMS interfaces has been studied within the standard 
drift-diffusion approach by Yu and Flatt\'{e}\cite{yuf02a}.  Comparing our 
description to that of these authors, we find that the field-dependent spin 
diffusion length $L_{s}^{\epsilon}$ given by  Eq.~(\ref{eq:80hui}) agrees with
the ``up-stream'' spin diffusion length $L_{u}$ given by Eq.~(2.23b) of 
Ref.~\onlinecite{yuf02a}, provided the ``intrinsic'' spin diffusion length $L$
of that reference is identified with the spin diffusion length $L_{s} = 
\sqrt{l l_{s}}$ of the present work.  Then, identifying in Eq.~(3.5) of 
Ref.~\onlinecite{yuf02a} (with the interface resistances set equal to zero) 
the spin-summed conductivity of the semiconductor, $\sigma_{s}$,  with 
$\sigma_{+}^{(0)}$, and the spin injection efficiency at the interface, 
$\alpha_{0}$, with $P_{J}(x_{1}^{+}) = \breve{A}_{1}/2 \epsilon 
L_{s}^{\epsilon}$, and using Eq.~(\ref{eq:57xyz}), we observe the equivalence
of the former equation with Eq.~(\ref{eq:69hlbv}). Consequently, the injected
current and density spin polarizations of either work are actually identical.

For arbitrary $l$, we now consider the injected current spin polarization in 
the zero-bias limit, when $|\breve{A}_{1}| \ll 1$.  Here, the spin 
accumulation function $A(x)$ is given by the exponentially decreasing solution
of Eq.~(\ref{eq:difeq18}), 
\begin{equation}
A(x) \propto e^{- (x - x_{1})/L}
\label{eq:69jozd}
\end{equation}
$(x \geq x_{1})$, where the generalized spin diffusion length $L$ is given by 
Eq.~(\ref{eq:difeq19}). From Eq.~(\ref{eq:82}), we then have 
\begin{equation}
A_{J}^{(1)} = \frac{2 \gamma}{1 + \gamma} \equiv \gamma^{\ast} ,
\label{eq:69opjf}
\end{equation}
with $\gamma$ given by Eq.~(\ref{eq:difeq24}), and hence from the zero-bias 
limit of Eq.~(\ref{eq:69rst}) 
\begin{equation}
\breve{A}_{1} =  \frac{2J}{v_{e} n_{+}^{(0)}(x_{1})}  
\frac{P_{l}}{G_{l}/\tilde{\cal G}_{1} + \gamma^{\ast}} ,
\label{eq:69ahoi}
\end{equation}
where
\begin{equation}
\tilde{\cal G}_{1} \equiv \frac{1}{\tilde{\rho}_{1}} = \beta e^{2} v_{e} 
n_{+}^{(0)}(x_{1})
\label{eq:70pfdc}
\end{equation}
is the Sharvin interface conductance [see Eq.~(\ref{eq:sharv05kl})]. Inserting 
this in Eq.~(\ref{eq:81ijk}), we find  
\begin{equation}
P_{J}(x_{1}^{+}) = \frac{P_{l}}{1 + G_{l}/ \gamma^{\ast} \tilde{\cal G}_{1} } ,
\label{eq:70xy}
\end{equation}
i.e., for zero bias, the injected current spin polarization is independent of
the total current $J$.

In the drift-diffusion regime, when $l/l_{s} \ll 1$, we have 
\begin{equation}
\gamma^{\ast} = 2 \sqrt{\frac{l}{l_{s}}} ,
\label{eq:70kbf}
\end{equation}
so that $P_{J}(x_{1}^{+}) $ reduces to
\begin{equation}
P_{J}(x_{1}^{+}) =  \frac{P_{l}}{1 + G_{l}/{\cal G}_{1}} ,
\label{eq:75}
\end{equation}
where 
\begin{equation}
{\cal G}_{1} = 2 \tilde{\cal G}_{1} \sqrt{\frac{l}{l_{s}}} 
= \frac{\sigma_{+}^{(0)}}{L_{s}} ,
\label{eq:75tlmc}
\end{equation}
with $\sigma_{+}^{(0)}$ defined by Eq.~(\ref{eq:75zqsp}), and $L_{s}$ by 
 Eq.~(\ref{eq:ballim26}).
The quantity ${\cal G}_{1}$ is seen to be the semiconductor analogue of the 
``interface conductance'' $G_{l}$ characterizing the ferromagnet [see
Eq.~(\ref{eq:57xyz})]. For typical values of the parameters of the FM/NMS/FM
system (see, e.g., Ref.~\onlinecite{lip05}), the values of $G_{l}$ exceed 
those of ${\cal G}_{1}$ by several orders of magnitude. This indicates a
``conductance mismatch'' between ferromagnet and 
semiconductor,\cite{ras00,sch02,sch05,sch00} which gives rise to very low 
values of the injected spin polarizations at FM/NMS interfaces. [To remedy  
this mismatch, one may introduce spin-selective interface resistances, for 
example, in the form of 
tunneling barriers.\cite{ras00,smi01,fer01,sch02,ras02,kra03}] 

In the ballistic regime, when $l/l_{s} \gg 1$, we have 
\begin{equation}
\gamma^{\ast} = \gamma = 1 , 
\label{eq:79oce}
\end{equation}
so that
\begin{equation}
P_{J}(x_{1}^{+})  = \frac{P_{l}}{1 + G_{l}/\tilde{\cal G}_{1}} .
\label{eq:79y}
\end{equation}
Here, the Sharvin interface conductance $\tilde{\cal G}_{1}$ takes the place of
the quantity ${\cal G}_{1}$ in Eq.~(\ref{eq:75}). As $\tilde{\cal G}_{1}$ is 
proportional to the density $n_{+}^{(0)}(x_{1})$, Eq.~(\ref{eq:79y}) yields 
values for $P_{J}(x_{1}^{+})$ close to the bulk spin polarization of the
feromagnet, $P_{l}$, if $n_{+}^{(0)}(x_{1})$ is sufficiently high. For typical 
parameter values,\cite{lip05} however, the high donor densities needed to 
obtain the required high electron densities implies very small values of the
momentum relaxation length $l$, such that ballistic transport is excluded as
the dominating transport mechanism. On the other hand, for densities so low 
that ballistic transport prevails, the injected spin polarization is confined
to very small values. This result corroborates previous estimates\cite{kra03}
according to which spin injection is suppressed even in the ballistic regime
unless spin-selective interface resistances are introduced.

\subsection{DMS/NMS/DMS heterostructures}

\label{sec:dmsnmshe}

\begin{figure}[t]
\includegraphics[width=0.45\textwidth]{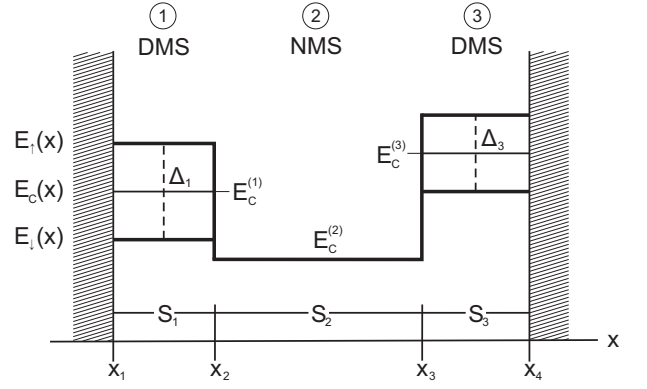}
\caption{Schematic zero-bias potential energy profiles for a DMS/NMS/DMS 
heterostructure composed of three homogeneous semiconducting layers enclosed
between nonmagnetic metal contacts.
}
\label{fig:7}
\end{figure}

As an example of particular interest from the point of view of applications,
we now consider spin-polarized electron transport in heterostructures involving
diluted magnetic semiconductors (DMS) in their paramagnetic phase (see, e.g.,
Refs.~\onlinecite{sch05} and 
\onlinecite{egu98,oes99,fie99,jon00,guo00,sch01,egu01,cha01,sch04,kha05,roy06,san07,slo07,cio09}). 
In structures of this kind, a
nonmagnetic semiconducting (NMS) layer is sandwiched between two DMS layers
which, in turn, are enclosed between nonmagnetic metal contacts (see
Fig.~\ref{fig:7}).

In a complete thermoballistic description, the semiconductor part of a 
DMS/NMS/DMS structure should be treated  as a {\em single} sample. Then, the 
ballistic intervals $[x',x'']$ covering the sample may contain one or both of
the DMS/NMS interfaces, at which the potential energy profiles must be expected
to change abruptly. The same holds for the material parameters, in 
particular the momentum and spin relaxation lengths, so that a description in 
terms of position-dependent parameters would become necessary. Here, we adopt a
simplified treatment by assuming the different layers in a DMS/NMS/DMS 
heterostructure to be {\em homogeneous} and requiring the interfaces to act
as {\em fixed points} of local thermodynamic equilibrium. This allows us to 
apply the thermoballistic description separately to the different layers 
(the potential energy profiles are then, in general, {\em discontinuous} at 
the interfaces). For each layer, we evaluate the spin accumulation function and
the zero-bias current spin polarization for a homogeneous semiconductor (see 
Sec.~\ref{sec:homsem}), and subsequently match the current spin polarization 
at the interfaces to obtain its full position dependence as well as the 
magnetoresistance. 

\subsubsection{Zero-bias current spin polarization}

\label{sec:zeropol}

Heterostructures of the kind depicted schematically in Fig.~\ref{fig:7} are 
parametrized here by attaching labels $j=1,2,3$ to quantities referring to the
left DMS layer, the NMS layer, and the right DMS layer, respectively. The
positions of the interfaces (including the interfaces between the
DMS layers and the contacts) are denoted by $x_{k}$ $(k=1,2,3,4)$; this 
notation deviates from the one used in the preceding sections.

Inside the left and right metal contacts, the position dependence of the 
current spin polarization  is given by Eqs.~(\ref{eq:57lala}) and 
(\ref{eq:57zzz}), respectively [with $x_{2}$ in the latter equation replaced 
with $x_{4}$], with $\mu_{-}(x_{1,4}) = \breve{A}_{1,4}/ \beta$ [see 
Eq.~(\ref{eq:spfra01hl}) for $|\beta \mu_{-}(x')| \ll 1$] and $P_{l} = P_{r} = 
0$,  
\begin{equation}
P_{J}(x) = \mp \frac{G_{l,r}}{2 \beta e^{2} J} \breve{A}_{1,4} e^{-|x - x_{1,4}
|/L_{s}^{(l,r)}} 
\label{eq:a1}
\end{equation}
for $x < x_{1}$ and  $x > x_{4}$.

Inside the semiconducting part of the heterostructure, $P_{J}(x)$ is found in 
terms of the static spin polarizations $P_{j}$ of the three layers, and of the 
boundary values of the spin accumulation function $A(x)$ at the interface 
positions $x_{k}$, $A(x_{k}) \equiv A_{k}$, which are all points of local 
thermodynamic equilibrium.  [The interface positions $x_{2,3}$ are points of 
local thermodynamic equilibrium in the same sense as are the interface 
positions $x_{1,4}$, except that there the interface is between metal contact 
and DMS, while here it is between DMS and NMS.] In each layer $j$, 
the function $A(x)$ has the Sharvin discontinuities 
$A(x_{j}^{+}) - A_{j}$ and $ A_{j+1} - A(x_{j+1}^{-})$ at the interface
positions $x_{j}$ and $x_{j+1}$, respectively [see Eqs.~(\ref{eq:difeq25}) and
 (\ref{eq:difeq26})]. Using now, for zero bias, Eq.~(\ref{eq:66ze}) together 
with Eq.~(\ref{eq:difeq20}) separately in each layer, we have
\begin{eqnarray}
P_{J}(x) &=& P_{j} + \frac{v_{e} N_{c} }{J} \, Q_{j} e^{- \beta
E_{c}^{(j)}}\gamma_{j} \nonumber \\ &\ & \times [C_{1}^{(j)} 
e^{-(x - x_{j})/L_{j}} - C_{2}^{(j)} e^{-(x_{j+1} - x)/L_{j}}]  \nonumber \\
\label{eq:a2}
\end{eqnarray}
for $x_{j} \leq x \leq x_{j+1}$ $(j=1,2,3)$. Here, 
\begin{equation}
\gamma_{j} = \frac{\bar{l}_{j}}{L_{j}} = \frac{L_{j}}{l_{s}^{(j)}} ,
\label{eq:a6}
\end{equation}
with $\bar{l}_{j}$ and $L_{j}$ defined by 
Eqs.~(\ref{eq:difeq14c}) and (\ref{eq:difeq19}) in terms of the momentum 
relaxation length, $l_{j}$, and spin relaxation length, $l_{s}^{(j)}$, of layer
$j$. Further,
\begin{equation}
C_{1}^{(j)} = \frac{1}{D_{j}} [(1 + \gamma_{j}) e^{S_{j}/L_{j}} A_{j} - (1 -
\gamma_{j}) A_{j+1}]
\label{eq:a4}
\end{equation}
and
\begin{equation}
C_{2}^{(j)} = - \frac{1}{D_{j}} [(1 - \gamma_{j}) A_{j} - (1 + \gamma_{j})
e^{S_{j}/L_{j}} A_{j+1}] ,
\label{eq:a5}
\end{equation}
where
\begin{equation}
D_{j} = 2 [(1 + \gamma_{j}^{2}) \sinh(S_{j}/L_{j}) + 2 \gamma_{j}
\cosh(S_{j}/L_{j})] ,
\label{eq:a7}
\end{equation}
and 
\begin{equation}
S_{j} = x_{j+1} - x_{j} 
\label{eq:aghsm}
\end{equation}
is the thickness of layer $j$. [Owing to a difference in normalization, 
expression (\ref{eq:a2}) for $P_{J}(x)$ differs from the corresponding 
expression given by Eq.~(141) of Ref.~\onlinecite{lip09};\ see the remark 
following Eq.~(\ref{eq:66ze}) above.]

In order to relate the quantities $\breve{A}_{k}$ to the total physical current
$J$ and the parameters of the contact-semi\-conductor system, we invoke the
continuity of the current spin polarization $P_{J}(x)$ at all interfaces,
setting $P_{2} = 0$ ($Q_{2} = 1$) in Eq.~(\ref{eq:a2}).  From 
Eqs.~(\ref{eq:a1}) and (\ref{eq:a2}), we then obtain a set of four coupled 
linear equations for $\breve{A}_{k}$,
\begin{equation}
(\tilde{G}_{l}^{(1)} + Q_{1} g_{1}) \breve{A}_{1} - Q_{1}  h_{1} \breve{A}_{2} 
= -  \frac{2 J}{v_{e} n_{+}^{(0;1)}} P_{1} ,
\label{eq:a8}
\end{equation}
\begin{eqnarray}
 - Q_{1} h_{1} \breve{A}_{1} &+& ( Q_{1} g_{1} + e^{\beta \delta_{12}} g_{2})
\breve{A}_{2} \nonumber \\ &-& e^{\beta \delta_{12}}  h_{2} \breve{A}_{3} 
=  \frac{2 J}{v_{e} n_{+}^{(0;1)}} P_{1} , 
\label{eq:a9}
\end{eqnarray}
\begin{eqnarray}
- e^{\beta \delta_{32}} h_{2} \breve{A}_{2} &+& (e^{\beta \delta_{32}} g_{2} + 
Q_{3} g_{3} ) \breve{A}_{3} \nonumber \\  &-& Q_{3}  h_{3} \breve{A}_{4} = -
\frac{2 J}{v_{e} n_{+}^{(0;3)}} P_{3} , 
\label{eq:a10}
\end{eqnarray}
\begin{equation}
- Q_{3} h_{3} \breve{A}_{3} + (\tilde{G}_{r}^{(3)} + Q_{3} 
g_{3}) \breve{A}_{4} =  \frac{2 J}{v_{e} n_{+}^{(0;3)}} P_{3} .
\label{eq:a11}
\end{equation}
Here,
\begin{equation}
\tilde{G}_{l,r}^{(1,3)} = \frac{{G}_{l,r}}{\beta e^{2} v_{e} n_{+}^{(0;1,3)}} ,
\label{eq:a987}
\end{equation}
where
\begin{equation}
n_{+}^{(0;j)} = N_{c} e^{-\beta [E_{c}^{(j)} - \mu_{1}]} 
\label{eq:a853}
\end{equation}
is the (constant) total equilibrium electron density in layer $j$, and 
\begin{equation}
\delta_{j,j'} = E_{c}^{(j)} -  E_{c}^{(j')}
\label{eq:a297}
\end{equation}
are the band offsets. Further,
\begin{equation}
g_{j} = h_{j} [\cosh(S_{j}/L_{j}) + \gamma_{j} \sinh(S_{j}/L_{j})] ,
\label{eq:a12}
\end{equation}
with
\begin{equation}
h_{j} = \frac{4 \gamma_{j}}{D_{j}} .
\label{eq:a13}
\end{equation}
The system of equations (\ref{eq:a8})--(\ref{eq:a11}) can be easily solved, so
that the complete position dependence of the zero-bias current spin
polarization is obtained in explicit form.

We now specialize to the case of a {\em symmetric} heterostructure, for which 
the parameters of the right DMS layer and metal contact are identical to those 
of the left DMS layer and metal contact. Assuming arbitrarily high 
conductivities of the contacts, $\sigma^{(l,r)}_{+} \rightarrow \infty$, so 
that $\tilde{G}_{l,r}^{(1,3)} \rightarrow \infty$, we have from 
Eqs.~(\ref{eq:a1}) and (\ref{eq:a987})
\begin{equation}
\breve{A}_{1,4} \rightarrow 0 , 
\label{eq:a14}
\end{equation}
such that $\tilde{G}_{l,r}^{(1,3)} \breve{A}_{1,4}$ remains nonzero and 
finite. Then, from Eqs.~(\ref{eq:a9}) and (\ref{eq:a10}) with $P_{1} = P_{3} =
P$ and $Q_{1} = Q_{3} = Q$,
\begin{equation}
\breve{A}_{2} = - \breve{A}_{3} = \frac{2 J }{v_{e} n_{+}^{(0;D)}} 
\frac{P}{{\cal Q}_{DN}} ,
\label{eq:a15}
\end{equation}
where
\begin{equation}
{\cal Q}_{DN} = Q g_{D} + (g_{N} + h_{N}) e^{\beta \delta_{DN}} .
\label{eq:a4711}
\end{equation}
Hence, from Eqs.~(\ref{eq:a8}) and (\ref{eq:a11}),
\begin{eqnarray}
\tilde{G}_{l}^{(1)} \breve{A}_{1} &=& - \tilde{G}_{r}^{(4)} \breve{A}_{4} = - 
\frac{2 J}{v_{e} n_{+}^{(0;D)}} P + Q  h_{D} \breve{A}_{2} \nonumber \\ &=& - 
\frac{2 J P}{v_{e} n_{+}^{(0;D)}} \left( 1 - \frac{Q  h_{D}}{{\cal Q}_{DN}} 
\right) . 
\label{eq:a16}
\end{eqnarray}
Here, the DMS parameters have been labeled by ``D", and the NMS parameters by
``N".

Thus, in the symmetric case, the zero-bias current spin polarization $P_{J}(x)$
is completely determined by the quantity $\breve{A}_{2}$. Explicitly, we 
obtain, setting $L_{s}^{(l)} = L_{s}^{(r)} = L_{s}^{(mc)}$ (where the 
superscript ``$mc$'' refers to the metal contacts) in Eq.~(\ref{eq:a1}) and 
using Eqs.~(\ref{eq:a987}) and (\ref{eq:a16}),
\begin{equation}
P_{J}(x) = P \left( 1 - \frac{Q  h_{D}}{{\cal Q}_{DN}} \right) 
 e^{-|x - x_{1,4}|/L_{s}^{(mc)}}  , 
\label{eq:aa1}
\end{equation}
if $x < x_{1}$ and  $x > x_{4}$, respectively. Further, from Eq.~(\ref{eq:a2}),
we have, using Eqs.~(\ref{eq:a6})--(\ref{eq:a7}) along with 
Eqs.~(\ref{eq:spfra01hl}), (\ref{eq:61ujo}), and (\ref{eq:a15}),
\begin{eqnarray}
P_{J}(x) &=& P \left\{ 1 - \frac{Q h_{D}}{{\cal Q}_{DN}} [ \cosh 
(|x - x_{1,4}|/L_{D}) \right. \nonumber \\ &\ & \left. + \gamma_{D} 
\sinh ( |x - x_{1,4}|/L_{D} ) ] \rule{0mm}{5mm} \right\} , 
\label{eq:aa2}
\end{eqnarray}
if $x_{1} \leq x \leq x_{2}$ and $x_{3} \leq x \leq x_{4}$, respectively, and
\begin{eqnarray}
P_{J}(x) &=& \frac{2e^{\beta \delta_{DN}} h_{N} P}{{\cal Q}_{DN}} \nonumber \\
&\ & \times [\cosh (S_{N}/2 L_{N}) + \gamma_{N} \sinh (S_{N}/2 L_{N}) ] 
\nonumber \\ &\ & \times \cosh \bm{(} [x - (x_{2} + x_{3})/2]/L_{N} \bm{)}  , 
\label{eq:aa3}
\end{eqnarray}
if $x_{2} \leq x \leq x_{3}$. 

In Refs.~\onlinecite{lip09} and \onlinecite{lip07}, we have presented numerical
results for $P_{J}(x)$ for values of the momentum relaxation length $l$
ranging from the drift-diffusion regime to the ballistic regime.

\subsubsection{Magnetoresistance}

\label{sec:magnhet}

The relative magnetoresistance $R_{m}$ of the DMS/NMS/DMS heterostructure 
follows by inserting in Eq.~(\ref{eq:sharv07xbs}) the reduced resistance 
$\tilde{\mathfrak{R}}$ and the corresponding zero-field resistance 
$\tilde{\mathfrak{R}}_{0}$  obtained by summing up the respective contributions
$\tilde{\mathfrak{R}}_{j}$ and $\tilde{\mathfrak{R}}_{0}^{(j)}$ of the 
(homogeneous) layers $j$. These contributions are obtained from the 
(appropriately labeled)
expressions (\ref{eq:66zick}) and (\ref{eq:66tfk}) by multiplication with a 
factor $B_{+}^{m}(x_{1}, x_{4})/B_{+}^{m}(x_{j}, x_{j+1})$, such that  
$\tilde{\mathfrak{R}}$ and $\tilde{\mathfrak{R}}_{0}$ are normalized in
conformance with the definition (\ref{eq:inteq04scp}) for a single sample
extending from $x_{1}$ to $x_{4}$.

Specializing immediately to the case of the symmetric structure considered 
above, we have, defining
\begin{equation}
{\cal S}_{j} = 1 + \frac{S_{j}}{2l_{j}} 
\label{eq:66zskx}
\end{equation}
$(j = D, N)$ and using Eqs.~(\ref{eq:a14}),
\begin{eqnarray}
\tilde{\mathfrak{R}} &=& 2 Q e^{\beta E_{c}^{(D)}} {\cal S}_{D}  +  
e^{\beta E_{c}^{(N)}} {\cal S}_{N}  + \frac{v_{e} N_{c} 
e^{\beta \mu_{1}}}{2 J} P (\breve{A}_{2} - \breve{A}_{3})  \nonumber \\
\label{eq:66fdlv}
\end{eqnarray}
(we have omitted here the overall factor $B_{+}^{m}(x_{1}, x_{4})$ which drops 
out when $R_{m}$is formed), from which $\tilde{\mathfrak{R}}_{0}$ follows by 
setting $P = 0$ and $Q = 1$. Then, using Eq.~(\ref{eq:a15}), we obtain
\begin{equation}
R_{m} = \frac{(Q-1) {\cal S}_{D} + P^{2}/{\cal Q}_{DN}}{{\cal S}_{D} 
+ e^{- \beta \delta_{DN}} {\cal S}_{N}/2}  .
\label{eq:a17}
\end{equation}
Here, we note that in Ref.~\onlinecite{lip09}, in contrast to its definition in
terms of the zero-field resistance $\tilde{\mathfrak{R}}_{0}$ via 
Eq.~(\ref{eq:sharv07xbs}), the relative magnetoresistance $R_{m}$ has been 
defined with respect to the {\em spin-equilibrium} resistance 
obtained by setting $\breve{A}_{2} = \breve{A}_{3} = 0$ in expression 
(\ref{eq:66fdlv}). The latter definition does not seem to correspond to a 
genuine magnetoresistance, and so one should not attach quantitative 
significance to the numerical results for $R_{m}$ shown in Fig.~4 of
Ref.~\onlinecite{lip09}.

In the limit of low external magnetic field, when the static spin 
polarization $P$ depends linearly on the field strength, we 
have from  Eq.~(\ref{eq:a17}), keeping terms of order $P^{2}$,
\begin{equation}
R_{m} = \frac{\{ 1/[g_{D} + (g_{N} + h_{N}) e^{\beta \delta_{DN}}]  - 
{\cal S}_{D}/2 \} P^{2}}{{\cal S}_{D} + e^{- \beta \delta_{DN}} 
{\cal S}_{N}/2} .
\label{eq:a17pdfg}
\end{equation}
While based on assumptions that differ from, and are more general, than those 
underlying the ''two-band model'' for the (transverse) 
magnetoresistance,\cite{zim65,rot94} expression (\ref{eq:a17pdfg}) exhibits the
quadratic dependence on the field strength characterizing the latter model in 
the low-field limit.

\subsubsection{Drift-diffusion and ballistic regimes}

\label{sec:diffball}

Considering first the thermoballistic description of spin-polarized 
transport in DMS/NMS/DMS hetero\-structures in the drift-diffusion  
regime, we can compare our results to those of Ref.~\onlinecite{kha05} 
obtained within the standard drift-diffusion approach.

In the drift-diffusion regime, when $l_{j} \ll S_{j}$ and $l_{j} \ll
l_{s}^{(j)}$, we have 
\begin{equation}
L_{j} \rightarrow L_{s}^{(j)} = \sqrt{l_{j} l_{s}^{(j)}}
\label{eq:a21ehsl}
\end{equation}
and 
\begin{equation}
\gamma_{j} \rightarrow \frac{l_{j}}{L_{s}^{(j)}} 
= \sqrt{\frac{l_{j}}{l_{s}^{(j)}}} \ll 1
\label{eq:a21oanb}
\end{equation}
[see Eqs.~(\ref{eq:ballim26}), (\ref{eq:difeq14c}), (\ref{eq:difeq19}), and 
(\ref{eq:a6})]. Hence, from Eqs.~(\ref{eq:a12}) and (\ref{eq:a13}),
\begin{equation}
g_{j} = h_{j} \cosh(S_{j}/L_{s}^{(j)})
\label{eq:a21}
\end{equation}
and
\begin{equation}
h_{j} = \frac{2 l_{j}}{L_{s}^{(j)} \sinh (S_{j}/L_{s}^{(j)})}  ,
\label{eq:a22}
\end{equation}
respectively. Then, specializing to the symmetric heterostructure, we find 
for the quantity ${\cal Q}_{DN}$ defined by Eq.~(\ref{eq:a4711})  
\begin{eqnarray}
{\cal Q}_{DN} &=& 2 \left[ Q \frac{l_{D}}{L_{s}^{(D)}} \coth \bm{(} 
S_{D}/L_{s}^{(D)} \bm{)} \right. \nonumber \\ &\ & + \left. 
\frac{l_{N}}{L_{s}^{(N)}} \coth \bm{(} S_{N}/2L_{s}^{(N)} \bm{)}
e^{\beta \delta_{DN}} \right]  . \nonumber \\
\label{eq:a23}
\end{eqnarray}
Introducing the (spin-summed) conductivity $\sigma_{+}^{(j)}$ for layer $j$ 
$(j = D,N)$,
\begin{equation}
\sigma_{+}^{(j)} = 2 \beta e^{2} v_{e} n_{+}^{(j)} l_{j}
\label{eq:a16a}
\end{equation}
[see Eq.~(\ref{eq:drude006f})], with
\begin{equation}
n_{+}^{(j)} = \frac{n_{+}^{(0;j)}}{Q_{j}}
\label{eq:a16hxsy}
\end{equation}
$(Q_{D} = Q, Q_{N} = 1)$, we can rewrite expression (\ref{eq:a23}) in the form
\begin{eqnarray}
{\cal Q}_{DN} &=& \frac{2}{Q} \frac{l_{D}}{L_{s}^{(D)}} \left[ 
\rule{0mm}{6mm} Q^{2} 
\coth \bm{(} S_{D}/L_{s}^{(D)} \bm{)} \right. \nonumber \\ &\ & +  \left.
\frac{\sigma_{+}^{(N)}}{\sigma_{+}^{(D)}} \frac{L_{s}^{(D)}}{L_{s}^{(N)}} 
\coth \bm{(}S_{N}/2L_{s}^{(N)} \bm{)} \right]  . 
\label{eq:a23cayk}
\end{eqnarray}
Now, inserting expression (\ref{eq:a23cayk}) in 
Eq.~(\ref{eq:a15}) and using the resulting expression for $\breve{A}_{2}$ in
Eq.~(\ref{eq:a2}), we find that the values of
the zero-bias current spin polarization $P_{J}(x)$ at the positions $x_{1,4}$,
$x_{2,3}$, and $(x_{2} + x_{3})/2$, respectively, agree with those given by
Eqs.~(17), (16), and(18) of Ref.~\onlinecite{kha05}, if we identify in the
latter equations the spin-flip lengths $\lambda_{D}$ and $\lambda_{N}$ with
$L_{s}^{(D)}$ and $L_{s}^{(N)}$, respectively, and the layer thicknesses $d$
and $2 x_{0}$ with $S_{D}$ and $S_{N}$, respectively.

For the magnetoresistance in the drift-diffusion regime, we obtain from
Eq.~(\ref{eq:a17}), setting ${\cal S}_{j} = S_{j}/2l_{j}$ and inserting 
expression  (\ref{eq:a23cayk}) for ${\cal Q}_{DN}$, 
%j
\begin{equation}
R_{m} =  \frac{S_{D} [ 1/\sigma_{+}^{(D)} - 1/\sigma_{+}^{(0;D)}
 ] + P^{2}/\mathfrak{Q}_{DN} }{S_{D}/ \sigma_{+}^{(0;D)} + S_{N}/2 
\sigma_{+}^{(0;N)}} ,
\label{eq:a17zglv}
\end{equation}
where $\sigma_{+}^{(0;j)}$ is the conductivity for zero spin splitting, which 
is given by Eq.~(\ref{eq:a16a}) with $n_{+}^{(j)} =n_{+}^{(0;j)}$, and 
\begin{eqnarray}
\mathfrak{Q}_{DN} &=& Q^{2}  \frac{\sigma_{+}^{(D)}}{L_{s}^{(D)}}  
\coth \bm{(} S_{D}/L_{s}^{(D)} \bm{)} \nonumber \\ &\ & + 
\frac{\sigma_{+}^{(0;N)}}{L_{s}^{(N)}} \coth \bm{(} S_{N}/ 2L_{s}^{(N)} 
\bm{)} . 
\label{eq:a17tajb}
\end{eqnarray}
Then, identifying the parameters as above, we find that $R_{m}$ agrees 
with expression (13) of Ref.~\onlinecite{kha05}.

Turning now to the ballistic regime, when $l_{j} \gg S_{j}$ and $l_{j} \gg 
l_{s}^{(j)}$, we 
have 
\begin{equation}
L_{j} \rightarrow l_{s}^{(j)}
\label{eq:a17fsap}
\end{equation}
and 
\begin{equation}
\gamma_{j} \rightarrow 1
\label{eq:a17gclb}
\end{equation}
[see Eqs.~(\ref{eq:ballim26}), (\ref{eq:difeq14c}), (\ref{eq:difeq19}), and 
(\ref{eq:a6})], so that 
\begin{equation}
g_{j} = 1
\label{eq:a17kxpu}
\end{equation}
and 
\begin{equation}
h_{j} = e^{- S_{j}/l_{s}^{(j)}} .
\label{a17zfdq}
\end{equation}
For the symmetric heterostructure, we then find from Eq.~(\ref{eq:a4711})
\begin{equation}
{\cal Q}_{DN} =  Q + [1 + e^{-S_{N}/l_{s}^{(N)}} ] e^{\beta \delta_{DN}} .
\label{eq:a21a}
\end{equation}
With this used in Eq.~(\ref{eq:a15}) for $\breve{A}_{2}$ as well as in 
Eq.~(\ref{eq:a16}) for $\tilde{G}_{l}^{(1)} \breve{A}_{1}$, we obtain the 
position dependence of the zero-bias current spin polarization from 
Eqs.~(\ref{eq:a1}) and (\ref{eq:a2}). 

For the magnetoresistance in the ballistic regime, we have from 
Eqs.~(\ref{eq:a17}), setting ${\cal S}_{j} = 1$ and inserting expression 
(\ref{eq:a21a}) for ${\cal Q}_{DN}$,  
\begin{equation}
R_{m} = \frac{Q - 1 + P^{2}/\{ Q + [ 1 + e^{-S_{N}/l_{s}^{(N)}}] 
e^{\beta \delta_{DN}} \} }{1 + e^{- \beta \delta_{DN}}/2} . 
\label{eq:22}
\end{equation}
Depending on the values assigned to the different parameters in this 
expression, the ballistic (relative) magnetoresistance is positive or negative.

\section{Summary and outlook}

\label{sec:summconc}

In this article, we have presented a comprehensive survey of the 
thermoballistic approach to charge carrier transport in semiconductors. The
principal aim has been to develop the basic physical concept underlying this 
approach in detail, and to give a coherent exposition of the ensuing formalism,
which unifies, and partly modifies, generalizes, and corrects, the formal 
developments presented in our previous publications.
 
To make the presentation self-contained and easy to follow, we have proceeded
step by step, starting with an account of Drude's model as the origin 
of all semiclassical transport models. We then reviewed the
standard drift-diffusion and ballistic transport models, basic features of 
which have been adopted to shape  the thermoballistic description 
of carrier transport in terms of averages over random configurations of 
ballistic transport intervals. The contributions of the individual ballistic 
intervals to the total carrier current are governed by collision 
probabilities involving the carrier mean free path, or momentum relaxation
length, as the determining parameter of the thermoballistic concept. 

This concept finds a first concrete expression in the prototype thermoballistic
model, which is based on the simplifying assumption of current conservation 
across the points of local thermodynamic equilibrium linking ballistic 
intervals. The implementation of the prototype model results in a 
current-voltage characteristic containing a reduced resistance, which can be
expressed explicitly in terms of the parameters of the semiconducting system.

In the full thermoballistic concept, current conservation across the 
equilibrium points is abandoned, and position-dependent, total and 
spin-polarized thermoballistic currents and densities are introduced in terms 
of an average chemical-potential function and a spin accumulation function 
related to the spin splitting of the chemical potential. The algorithms 
for determining these dynamical functions follow from two physical 
conditions. First, the average of the total thermoballistic current over the
length of the semiconducting sample (as well as over the range between either 
end and an arbitrary point of local thermodynamic equilibrium) is
required to equal the conserved physical current, whereby one is able 
to set up a scheme for obtaining the average chemical potential. The 
explicit calculation of this function is implemented in terms of 
resistance functions, for which Volterra-type integral equations are 
derived. Second, spin relaxation is assumed to act only inside the 
ballistic transport intervals.  As a result, the thermoballistic 
spin-polarized current and density are connected by a spin balance equation, 
from which one obtains an inhomogeneous Fredholm-type integral equation 
for the spin accumulation function.      
In the general case of arbitrarily shaped potential energy profiles, 
considerable numerical effort is needed for solving the integral equations
for the resistance functions and the spin accumulation function. In a
number of important special cases, however, these equations reduce to a form
which greatly facilitates their solution. Examples are the equations 
for the resistance functions when the spin splitting of the band edge profile 
is independent of
position, or the equation for the spin accumulation function for a homogeneous
sample in an external electric field. In the latter case, the integral equation
can be converted into an easily tractable second-order differential equation.
For homogeneous semiconductors at zero bias, the solutions of the equations 
for the dynamical functions can be obtained in closed form throughout.
 
For the purpose of demonstrating the potentialities of the thermoballistic 
approach in
present-day semiconductor and spintronics research, we have summarized in this 
article the treatment of a number of specific examples. The prototype model 
is employed to describe electron transport across potential energy profiles 
exhibiting an arbitrary number of barriers, where effects of tunneling and 
degeneracy are included. This approach proves to be of particular relevance 
to the description of grain boundary effects in electron transport in 
polycrystalline semiconductors and has already been applied with promising
results in the analysis of experimental data. The 
full thermoballistic approach is used in the treatment of spin-polarized
transport in heterostructures. For the prototype problem of semiconductor 
spintronics, {\em viz.}, the injection of spin polarization from ferromagnetic 
contacts into a semiconducting sample, the thermoballistic description extends
the standard drift-diffusion description so as to allow for arbitrary values of
the momentum relaxation length. The same holds for transport in 
heterostructures composed of layers of diluted magnetic and nonmagnetic 
semiconductors at zero bias, where the position dependence of the current spin
polarization as well as the magnetoresistance are obtained in closed form.   
         
In our previous publications, we have presented results of explicit 
calculations for a variety of specific cases. These calculations were 
mostly of exploratory character, with the aim to reveal qualitative 
trends in the parameter dependence of the relevant transport properties.  In
future work, emphasis should be placed on the application of the 
thermoballistic approach in quantitative studies, as demanded for the analysis
of specific experimental results. For these studies to become successful,  the
careful evaluation of the potential energy profiles is prerequisite, and
advanced numerical techniques for solving the integral equations for 
the dynamical functions are to be employed. With these goals achieved, the
thermoballistic approach will certainly prove useful as a practical tool,  
beyond its basic theoretical relevance as the bridge between the 
drift-diffusion and ballistic descriptions of charge carrier transport in 
semiconductors. 

\acknowledgments 

We are indebted to the Board of Directors and the staff of the
Helmholtz-Zentrum Berlin f\"{u}r Materialien und Energie for generously 
granting access to its premises and facilities beyond the date of our 
retirement, thereby allowing us to continue and complete our research work on 
charge carrier transport in semiconductors.


\begin{thebibliography}{999}
\bibitem{foot1} After having introduced the term ''thermoballistic'' into the
theory of carrier transport in semiconductors, we became aware of
Ref.~\onlinecite{has71}, in which this term is used in quite a different
context.
\bibitem{chempo} Throughout this article, we use the term 
``chemical potential'' as defined in Ref.~\onlinecite{ash76}.
\bibitem{semcl} We use the term ``semiclassical'' essentially as defined in 
Chap.~12 of Ref.~\onlinecite{ash76}, where the band structure of the 
unperturbed system is assumed to be generated from quantum-mechanical 
calculations, whereas the effect of an externally applied electric field is 
treated classically (the effect of a static external magnetic field is assumed 
here to be included in the band edge potential). We go beyond this scope when 
considering corrections to the ballistic currents due to tunneling and 
degeneracy.
\end{thebibliography}
\end{document}